\documentclass[aps,prd,preprintnumbers,showpacs,showkeys,nofootinbib,
superscriptaddress,fleqn,floatfix,tightenlines,10pt]{revtex4-2}
\usepackage{amsmath,amsfonts,amssymb,amscd,amsxtra,amsthm}
\usepackage{graphicx}  
\usepackage{epstopdf}
\usepackage{dcolumn}  
\usepackage{bm}          
\usepackage{slashed}
\usepackage{cancel}
\usepackage{subfigure}
\usepackage{float} 
\usepackage{mathtools}
\usepackage{amsbsy}
\usepackage{amstext}

\usepackage[utf8]{inputenc} 
\usepackage{booktabs}
\usepackage[normalem]{ulem} 
\usepackage[dvipsnames]{xcolor} 
\usepackage{tabularx}
\usepackage{enumitem}  
\usepackage{array} 
\usepackage{slashed}
\usepackage{tikz}
\usepackage{float}
\usepackage{multirow}
\renewcommand\sout{\bgroup \color{red} \ULdepth=-.5ex \ULset}

\makeatletter

\begin{document}  
\preprint{INHA-NTG-01/2023}
\title{Two-dimensional charge distributions of the $\Delta$ baryon:
  Interpolation between the nonrelativistic and ultrarelativistic limit} 

\author{Ki-Hoon Hong}
\email[E-mail: ]{kihoon@inha.edu}
\affiliation{Department of Physics, Inha University, Incheon 22212,
Republic of Korea}

\author{June-Young Kim}
\email[E-mail: ]{jykim@jlab.org}
\affiliation{Theory Center, Jefferson Lab, Newport News, VA 23606, USA}

\author{Hyun-Chul Kim}
\email[E-mail: ]{hchkim@inha.ac.kr}
\affiliation{Department of Physics, Inha University, Incheon 22212,
Republic of Korea}
\affiliation{School of Physics, Korea Institute for Advanced Study
  (KIAS), Seoul 02455, Republic of Korea}

\date{\today}
\begin{abstract}
We investigate how the charge distributions of both the
unpolarized and transversely polarized $\Delta$ baryon 
change as the longitudinal momentum~($P_{z}$) of the $\Delta$ baryon
increases from $P_{z}=0$ to $P_{z}=\infty$ in a Wigner phase-space 
perspective. When the $\Delta$ baryon is longitudinally polarized, its 
two-dimensional charge distribution is kept to be spherically
symmetric with $P_{z}$ varied, whereas when the $\Delta$ baryon
is transversely polarized along the $x$-axis, the quadrupole
contribution emerges at the rest frame ($P_{z}=0$). When $P_{z}$
grows, the electric dipole and octupole moments are induced. The
induced dipole moment dominates over other higher multipole
contributions and governs the deformation of the charge distribution
of the $\Delta$ baryon.   
\end{abstract}
\maketitle

\section{Introduction}
The electromagnetic~(EM) form factors of the nucleon have been one of
the essential observables in understanding its structure well over 
decades. They provide crucial information on the charge and
magnetization distributions inside a nucleon. This interpretation assumes
that the nucleon is at rest in the Breit frame
(BF)~\cite{Sachs:1962zzc}. This assumption is valid only 
if the nucleon's spatial size $R_N$ were larger than the Compton
wavelength $1/M_N$, so the spatial wave 
functions could have been well defined. In reality, however,
the size of the nucleon is comparable to $1/M_N$, so the
nucleon wave function is no more localized below the Compton
wavelength. It causes ambiguous relativistic corrections that mar the
probabilistic interpretation of the 3D EM
distributions in the BF~\cite{Burkardt:2000za, Burkardt:2002hr, 
  Belitsky:2005qn}. This flaw of the 3D charge and
magnetization distributions was already pointed out in the
1950s~\cite{Yennie:1957}. To understand the EM distributions of the
nucleon without any ambiguity, one needs to view the nucleon from the 
light-front (LF) or, equivalently, the infinite momentum frame (IMF),
where the relativistic corrections are kinematically suppressed. Then,
the charge distribution emerges in the 
two-dimensional (2D) plane transverse to the nucleon momentum with the
probabilistic meaning properly borne~\cite{Burkardt:2000za,
  Burkardt:2002hr}.  It is obtained as the 2D Fourier transform of the
EM form factors and called the transverse charge distribution of the 
nucleon~\cite{Miller:2007uy, Carlson:2007xd}. The only problem with
the transverse charge distribution is that we lose information along
the infinite momentum direction. Since then, the transverse charge
distributions of the nucleon, $\Delta$ baryon, deuteron, pion, kaon,
and $\rho$ meson have been extensively studied~\cite{Strikman:2010pu,
  Venkat:2010by, Yakhshiev:2013goa, Granados:2013moa,
  Chakrabarti:2014dna, Silva:2013laa, Jung:2015piw, Mondal:2016xpk,
  Gramolin:2021gln, Alarcon:2022adi, Panteleeva:2022khw, 
  Alexandrou:2009hs, Chakrabarti:2016lod, Alharazin:2022xvp,
  Carlson:2009ovh, Mondal:2017lph, Huang:2017gih, Lorce:2022jyi,
  Miller:2009qu, Polyakov:2009je, Miller:2010tz, Nam:2011yw,
  Mecholsky:2017mpc, Kumar:2019eck, Fu:2022rkn, Epelbaum:2022fjc} (see also a
review~\cite{Alexandrou:2012da}).

When the transverse charge density of the neutron was
reported~\cite{Miller:2007uy}, many were perplexed by the result:
While the positive charge is centered in the 3D charge distribution of
the neutron, the negative one was situated in the center of the
neutron for the neutron 2D transverse charge density. 
Recently, Lorcé resolved the discrepancy by showing that 
when the longitudinal momentum increases from the rest to infinity, the
charge distribution in the transverse plane undergoes drastic changes
from the positive center value to the negative one~\cite{Lorce:2020onh}. 
As the longitudinal momentum grows, a Wigner rotation and a mixing of
the four-current components under Lorentz boost give rise to a
magnetization contribution~\cite{Lorce:2020onh,
  Chung:1988my,Rinehimer:2009sz}, which makes the sign of
the neutron transverse charge density is changed to be negative.  
In doing so, Lorc\'e introduced the elastic frame (EF) to interpolate
from the BF to the IMF in the Wigner phase-space perspective, which
makes it possible to observe the change in the charge distribution
explicitly as the longitudinal momentum increases. This approach was
extended to the case of the polarized nucleon~\cite{Kim:2021kum},
where the Abel tomography was emphasized. It was recently elaborated
and enlarged by considering
the EM distributions for the spin-0 and spin-1/2
particles~\cite{Chen:2022smg} and the EM and energy-momentum tensor
(EMT) distributions of the spin-1
particle~\cite{Lorce:2022jyi}. Compared to the spin-0 and -1/2
particles~\cite{Lorce:2020onh, Chen:2022smg, Kim:2022bia}, the 
spin-1 particle reveals rather complicated multipole
sturctures~\cite{Carlson:2009ovh, Lorce:2022jyi, Kim:2022bia}. In 
this work, we want to investigate the multipole structure of the EM
distributions for the spin-3/2 $\Delta$ baryon and see how they are
altered under Lorentz boost.  

The EM form factors of the $\Delta$ baryon can be parametrized in
terms of four multipole form factors~\cite{Nozawa:1990gt}~(see also a 
review~\cite{Pascalutsa:2006up}): electric monopole (E0), quadrupole
(E2), magnetic dipole (M1), and octupole (M3) ones. While it is
extremely difficult to measure them experimentally because of the
$\Delta$'s ephemeral nature, the $\Delta$ EM form factors and
corresponding transverse charge densities were computed in lattice
QCD~\cite{Nozawa:1990gt, Alexandrou:2009hs}. In the rest BF, we can
define four frame-dependent functions, which are respectively related
to the EM multipole form factors. The Lorentz boost induces
the electric dipole (E1) and octupole (E3) contributions to the
transverse charge densities of the $\Delta$ baryon. In the IMF, all
the frame-dependent functions coming from the third spatial component
of the EM four current become equivalent to those from the temporal
one. It is crucial to analyze these consequences arising
from the Lorentz boost. In this work, thus, we examine the expressions
for the $\Delta$ baryon matrix elements of the EM current in terms of
the frame-dependent functions defined in the 2D EF. They are given as
the functions of the momentum transfer $t$ and the longitudinal
momentum $P_{z}$. For any values of $P_{z}$, we are able to define the
frame-dependent functions and their 2D Fourier transforms, so each
contribution to the transverse $\Delta$ charge densities can be
examined with the $P_{z}$ given. If we take $P_{z}=0$, the
frame-dependent functions are reduced  
to the EM multipole form factors. To investigate
the transverse charge densities of the moving $\Delta$ baryon, we need 
information on the EM form factors. In the present work, we will take
the numerical results of the EM form factors obtained in the SU(3)
chiral quark-soliton model~\cite{Kim:2019gka}. We will then 
visualize  in the 2D space how the charge distributions are deformed
under the Lorentz boost. 

The present work is organized as follows: In Section II, we construct
the formalism for the multipole structure of the transverse charge
densities of the $\Delta$ baryon. In Section~III, we present the
numerical results for the transverse charge distributions
interpolating from the BF to the IMF. We also examine each
contribution of the multipole components to the transverse charge
distributions of the moving $\Delta^+$ and $\Delta^0$ and discuss it. 
In Section~IV, we summarize and draw conclusions of the current
work. In Appendix, we list the explicit expressions for the
frame-dependent functions.  
\section{Multipole structure of the transverse charge densities}
The matrix element of the EM current is defined as
\begin{align}
  \label{eq:LHcurrent}
J^\mu (x) = \bar{\psi} (x) \gamma^\mu \hat{\mathcal{Q}} \psi(x),   
\end{align}
where $\psi(x)$ denotes the quark field. The charge operator of the
quarks $\hat{\mathcal{Q}}$ is written in terms of the flavor SU(3)
Gell-Mann matrices $\lambda_3$ and $\lambda_8$ 
\begin{align}
 \label{eq:chargeOp}
\hat{\mathcal{Q}} =
  \begin{pmatrix}
   \frac23 & 0 & 0 \\ 0 & -\frac13 & 0 \\ 0 & 0 & -\frac13
  \end{pmatrix} = \frac12\left(\lambda_3 + 
                        \frac1{\sqrt{3}}  \lambda_8\right).
\end{align}
The matrix elements of the EM current between the $\Delta$ baryon
states with spin 3/2 can be parametrized in terms of four form factors
$F^{*}_i$ ($i=1,\cdots,4$) as follows:  
\begin{align}
\langle \Delta (p',\sigma') | e  J^{\mu}(0) | \Delta (p,\sigma) \rangle 
&= - e_{B} \overline{u}^{\alpha}(p',\sigma') \left[ \gamma^{\mu} \left \{
  F^{*}_{1}(t) g_{\alpha \beta} + F^{*}_{3}(t) \frac{ q_{\alpha} q_{\beta}
  }{4M_{\Delta}^{2}}  \right \}\right. \cr
&\hspace{2cm} \left.  + \,i\frac{\sigma^{\mu \nu} q_{\nu}}{2M_{\Delta}}
  \left \{ F^{*}_{2}(t) g_{\alpha \beta} + F^{*}_{4} (t)\frac{q_{\alpha}
  q_{\beta}}{4 M_{\Delta}^2}  \right \}  \right ]{u}^{\beta}(p,\sigma), 
\label{eq:4current}
\end{align}
where $M_{\Delta}$ denotes the mass of the $\Delta$ baryon, and $e_{B}$
stands for the corresponding electric charge in unit of $e$. $q$
designates the momentum transfer $q=p'-p$ and its square is given
as $q^2=t$ with $-t >0$. $u^\alpha (p,\,\sigma)$ represents the  
Rarita-Schwinger spinor, carrying the momentum $p$ and the spin
component $\sigma$ projected along the direction of the momentum. The
explicit expression for the Rarita-Schwinger spinor is given by 
\begin{align}
    u^\mu(p,\sigma)=\sum_{\lambda,s}
  C_{1\lambda\frac{1}{2}s}^{\frac{3}{2}\sigma}u_s(p)\epsilon_\lambda^\mu(p),\quad
  {\rm and}\quad u_s(p)=\sqrt{M_{\Delta}+p_0}
\left(\begin{array}{c}  
1 \\ 
\frac{\vec{\sigma}\cdot\vec{p}}{M_{\Delta}+p_0}
    \end{array}\right)\phi_s,
\end{align}
where $u_{s}(p)$ and $\phi_s$ stand for the Dirac and Pauli spinors
with its spin polarization $s$, respectively. Here, we 
choose the canonical spin states (see relevant
discussions~\cite{Polyzou:2012ut, Lorce:2019sbq, Lorce:2020onh}). By 
coupling the Dirac spinor to the spin-one polarization vector, one can
construct the Rarita-Schwinger spinor. The spin-one vector
$\epsilon^{\mu}$ in any frame is expressed by  
\begin{align}
    \epsilon_\lambda^\mu(p)=\left(\frac{\hat{e}_\lambda\cdot\vec{p}}{M_\Delta}, 
  \hat{e}_\lambda+\frac{\vec{p}(\hat{e}_\lambda\cdot\vec{p})}{M_\Delta
  (M_\Delta+P_0)}\right),
\end{align}
where $\hat{e}$ is the polarization vector in the rest frame.
\begin{align}
    &\hat{e}_{+1}=\frac{1}{\sqrt{2}}\left(-1,-i,0\right), 
\quad \hat{e}_{0}=(0,0,1), \quad
      \hat{e}_{-1}=\frac{1}{\sqrt{2}}\left(1,-i,0\right). 
\end{align}

In order to discuss the multipole structure of the EM form factors in
a systematical way, it is convenient to introduce the  rank-$n$
irreducible tensors and multipole operators in 2D space. The rank-$n$
irreducible tensors in coordinate (or momentum) space are defined by 
\begin{align}
    &X_{0}:=1, \quad X_n^{i_1\cdots i_n} := \frac{(-1)^{n+1}}{(2n-2)!!}
      x^n_\perp\partial^{i_1} \cdots \partial^{i_n} \ln x_\perp 
      \mbox{ with $n>0$, $i_n=1,2$}. 
\end{align}
For a spin-3/2 baryon, the quadrupole- and octupole-spin operators
$\hat{Q}^{ij}$(rank-2 tensor) and $\hat{O}^{ijk}$(rank-3 tensor)
appear in the matrix element of the 
EM current and are respectively defined in terms of the spin operator
$\hat{S}^{i}$ as follows: 
\begin{align}
\hat{Q}^{ij} & := \frac{1}{2}\left[ \hat{S}^{i}\hat{S}^{j}
               +\hat{S}^{j}\hat{S}^{i}
               -\frac{2}{3}S(S+1)\delta^{ij}\right], \cr 
\hat{O}^{ijk} & := \frac{1}{6}\bigg{[}
                \hat{S}^{i}\hat{S}^{j}\hat{S}^{k} + \hat{S}^{j}
                \hat{S}^{i} \hat{S}^{k} + \hat{S}^{k} \hat{S}^{j}
                \hat{S}^{i} + \hat{S}^{j} \hat{S}^{k} \hat{S}^{i} +
                \hat{S}^{i}
                \hat{S}^{k}\hat{S}^{j}+\hat{S}^{k}\hat{S}^{i}\hat{S}^{j} 
                \cr 
&\hspace{0.1cm}-\frac{6S(S+1)-2}{5}(\delta^{ij} \hat{S}^{k} +
  \delta^{ik} \hat{S}^{j}+\delta^{kj}\hat{S}^{i})\bigg{]}.
\end{align}
Since the tensor operators are irreducible, so they are fully
symmetrized under the exchanges of the indices $i,j,k=1,2,3$ and
traceless ($\hat{Q}^{ii}=0$ and $\delta_{ij}\hat{O}^{ijk}=0$). The
spin operators can be expressed in terms of SU(2) Clebsch-Gordan
coefficients in the spherical basis  
\begin{align}
\hat{S}^{a}_{\sigma'\sigma} = \sqrt{S(S+1)}C^{S \sigma'}_{S \sigma 1
  a} \ \ \ \mathrm{with} \ \ \ (a=0,\pm1. \  \  \sigma,\sigma'=0,
  \cdot\cdot\cdot,\pm S). 
\end{align}

To see how the matrix element of the EM current is changed under 
Lorentz boost, we need to employ the EF, where space-like momentum
transfer $\bm{q}$ lies in the transverse plane with conditions
$q^{0}=0$ and $\bm{P}\neq0$, as suggested in
Ref.~\cite{Lorce:2020onh}. Without loss of generality, in the EF, the
average momentum $P=(p'+p)/2$ and momentum transfer $q$ with the
on-shell constraint are taken to be  
\begin{align}
    &P=(P_0,\vec{0},P_z)\quad q=(0,\vec{q}_\perp,0), \quad
      P_0=\sqrt{(1+\tau)M_\Delta^2 +P_z^2},
\end{align}
with $\tau=-t/(4M_\Delta^2)$. Then, the matrix element of the temporal
component of the EM current $J^0$ in the EF from Eq.~\eqref{eq:4current}
is written in terms of the multipole $n$-rank irreducible tensors in
momentum space and in spin polarization together with the
frame-dependent scalar functions $G_{E0,E1,E2,E3}$: 
\begin{align}
            \frac{\langle J^0\rangle_{\sigma'\sigma}}{2P^0}
& =  \left\{G_{E0}  (t;P_{z}) - \frac23 \tau G_{E2}(t;P_{z}) \right\}
   \delta_{\sigma'\sigma}+\left\{        
G^{a'a}_{E0}(t;P_{z})+\frac{4}{3}\tau   G_{E2}(t;P_{z})\right\} \delta_{a'a}
   \cr 
& + 2\sqrt{\tau} \left\{ G_{E1}(t;P_{z}) - \frac25 \tau G_{E3}
  (t;P_{z}) \right\}i\epsilon^{ij3}
S_{\sigma'\sigma}^iX_1^j(\theta_{q_\perp})\cr
& + 2\sqrt{\tau}\left\{G^{a'a}_{E1}(t;P_{z})+\tau G_{E3}(t;P_{z})
  \right\} i\epsilon^{ij3}S_{a'a}^iX_1^j(\theta_{q_\perp}) \cr
& +\frac{4}{3}\tau G_{E2}(t;P_{z})Q_{\sigma'\sigma}^{ij}
X_2^{ij}(\theta_{q_\perp})+8\tau^{3/2}
  G_{E3}(t;P_{z})i\epsilon^{3jk}
O_{\sigma'\sigma}^{jml}X_3^{klm}(\theta_{q_\perp}),
\label{eq:j0}
\end{align}
where we introduce the following short-handed notation
$\delta_{a'a}=\delta_{\sigma'a'}\delta_{\sigma a}\delta_{\sigma'
  \sigma}$ with $a',a=-\frac{1}{2},\frac{1}{2}$ and $\langle
J^\mu\rangle_{\sigma'\sigma}:=\langle  \Delta (p', \sigma') |
\hat{J}^{\mu}(0) | \Delta(p, \sigma) \rangle$. Here, 
$\theta_{q_{\perp}}$ denotes the 2D angle of the
$q_{\perp}^i$ variable. 

Before discussing the EM form factor in the 2D EF, we want
briefly to mention the 2D and 3D BFs. To study the 3D spatial
distributions, the 3D BF is adopted, where $q^{0}=0$ and
$\bm{P}=0$. It yields the well-known Sach-type or multipole form
factors. Since we interpolate the 2D BF to 2D IMF distributions in
this work, we introduce the 2D BF, where each component of the four
momenta $P$ and $q$ are taken to be the same as in the 3D BF. It can
simply be achieved by taking $q_{z}=0$.

In the 3D BF, $\langle
J^{0}\rangle_{\sigma'\sigma}$ yields normally two contributions: the
electric monopole~($E0$) and electric quadrupole~($E2$) ones. However,
the projection from the 3D BF to the 2D one and the Lorentz boost
induce various contributions. Firstly, in the presence of the $E2$ contribution, the projection from the 3D
BF to the 2D one induces the monopole contribution. In addition, it is split into the spin-polarizations $\sigma=-\frac{3}{2},\cdots,\frac{2}{3}$ and   
its subsystem $a=-\frac{1}{2},\frac{1}{2}$. Secondly, under the Lorentz boost, the matrix
element of the temporal component of the EM current $J^{0}$ is subject
to the Wigner spin rotation and the admixture with the spatial
component of the EM current. It results in the induced electric
dipole~($E1$) and induces the $E3$ contributions, and the Lorentz
boost brings about the frame dependence on $P_{z}$ in the matrix
element of $J^{0}$. These effects from the Lorentz boost are conveyed
to the frame-dependent $G_{E0,E1,E2,E3}$ given as the functions of
$P_{z}$ and $t$. The explicit expressions for them are listed in   
Appendix~\ref{app:a}. In the BF ($P_{z}=0$), these frame-dependent
functions are reduced to the 2D BF expressions: 
\begin{align}
    \frac{\langle J^0
  \rangle_{\sigma'\sigma}}{2P^{0}}\;{\buildrel{P_z\rightarrow
  0}\over=}
  \frac{1}{\sqrt{1+\tau}}\left[\left(G_{E0}(t)+\frac{1}{3}\tau
  G_{E2}(t)\right)\delta_{\sigma'\sigma}-\frac{2\tau}{3}
  G_{E2}(t)\delta_{a'a}-\frac{2\tau}{3}
  G_{E2}(t)Q_{\sigma'\sigma}^{ij}X_2^{ij}(\theta_{\Delta_\perp})\right]. 
\end{align}
In the 2D BF limit, we recover the traditional definitions of
the Sach-type EM form factors together with the relativistic factor
$1/\sqrt{1+\tau}$, which comes into play when interpolating the BF
expressions to the IMF ones: 
\begin{align}
&G_{E0}(t;P_{z}=0) = \frac{1}{\sqrt{1+\tau}}G_{E0}(t), \quad
  G^{a'a}_{E0}(t;P_{z}=0) = 0,\quad G_{E1}(t;P_{z}=0) = 0, \cr 
 &G^{a'a}_{E1}(t;P_{z}=0) = 0, \quad G_{E2}(t;P_{z}=0)
   =-\frac{1}{2\sqrt{1+\tau}}G_{E2}(t), \quad G_{E3}(t;P_{z}=0) = 0, 
\end{align}
where the EM multipole form factors are expressed in terms of
$F^{*}_{i}$:  
\begin{align}
&G_{E0}(t) = \left(1 + \frac{2}{3}
  \tau\right)\bigg{[}F^{*}_{1}(t)-\tau F^{*}_{2}(t)\bigg{]} - 
  \frac{1}{3} \tau (1+ \tau) \bigg{[}F^{*}_{3}(t) - \tau
  F^{*}_{4}(t)\bigg{]}, \cr  
&G_{E2}(t) = \bigg{[}F^{*}_{1}(t)-\tau F^{*}_{2}(t)\bigg{]}-
  \frac{1}{2} (1+ \tau) 
  \bigg{[}F^{*}_{3}(t) - \tau F^{*}_{4}(t)\bigg{]},  \cr 
&G_{M1}(t) = \left(1+\frac{4}{5}
  \tau\right)\bigg{[}F^{*}_{1}(t)+F^{*}_{2}(t)\bigg{]}  - 
  \frac{2}{5}  \tau (1+
  \tau)\bigg{[}F^{*}_{3}(t)+F^{*}_{4}(t)\bigg{]}, \cr  
&G_{M3}(t) =  \bigg{[}F^{*}_{1}(t)+F^{*}_{2}(t)\bigg{]}
  -\frac{1}{2}(1+\tau) \bigg{[}F^{*}_{3}(t)+F^{*}_{4}(t)\bigg{]}.  
  \label{eq:sach}
\end{align}
They are called, respectively, the electric monopole~($E0$), electric
quadrupole~($E2$), magnetic dipole~($M1$), and magnetic
octupole~($M3$) form factors. The $M1$ and $M3$ form 
factors will be obtained in the matrix element of the spatial
component of the EM current $J^{i}$. By taking $P_{z}\to\infty$ in
Eq.\eqref{eq:j0}, we can naturally recover the results 
from the LF formalism~\cite{Alexandrou:2009hs}: 
\begin{align}
&G_{E0}(t;P_{z}\to\infty) =
  \frac{1}{1+\tau}\left(G_{E0}(t)+\frac{1}{3}\tau
  G_{M1}(t)-\frac{4}{15}\tau^2G_{M3}(t)\right),\cr 
&G^{a'a}_{E0}(t;P_{z}\to\infty) =
  -\frac{4\tau}{(1+\tau)^2}\left[G_{E0}(t)+\frac{1}{3}\tau
  G_{E2}(t)-\frac{1}{3}\left(2-\tau\right)G_{M1}(t)
  -\frac{\tau}{15}(2-\tau)G_{M3}(t)\right],\cr  
&G_{E1}(t;P_{z}\to\infty) =
  -\frac{1}{(1+\tau)^2}\left[\left(1+\frac{\tau}{15}\right)G_{E0}(t)
+\frac{2\tau}{15}\left(2-\frac{\tau}{3}\right)G_{E2}(t)
-\frac{1}{3}\left(1-\frac{9}{5}\tau\right)G_{M1}(t)
+\frac{14}{75}\tau^2G_{M3}(t)\right],\cr 
&G^{a'a}_{E1}(t;P_{z}\to\infty) = \frac{2\tau}{3(1+\tau)^2}
\left[G_{E0}(t)-G_{M1}(t)+\frac{\tau}{3}G_{E2}(t)-
\frac{\tau}{5}G_{M3}(t)\right], \cr 
&G_{E2}(t;P_{z}\to\infty) = \frac{1}{2(1+\tau)^2}
\left[3G_{E0}(t)-G_{E2}(t)-(2-\tau)G_{M1}(t)-\frac{\tau}{5}
\left(7+4\tau\right)G_{M3}(t)\right],\cr 
&G_{E3}(t;P_{z}\to\infty) = -\frac{1}{6(1+\tau)^2}
\left[G_{E0}(t)-G_{E2}(t)\left(1+\frac{2\tau}{3}\right)-
G_{M1}(t)+\left(1+\frac{4\tau}{5}\right)G_{M3}(t)\right].
\label{eq:j0_IMF}
\end{align}
At the zero momentum transfer $t=0$ in Eq.\eqref{eq:j0_IMF}, we have
\begin{align}
&G_{E0}(t;P_{z}\to\infty) = G_{E0}(t),\quad 
G^{a'a}_{E0}(t;P_{z}\to\infty) = 0, \quad 
G_{E1}(t;P_{z}\to\infty) =
  -\left[G_{E0}(0)-\frac{1}{3}G_{M1}(0)\right], \cr 
&G^{a'a}_{E1}(t;P_{z}\to\infty) = 0, \quad 
G_{E2}(t;P_{z}\to\infty) = \frac{1}{2}\bigg{[}3G_{E0}(0)
-G_{E2}(0)-2G_{M1}(0)\bigg{]},\cr 
&G_{E3}(t;P_{z}\to\infty) = -\frac{1}{6}\bigg{[}
G_{E0}(0)-G_{E2}(0)-G_{M1}(0)+G_{M3}(0)\bigg{]}.
\end{align}
The above results are consistent with those in
Ref.~\cite{Alexandrou:2009hs}. It is also interesting to study how the 
spatial components of the EM current varies under the Lorentz
boost. Since we take the $z$-axis as a boost direction, $J^{3}$ and
$J^{i}_{\perp}$ with $i=1,2$ will behave differently under the Lorentz
boost. In the 2D EF, the matrix element of the transverse component of
the EM current $J^{i}_{\perp}$ is given by 
\begin{align}
    \frac{\langle J^i_\perp\rangle_{\sigma' \sigma} }{2P_0}
&=2\sqrt{\tau}\left[G_{M1}^\perp(t;P_{z})-\frac{1}{5}\tau 
G_{M3}^\perp (t;P_{z})\right]i\epsilon^{i3k}S^3 _{\sigma'\sigma}
X_1^k (\theta_{\Delta_{\perp}}) \cr
    &+2\sqrt{\tau}\left[G_{M1}^{\perp,a'a}(t;P_{z})+
2\tau G_{M3}^\perp (t;P_{z})\right]i\epsilon^{i3k}S_{a'a}^3 
X_1^k (\theta_{\Delta_{\perp}})+4\tau G_{M2}^\perp (t;P_{z}) 
Q_{\sigma'\sigma}^{l3}X_2^{li}(\theta_{\Delta_{\perp}}) \cr
    &-2\tau G_{M2}^\perp (t;P_{z}) Q_{\sigma'\sigma}^{i3}+
4\tau\sqrt{\tau}G_{M3}^\perp (t;P_{z}) i\epsilon^{i3k}
\left(2O^{3ml}_{\sigma'\sigma}X_3^{klm} (\theta_{\Delta_{\perp}})+
O^{3kl}_{\sigma'\sigma}X_1^l (\theta_{\Delta_{\perp}})\right).
\end{align}
The frame-dependent functions from the transverse
components of the EM current are labeled by $\perp$ in the
superscript. The matrix element of the transverse EM current 
yields the magnetic dipole and octupole contributions together
with the induced magnetic quadrupole one. In the 2D BF limit, the
frame-dependent functions
$G^{\perp}_{M1},G^{a'a,\perp}_{M1},G^{\perp}_{M2},G^{\perp}_{M3}$ are 
reduced to the Sach-type magnetic dipole (M1) and magnetic octopole
(M1) form factors given in Eq.~\eqref{eq:sach}: 
\begin{align}
  \frac{\langle J_\perp^i\rangle_{\sigma' \sigma}}{2P_0}
&\;{\buildrel{P_z\rightarrow 0}\over=}\frac{2}{3}
\sqrt{\frac{\tau}{1+\tau}}\left(G_{M1}(t)-
\frac{\tau}{10}G_{M3}(t)\right)i\epsilon^{i3k}
S_{\sigma'\sigma}^3 X_1^k (\theta_{\Delta_{\perp}})-
\frac{2}{3}\tau\sqrt{\frac{\tau}{1+\tau}}G_{M3}(t)
i\epsilon^{i3k}S^3_{a'a}X_1^k (\theta_{\Delta_{\perp}})\cr
  &-\frac{2}{3}\tau\sqrt{\frac{\tau}{1+\tau}}G_{M3}(t)
i\epsilon^{i3k}\bigg{(}2O^{3ml}_{\sigma'\sigma}
X_3^{klm} (\theta_{\Delta_{\perp}})+O^{3kl}_{\sigma'\sigma}
X_1^l (\theta_{\Delta_{\perp}})\bigg{)},
  \label{eq:Jperp_BF}
\end{align}
where 
\begin{align}
  &G_{M1}^\perp(t;P_z=0)=\frac{1}{3\sqrt{1+\tau}} G_{M1}(t),\quad
    G_{M1}^{\perp,a'a}(t;P_z=0)=0,\cr 
  &G_{M2}^\perp(t;P_z=0)=0,\quad 
G_{M3}^\perp(t;P_z=0) =-\frac{1}{6\sqrt{1+\tau}}G_{M3}(t).
\end{align}
Note that the induced magnetic dipole contribution $G_{M2}^\perp$
vanishes. Thus, we can regain the results from the 
LF formalism~\cite{Alexandrou:2009hs} as in the electric case. We
observe that the unusual structure $O^{3kl} X^{l}_{1}$ in the last term in 
Eq.~\eqref{eq:Jperp_BF} is induced by the projection from the 3D space
to 2D one. As shown in the case of the nucleon~\cite{Chen:2022smg},
all the relevant frame-dependent functions go to zero in
the IMF due to the $P_{z}$ suppression: 
\begin{align}
  &G_{M1}^\perp(t;P_z\rightarrow \infty)=0,\quad
    G_{M1}^{\perp,a'a}(t;P_z\rightarrow \infty)=0,\quad
    G_{M2}^\perp(t;P_z\rightarrow \infty)=0,\quad
    G_{M3}^\perp(t;P_z\rightarrow \infty)=0, 
\end{align}
so that the matrix element of the transverse components of the EM
current $J^{i}_{\perp}$ becomes zero in the IMF, i.e., $\frac{\langle
  J_\perp^i\rangle}{2P_0} \;{\buildrel{P_z\rightarrow
    \infty}\over=}0$. Lastly, we obtained the expression of the matrix
element of the $z$-component of the EM current as follows 
\begin{align}
\frac{\langle J^3\rangle_{\sigma' \sigma}}{2P_0}
&=\bigg{[}G^{3}_{M0}(t;P_{z})-\frac{2}{3}\tau G_{M2}^3-4\tau
  G_{M2}^\perp   \bigg{]}\delta_{\sigma'\sigma}+
\bigg{[}G_{M0}^{3,a'a}(t;P_{z})+\frac{4}{3}\tau
  G_{M2}^3(t;P_{z})+8\tau G_{M2}^\perp (t;P_{z})
  \bigg{]}\delta_{a'a}\cr  
&+2\sqrt{\tau}\left[G_{M1}^3-\frac{2}{5}\tau G_{M3}^3\right]
i\epsilon^{3jk}S^j_{\sigma'\sigma}X_1^k (\theta_{\Delta_{\perp}})+
2\sqrt{\tau}\left[G_{M1}^{3,a'a}+\tau G_{M3}^3\right]
i\epsilon^{3jk}S_{a'a}^j X_1^k (\theta_{\Delta_{\perp}})\cr
&+\frac{4}{3}\tau G_{M2}^3 Q_{\sigma'\sigma}^{lm}
X_2^{lm}(\theta_{\Delta_{\perp}})+8\tau\sqrt{\tau} 
G_{M3}^3i\epsilon^{3jk}O^{jml}_{\sigma'\sigma}
X_3^{klm}(\theta_{\Delta_{\perp}}).
\end{align}
The frame-dependent functions from the $z$-component of the EM current
are labeled by the $3$ in the superscript. The $J^3$ matrix element
produces the M1 and M3 contributions together with
the induced M0 and M2 contributions. Similar to
the transverse component of the EM current $J^{i}_{\perp}$, the
$z$-component of the EM current is reduced to the Sach-type magnetic
dipole and octupole form factors at $P_z=0$, and the other
frame-dependent functions vanish: 
\begin{align}
  \frac{\langle J^3\rangle_{\sigma' \sigma}}{2P_0}
&\;{\buildrel{P_z\rightarrow 0}\over=}\frac{2}{3}
\sqrt{\frac{\tau}{1+\tau}}\left(G_{M1}+\frac{\tau}{5}G_{M3}
\right)i\epsilon^{3jk}S_{\sigma'\sigma}^j X_1^k-
\frac{1}{3}\tau\sqrt{\frac{\tau}{1+\tau}}G_{M3}
i\epsilon^{3jk}S_{a'a}^j X_1^k\cr
  &-\frac{4}{3}\tau\sqrt{\frac{\tau}{1+\tau}}
G_{M3}i\epsilon^{3jk}O_{\sigma'\sigma}^{jml}X_3^{kml},
\end{align}
where
\begin{align}
  &G^{3}_{M0}(t;P_z=0)=0,\quad 
G_{M1}^3(t;P_z=0)=\frac{1}{3\sqrt{1+\tau}}G_{M1},\quad 
G_{M1}^{3,a'a}(t;P_z=0)=0,\cr
  &G_{M0}^{3,a'a}(t;P_z=0)=0,\quad 
G_{M2}^3(t;P_z=0)=0,\quad 
G_{M3}^3(t;P_z=0)=-\frac{1}{6\sqrt{1+\tau}}G_{M3}.
\end{align}
In the IMF, all the frame-dependent functions of the $J^{3}$ turn out
to be equivalent to those of the $J^{0}$: 
\begin{align}
  G_{M0}^{3}(t;P_z\rightarrow\infty)&
=G_{E0}(t;P_z\rightarrow\infty),   \quad
   G_{M0}^{3,a'a}(t;P_z\rightarrow\infty)=
G_{E0}^{a'a}(t;P_z\rightarrow\infty),\cr 
  G_{M1}^3(t;P_z\rightarrow\infty)
&=G_{E1}(t;P_z\rightarrow\infty), \quad 
G_{M1}^{3,a'a}(t;P_z\rightarrow\infty)=
G_{E1}^{a'a}(t;P_z\rightarrow\infty),\cr
  G_{M2}^3(t;P_z\rightarrow\infty)
&=G_{E2}(t;P_z\rightarrow\infty), \quad 
G_{M3}^3(t;P_z\rightarrow\infty)=G_{E3}(t;P_z\rightarrow\infty), 
\end{align}
so we have $  \frac{\langle
  J^3\rangle}{2P_0}\;{\buildrel{P_z\rightarrow 
    \infty}\over=}\frac{\langle J^0\rangle}{2P_0}$. A similar relation
for the nucleon was first derived in Ref.~\cite{Chen:2022smg}, and we
see that such a relation is also satisfied for the $\Delta$ baryon as
shown in the current work. 

We are now in a position to define the transverse charge
distributions. In this work, we will consider the temporal component 
of the EM current only, i.e., $J^{0}$. In the BF, the 3D distribution
is traditionally defined as a 3D Fourier transformation of the
corresponding form factor. As mentioned in the Introduction, the
baryon cannot be localized below the Compton wavelength, which causes
ambiguous relativistic corrections. Recently, these 3D distributions
in the BF and the 2D distributions of the moving baryon in the EF were
understood as quasi-probabilistic distributions in the phase space or
the Wigner distributions~\cite{Lorce:2018zpf, Lorce:2018egm,
  Lorce:2020onh, Lorce:2022jyi}. We will first construct the
transverse charge distribution of the moving $\Delta$ baryon by
introducing the EF and will show the connection between the 2D BF and
2D IMF distributions.   

In the Wigner phase-space perspective, the Fourier transform of the
matrix element of the EM current conveys information on the internal
structure of the particle. Since the average momentum and
momentum transfer of the initial and final states are 
respectively given by $P= (P_{0}, \bm{0}_{\perp}, P_{z})$ and $q=
(0, \bm{q}_{\perp}, 0)$ in EF, the EF distributions depend
on the impact parameter $\bm{x}_{\perp}$ and momentum
$\bm{P}=(\bm{0},P_{z})$, where the $\Delta$ baryon moves along the
$z$-direction without loss of generality. Thus, the charge
distribution can be expressed as the 2D Fourier transform of the
matrix element $\langle  \Delta (p', \sigma') | \hat{J}^{\mu}(0) |
\Delta (p,\sigma) \rangle$:    
\begin{align}
\rho_{\rm ch}(\bm{x}_\perp,\sigma',\sigma;P_{z})
&=\int\frac{d^2q_{\perp}}{(2\pi)^2}
\frac{\langle J^0\rangle_{\sigma'\sigma}}{2P^0}
e^{-i\vec{q}_{\perp}\cdot \vec{x}_{\perp}}\cr
&=\rho_{0}(x_\perp;P_{z})\delta_{\sigma'\sigma}+
\rho^{a'a}_{0}(x_\perp;P_{z})\delta_{a'a}\cr
            &+\rho_{1}(x_\perp;P_{z})\epsilon^{ij3}
X_1^j(\theta_{x_\perp})S_{\sigma'\sigma}^i+
\rho^{a'a}_{1}(x_\perp;P_{z}) \epsilon^{ij3}
X_1^j(\theta_{x_\perp})S_{a'a}^i\cr
            &+\rho_{2}(x_\perp;P_{z})Q_{\sigma'\sigma}^{ij}
X_2^{ij}(\theta_{x_\perp})+\rho_{3}(x_\perp;P_{z})
\epsilon^{3jk}O^{jml}_{\sigma'\sigma}X^{klm}(\theta_{x_\perp}),
\label{eq:chargedist}
\end{align}
where
\begin{align}
&\rho_{0}(x_\perp;P_{z})=\tilde{G}_{0}(x_\perp;P_{z}), \quad 
\rho^{a'a}_{0}=\tilde{G}^{a'a}_{0}(x_\perp;P_{z}), \quad 
\rho_{1}(x_\perp;P_{z})=-\frac{1}{M_\Delta}\frac{d}{dx_\perp}
\tilde{G}_{1}(x_\perp;P_{z}), \cr
  &\rho^{a'a}_{1}(x_\perp;P_{z})=-\frac{1}{M_\Delta}
\frac{d}{dx_\perp}\tilde{G}^{a'a}_{1}(x_\perp;P_{z}), \quad 
\rho_{2}(x_\perp;P_{z})=-\frac{1}{3M_\Delta^2}x_\perp
\frac{d}{dx_\perp}\frac{1}{x_\perp}\frac{d}{dx_\perp}
\tilde{G}_{2}(x_\perp;P_{z}), \cr
  &\rho_{3}(x_\perp;P_{z})=\frac{1}{M_\Delta^3}x_\perp^2
\frac{d}{dx_\perp}\frac{1}{x_\perp}\frac{d}{dx_\perp}
\frac{1}{x_\perp}\frac{d}{dx_\perp}\tilde{G}_{3}(x_\perp;P_{z}).
\end{align}
The variable $\theta_{x_{\perp}}$ denotes the 2D angle of the
$x^{i}_{\perp}$. Here we have used the 2D Fourier transform of the
generic function
$F=\{G_{0},G^{a'a}_{0},G_{1},G^{a'a}_{1},G_{2},G_{3}\}$:   
\begin{align}
&\int
  \frac{d^2q_{\perp}}{(2\pi)^2}e^{-i\vec{q}_{\perp}\cdot\vec{x}_\perp}
F(t;P_{z})=\tilde{F}(x_\perp;P_{z}),
\end{align}
where we define the following functions for convenience
\begin{align}
&G_{0}(t;P_{z})=\left\{G_{E0}(t;P_{z})-\frac{2}{3}\tau G_{E2}(t;P_{z})
  \right\},  
\quad G^{a'a}_{0}(t;P_{z})=\left\{ G^{a'a}_{E0}(t;P_{z})+
\frac{4}{3}\tau G_{E2}(t;P_{z}) \right\},\cr
&G_{1}(t;P_{z})=\left\{G_{E1}(t;P_{z})-\frac{2}{5}\tau 
G_{E3}(t;P_{z}) \right\}, \quad G^{a'a}_{1}(t;P_{z})=
\left\{G^{a'a}_{E1}(t;P_{z})+\tau G_{E3}(t;P_{z}) \right\}, \cr
&G_{2}(t;P_{z})=G_{E2}(t;P_{z}), \quad 
G_{3}(t;P_{z})=G_{E3}(t;P_{z}).
\end{align}
Thus, one can clearly see that the multipole patterns $\rho_{\mathrm{mon}},\rho_{\mathrm{dip}},\rho_{\mathrm{quad}}$, and $\rho_{\mathrm{oct}}$ of the charge distributions are given by the combinations of the $\rho_{0},\rho_{1},\rho_{2}$, and $\rho_{3}$.

\section{Numerical Results and discussions}
In this Section, We present the numerical results of the transverse
charge distribution of the spin-3/2 baryon and discuss them. 
We consider those of $\Delta^{+}$ and $\Delta^{0}$, regarding them as 
representatives for a spin-3/2 baryon. To study the charge
distribution in the Wigner phase-space perspective, we need
information on the $\Delta$ EM form factors. While there is a plenty
of available experimental data on the EM form factors of the nucleon,
those of the $\Delta$ baryons are almost unexplored on the
experimental side due to their short-lived nature. One
could import the lattice data~\cite{Alexandrou:2009hs} but they did
not consider the EM form factors of the $\Delta^{0}$, of which the
transverse charge distribution undergo a remarkable change under the
Lorentz boost as in the case of the neutron~\cite{Miller:2007uy,
  Lorce:2020onh}. Thus, we will take the results from the SU(3)
chiral quark-soliton model ($\chi$QSM)~\cite{Kim:2019gka}, where the available data of the EM form factors of the baryon decuplet exist. Note that the electric
monopole, quadrupole, and magnetic dipole were calculated in the
$\chi$QSM, but the magnetic octopole was ignored. This form factor is
strongly suppressed in the large $N_{c}$ expansion, which is
consistent with the lattice QCD data on the $G_{M3}$ form
factor~\cite{Alexandrou:2009hs}. It is compatible with zero within
the statistical accuracy.

\begin{figure}[htpb]
    \centering
    \subfigure[$s_z=3/2$]{\includegraphics[width=0.48\linewidth]{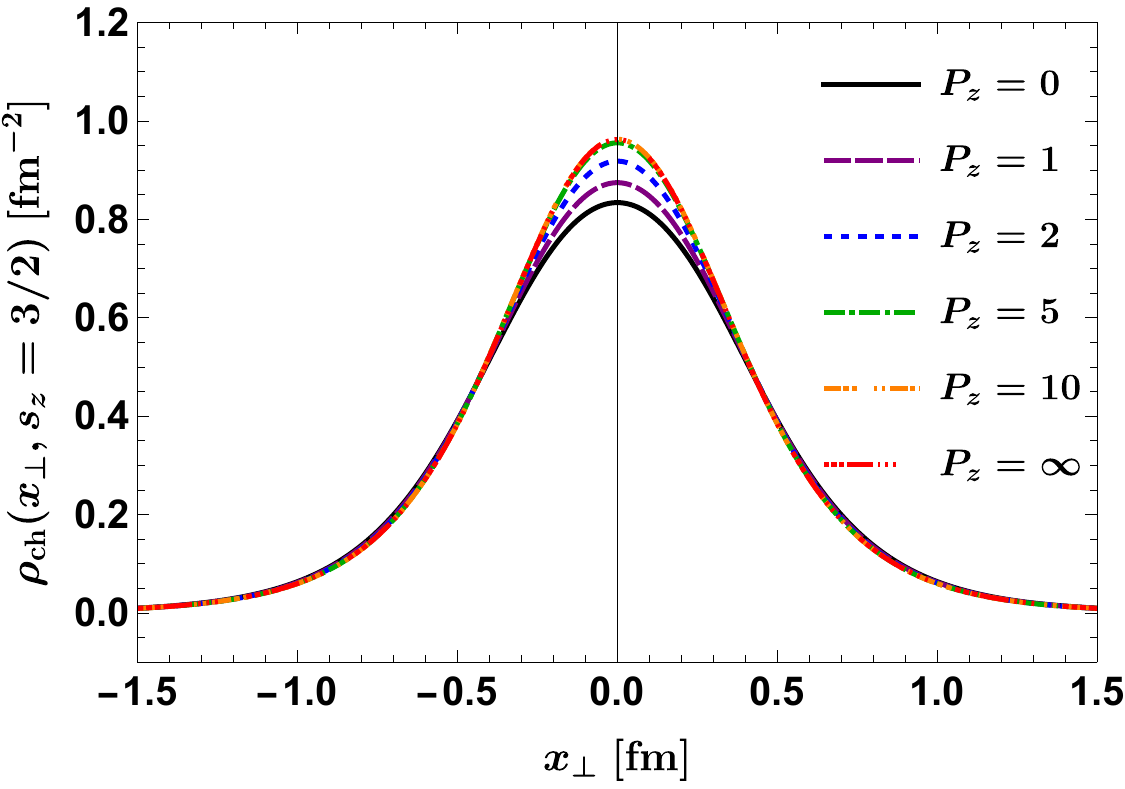}} 
    \subfigure[$s_z=1/2$]{\includegraphics[width=0.48\linewidth]{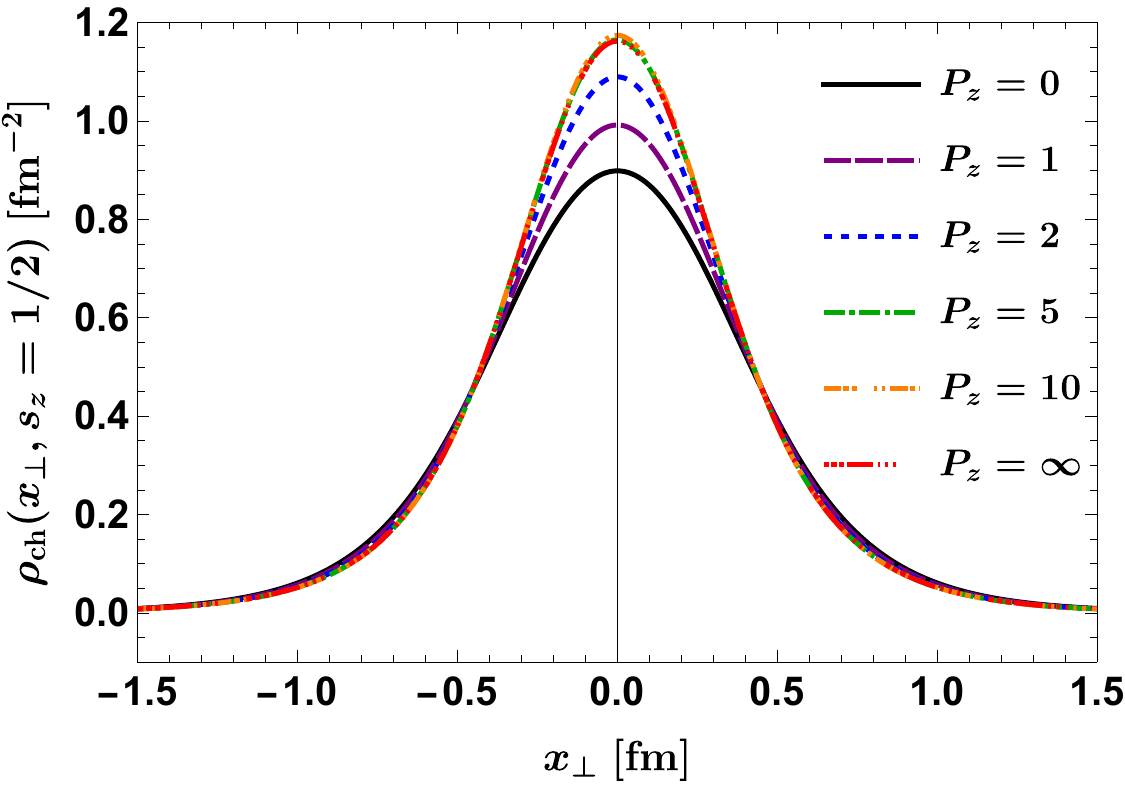}}
    \caption{The $y-$axis profiles of the transverse charge
      distributions of the moving $\Delta^+$ baryon as the 
      longitudinal momentum $P_{z}$ increases from $P_z=0$ to
      $P_z=\infty$. Its spin is polarized along the $z$-axis with
      $s_{z}=3/2$ and $s_{z}=1/2$, respectively. In the left (right) panel,
      $\rho_{\mathrm{ch}}$ with $s_z=3/2$ ($s_z=1/2$) is
      depicted.} 
\label{fig:1}  
\end{figure}

In Fig.~\ref{fig:1}, we show the $y-$axis profiles of the transvere
charge distribution of the moving $\Delta^+$ baryon with the
longitudinal momentum $P_{z}$ varied from $P_z=0$ to $P_z=\infty$. Its
spin is polarized along the $z$-axis with $s_{z}=3/2$ and $s_{z}=1/2$,
respectively. Taking $P_{z}=0$, we obtain the 2D BF charge
distribution. Here one should keep in mind that the 2D BF distribution
is distinctive from the 3D one~\cite{Kim:2022bia}. By carrying out the
Abel transformation, one can project out the 2D distribution from the
3D one. In the projection, the quadrupole structure induces the 
monopole contribution, so the monopole charge distribution is
subjected to the quadrupole contribution~\cite{Kim:2022bia,
  Kim:2022wkc}. In addition, this projection brings about the
spin-polarization dependence of the monopole charge distribution. In
addition, under the Lorentz boost, the transverse charge distributions
with $s_{z}=1/2$ and $s_{z}=3/2$ are altered in a different
manner. As demonstrated in Fig.~\ref{fig:1}, $\rho_{\mathrm{ch}}$ with
  $s_{z}=1/2$ changes stronger than that with $s_{z}=3/2$ as $P_z$
increases.  

\begin{figure}[htpb]
    \centering
    \subfigure[$s_z=3/2$]{\includegraphics[width=0.48\linewidth]{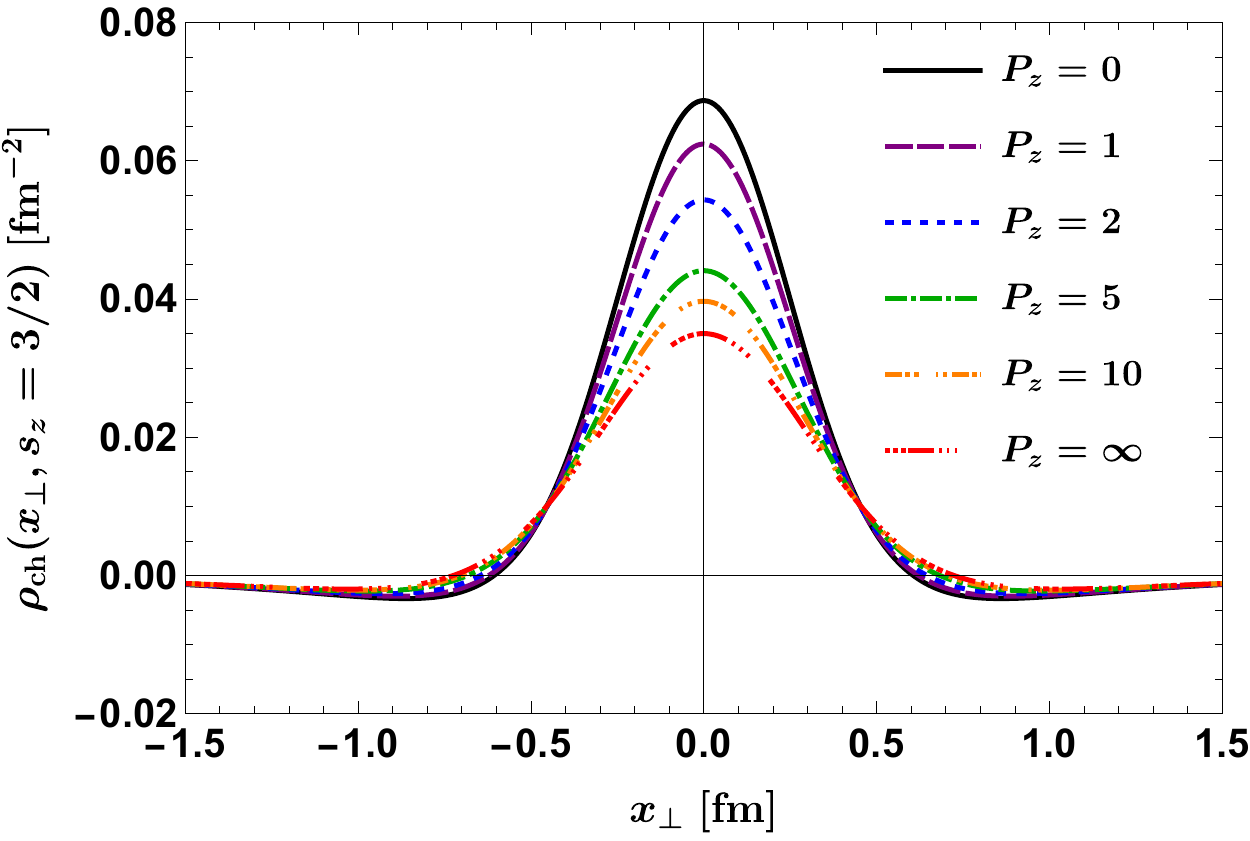}}
    \subfigure[$s_z=1/2$]{\includegraphics[width=0.48\linewidth]{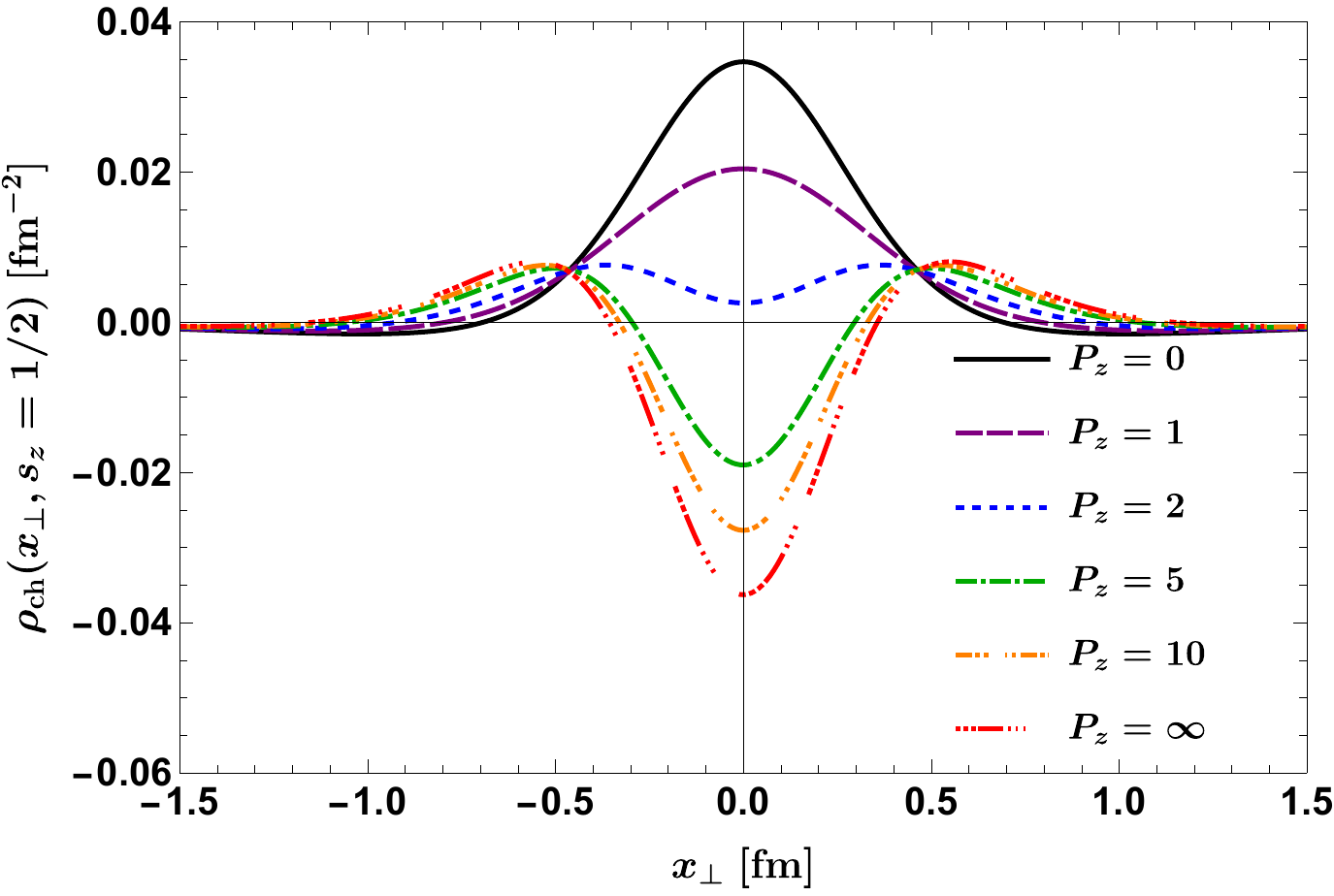}}
    \caption{The $y-$axis profiles of the transverse charge
      distributions of the moving $\Delta^0$ baryon as the 
      longitudinal momentum $P_{z}$ increases from $P_z=0$ to
      $P_z=\infty$. Its spin is polarized along the $z$-axis with
      $s_{z}=3/2$ and $s_{z}=1/2$, respectively. In the left (right) panel,
      $\rho_{\mathrm{ch}}$ with $s_z=3/2$ ($s_z=1/2$) is
      depicted.}
\label{fig:2} 
\end{figure}

In Fig.~\ref{fig:2}, we draw the $y-$axis profiles of the transverse
charge distributions of the moving $\Delta^0$ baryon as $P_{z}$
increases from $P_z=0$ to $P_z=\infty$. Again, its spin is polarized
along the $z$-axis with $s_{z}=3/2$ and $s_{z}=1/2$, respectively. We
observe that the relativistic effects (or Lorentz-boost effects) are
prominent in the neutral $\Delta^{0}$ baryon. Note that the
transverse charge distribution of the neutral $\Delta^0$ is normalized
to its zero charge. It indicates that $\rho_{\mathrm{ch}}$ must 
at least have one nodal point. For the $s_{z}=3/2$, the charge
distribution spread widely and its nodal point 
is placed at a distance. As $P_{z}$ increases from $P_{z}=0$ to
$P_{z}=\infty$, the core part of the charge distribution
gets weaker, whereas the tail part slowly gets lessened. So, the nodal
point moves away to the outer part of the baryon. When it comes to
$s_{z}=1/2$, the configuration of the transverse charge distribution
is dramatically changed under the Loretz boost. In the rest frame
($P_{z}=0$), the center of the $\Delta^{0}$ baryon is positively
charged, whereas the outer part is negatively charged. When the system
is boosted, the positive core gets weaker and then turns negative at around
$P_{z}\sim2.8\,\mathrm{GeV}$. It is very similar to the behavior of
the transverse neutron charge distribution under the Lorentz
boost~\cite{Lorce:2020onh}.  

If the baryon is longitudinally polarized, then one can get 
access to the electric monopole and quadrupole form factors only. To see the
additional contributions from the other form 
factors, the spin of the $\Delta$ baryon should be polarized
transversely. We express the transverse spin basis $s_{x}$ in
terms of the $s_{z}$ basis. Then the spin states $s_{x}=1/2$ and
$s_{x}=3/2$ are given~\cite{Alexandrou:2009hs} by
\begin{align}
    &|s_x=3/2\rangle=\frac{1}{\sqrt{8}} \left(
|s_{z}=3/2\rangle+\sqrt{3}|s_{z}=1/2\rangle+\sqrt{3}|
s_{z}=-1/2\rangle+|s_{z}=-3/2\rangle\right), \cr
    &|s_x=1/2\rangle=\frac{1}{\sqrt{8}}\left(\sqrt{3}|
s_{z}=3/2\rangle+|s_{z}=1/2\rangle-|s_{z}=-1/2\rangle-
\sqrt{3}|s_{z}=-3/2\rangle\right).
\end{align}

\begin{figure}[htp]
    \centering  
    \subfigure[Monopole,
    $s_x=3/2$]{\includegraphics[width=0.48\linewidth]{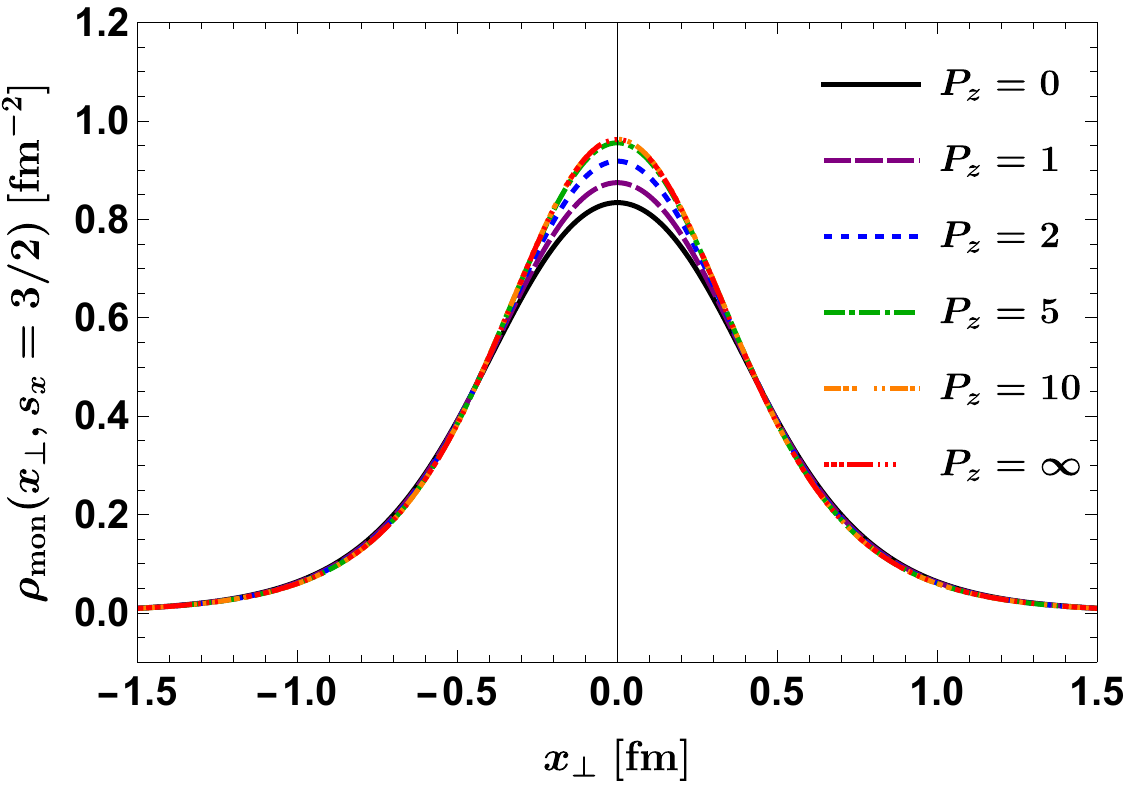}} 
    \subfigure[Dipole,
    $s_x=3/2$]{\includegraphics[width=0.48\linewidth]{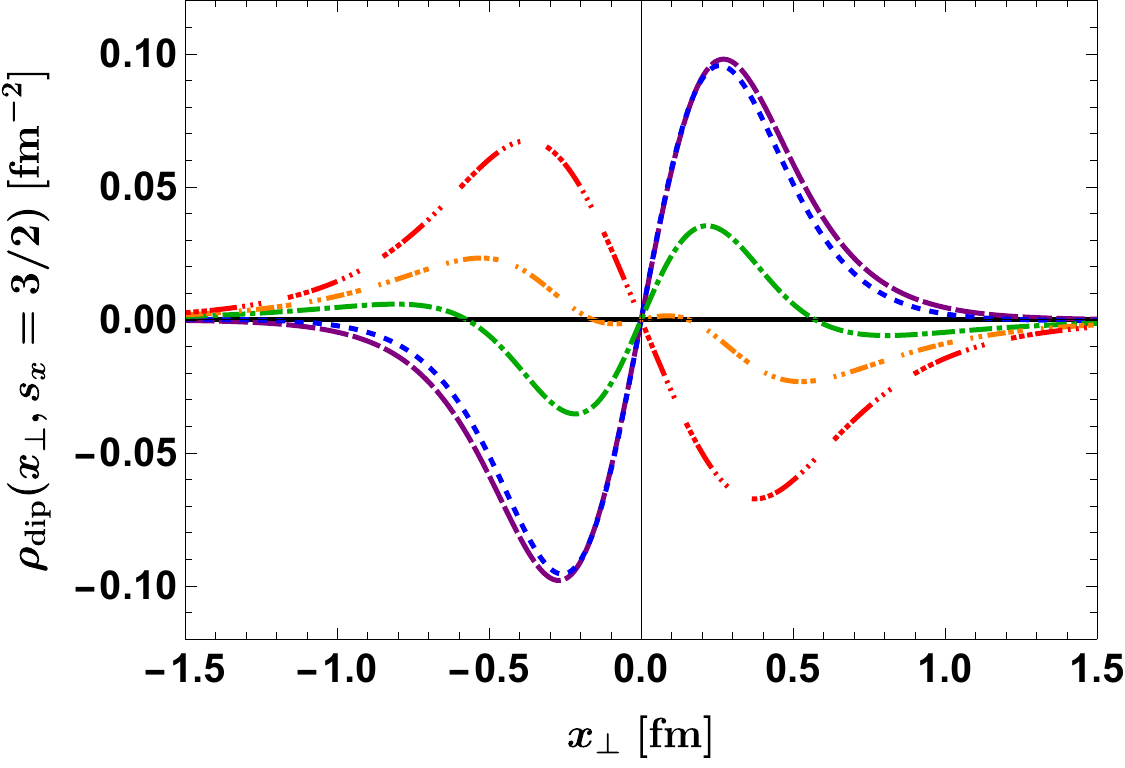}}\\ 
    \noindent
    \subfigure[Quadrupole,
    $s_x=3/2$]{\includegraphics[width=0.48\linewidth]{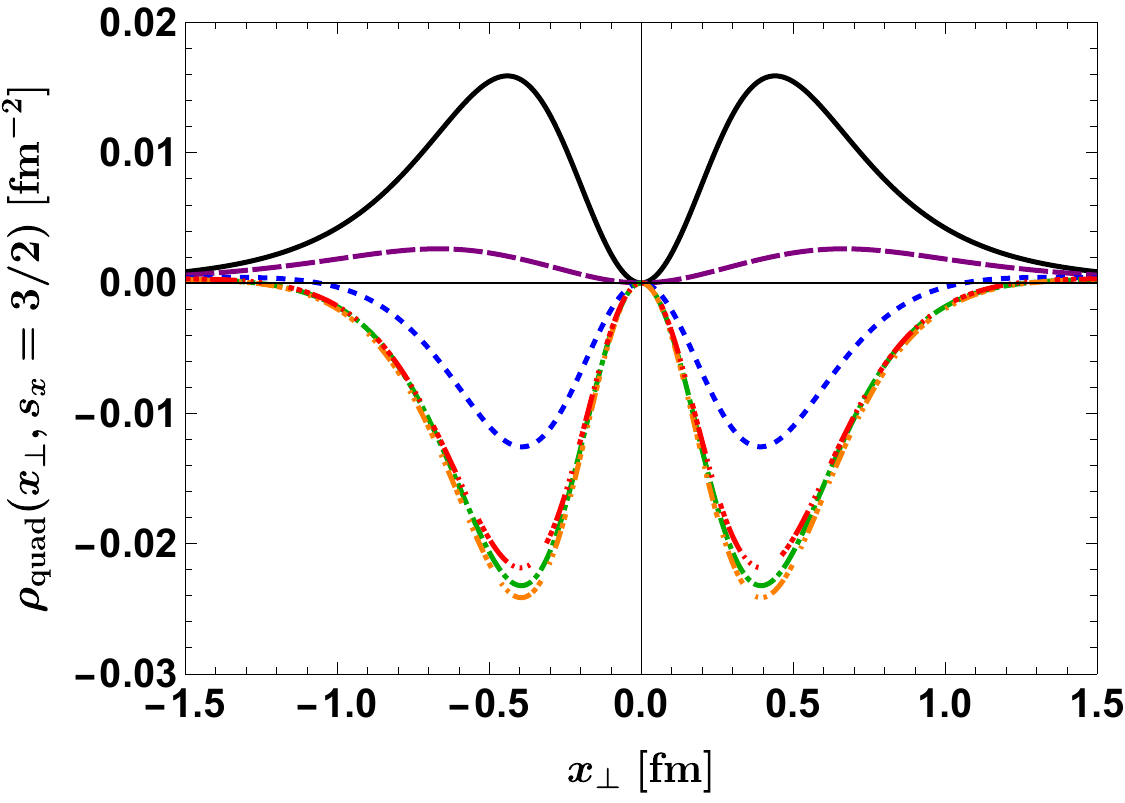}} 
    \subfigure[Octupole,
    $s_x=3/2$]{\includegraphics[width=0.48\linewidth]{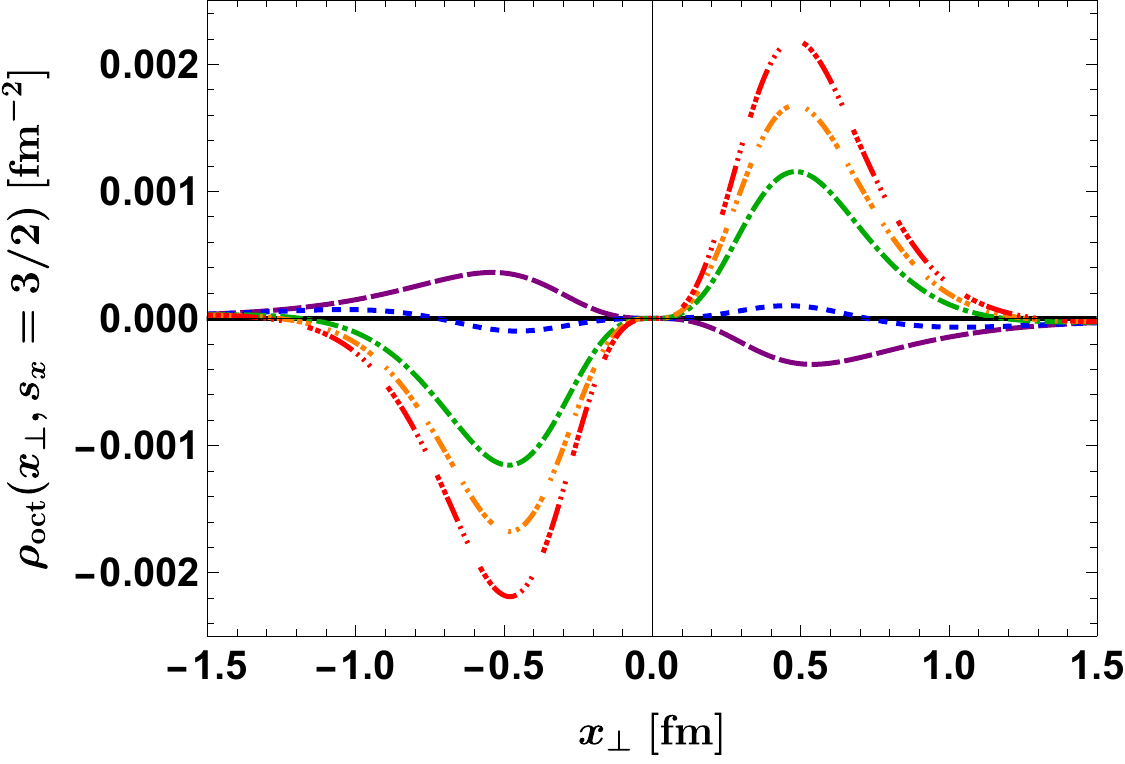}} 
    \subfigure[$s_x=3/2$]{\includegraphics[width=0.55\linewidth]{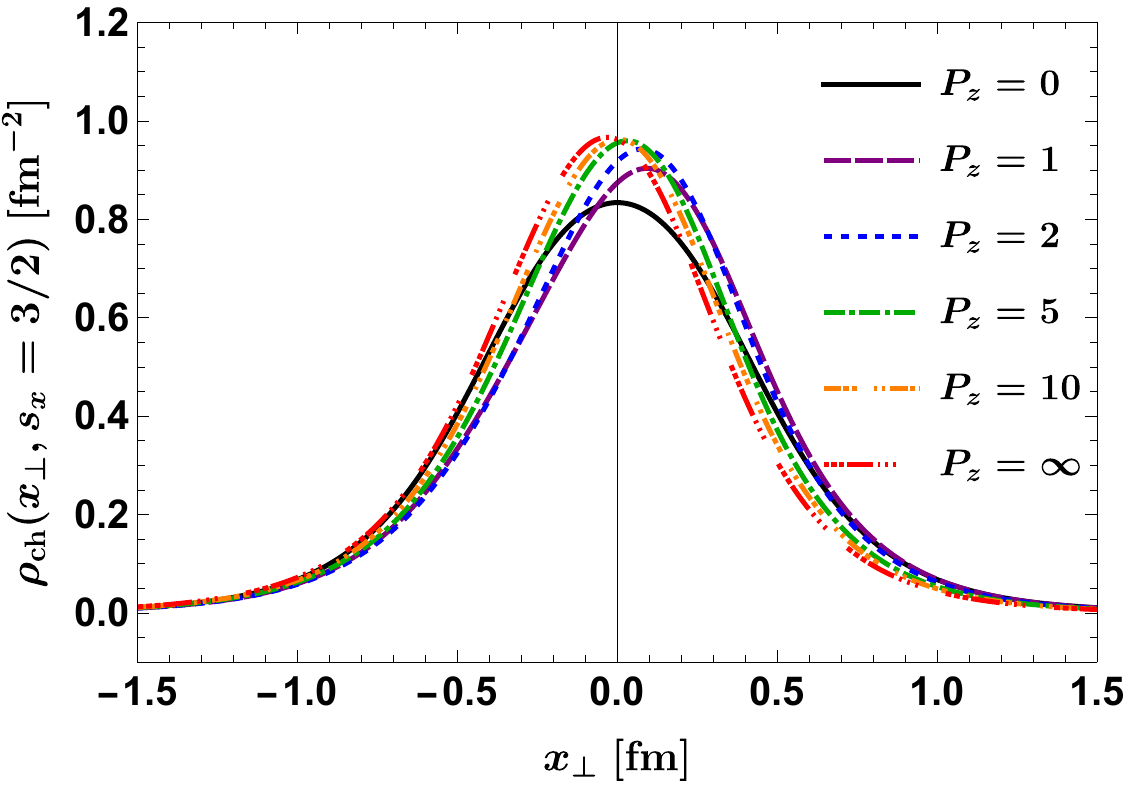}}\\
    \caption{(a) Monopole, (b) dipole, (c) quadrupole, and (d)
      octupole contributions to the $y$-axis profiles of the (e)
      transverse charge distributions of the $\Delta^+$ baryon when
      its spin is polarized along the $x$-axis with $s_{x}=3/2$.} 
    \label{fig:3}
\end{figure}

When the $\Delta$ baryon is transversely polarized along the $x$-axis,
its transverse charge distribution starts to get deformed as $P_{z}$
increases. In the presence of the external magnetic field $B$, 
the electric dipole moment is induced by the moving $\Delta$ baryon,
which produces the electric field $\bm{E}'$ depending on the
velocity $v$ of the moving $\Delta$, i.e., $\bm{E}'= \gamma
(\bm{v}\times \bm{B})$. A similar feature was also observed in the
case of the neutron~\cite{Carlson:2007xd}. In addition, the induced
electric octupole moment is also caused by this relativistic motion
and results in the deformed charge distribution with the octupole
pattern, unlike the nucleon. Figures~\ref{fig:3}(a)-(d)
depict the numerical results of the monopole, dipole, quadrupole, and
octupole patterns of the $\Delta$ baryon charge distribution,
respectively, when the $\Delta$ is polarized along the $x$-axis with
$s_{x}=3/2$. One can obviously see that while the higher multipole
contributions are found to be marginal, the dipole contribution arises
as the most dominant one to deform the transverse charge distribution.
At the rest frame, the dipole contribution is null, so the charge
distribution is symmetric with respect to $y=0$. Once the 
$\Delta$ is boosted, the dipole contribution starts to increase and
reaches its maximum value at around $P_{z}\sim 1.4\,\mathrm{GeV}$.
Then it diminishes gradually. At $P_{z} \sim 10\,\mathrm{GeV}$, the
size of the dipole contribution arrives at the minimum and then 
it start to increase again but its sign is reversed (see
Fig.~\ref{fig:3}(b)). 

In the rest frame, the quadrupole contribution survives and makes the
transverse charge distribution broaden. If the $\Delta$ is boosted,
the positive quadrupole contribution turns negative at around
$P_{z}\sim1.4\,\mathrm{GeV}$. Figure~\ref{fig:3}(e) draws the charge
distribution of the $\Delta^{+}$ baryon with $s_{x}=3/2$, which 
is the sum  of Figs.~\ref{fig:3}(a)-(d). As shown in
Fig.~\ref{fig:3}(e), the transverse charge distribution starts to be
tilted to the positive $x_\perp$-direction till
$P_{z}=1.4\,\mathrm{GeV}$ and becomes symmetric with respect to
$x_\perp=0$ at around $P_{z}\sim 10\,\mathrm{GeV}$ again. When $P_z$
increases more, the charge distribution starts to move to the left
$x_\perp$-direction. In the IMF ($P_{z}=\infty$), we obtain the  
$\Delta^{+}$ charge distribution with $s_{x}=3/2$ shifted to the left
direction, which is consistent with the results from the lattice
QCD~\cite{Alexandrou:2009hs}. Note that the induced electric dipole
moment of the proton is defined as $G^{N}_{M1}(0)-G^{N}_{E0}(0) > 0$,
whereas that of the $\Delta$ baryon is proportional to
$G^{\Delta}_{M1}(0)-3G^{\Delta}_{E0}(0) < 0$, so that charge
distribution of the $\Delta^{+}$ is shifted to the left, which 
is opposite to the transverse proton charge distribution (see also
Ref.~\cite{Alexandrou:2009hs}). 

\begin{figure}[htp]
    \subfigure[Monopole,
    $s_x=1/2$]{\includegraphics[width=0.48\linewidth]{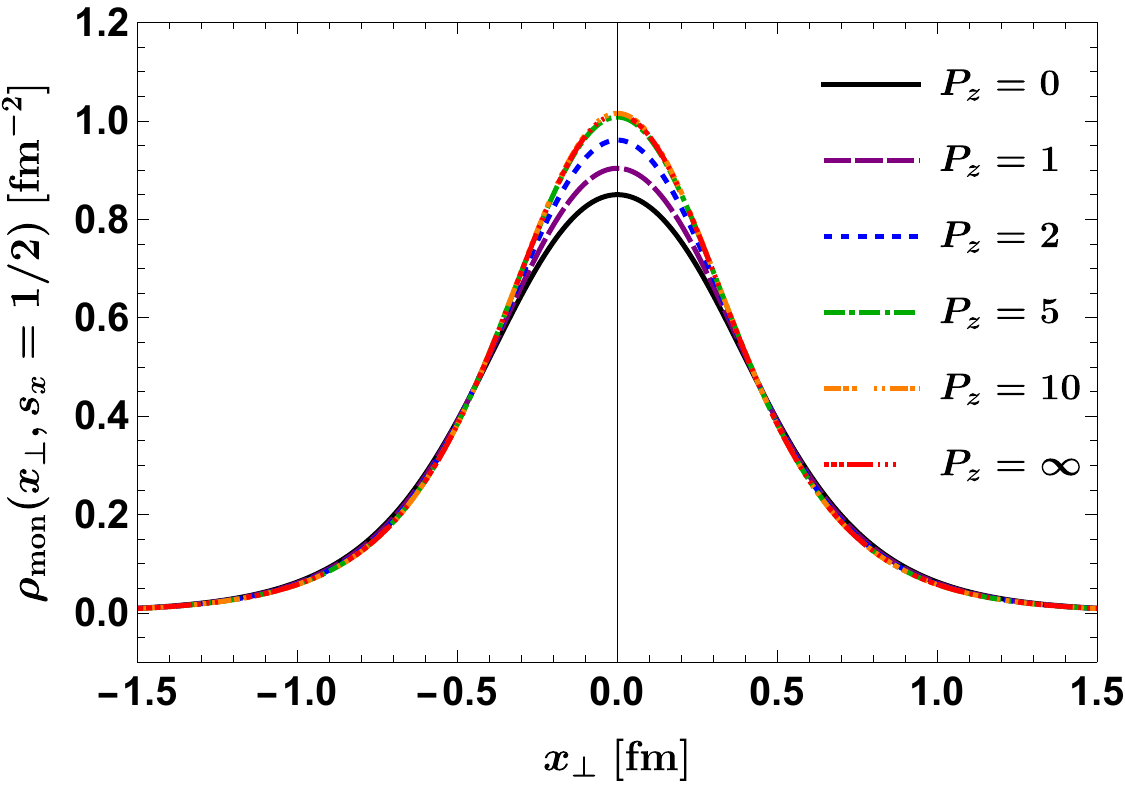}} 
    \subfigure[Dipole,
    $s_x=1/2$]{\includegraphics[width=0.48\linewidth]{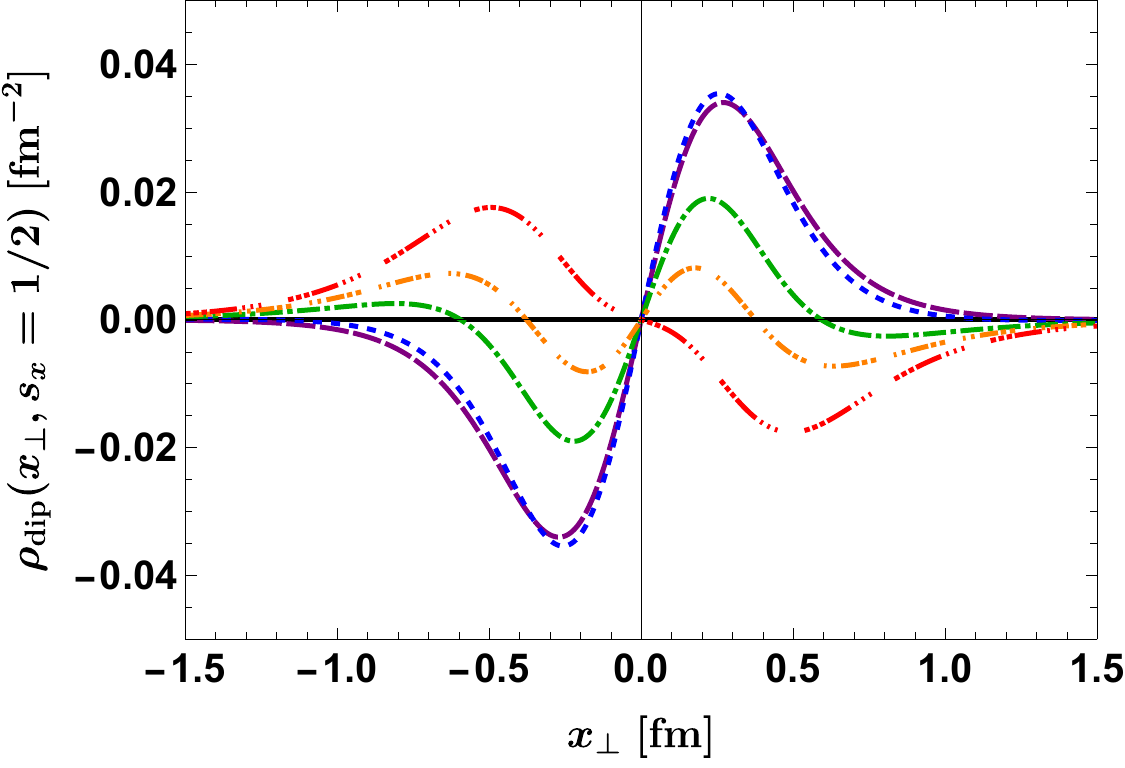}}\\ 
    \noindent
    \subfigure[Quadrupole,
    $s_x=1/2$]{\includegraphics[width=0.48\linewidth]{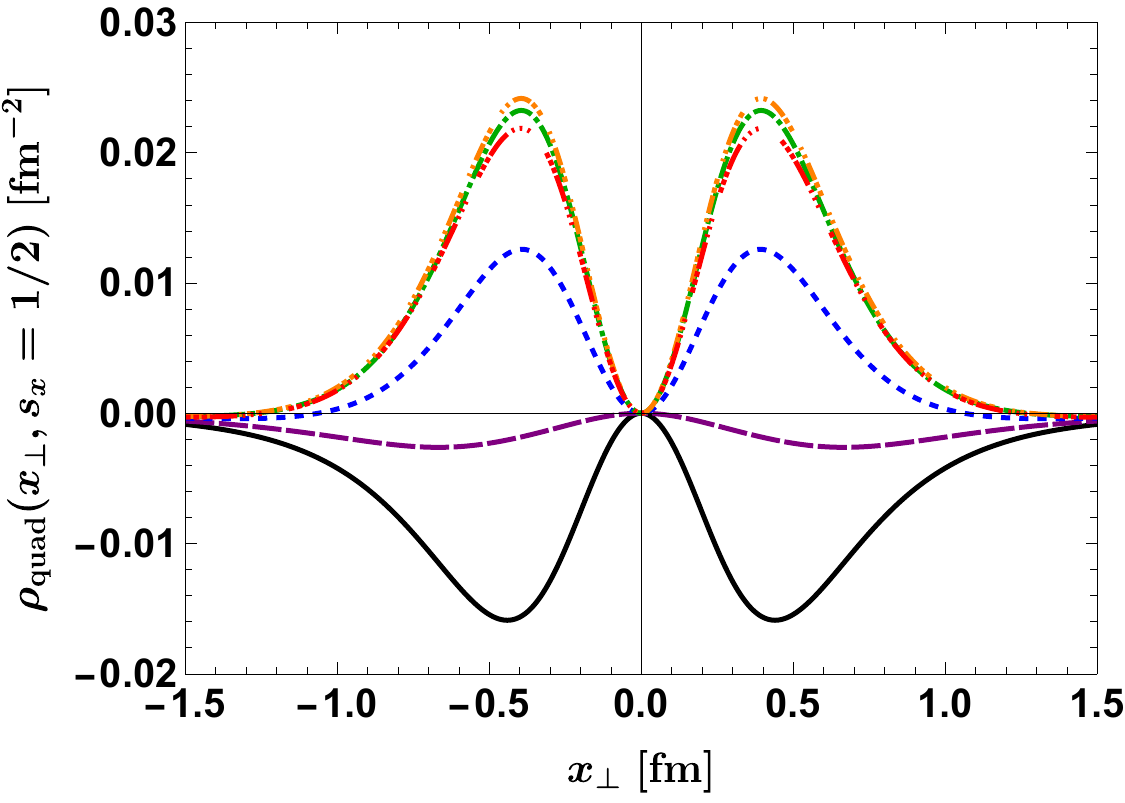}} 
    \subfigure[Octupole,
    $s_x=1/2$]{\includegraphics[width=0.48\linewidth]{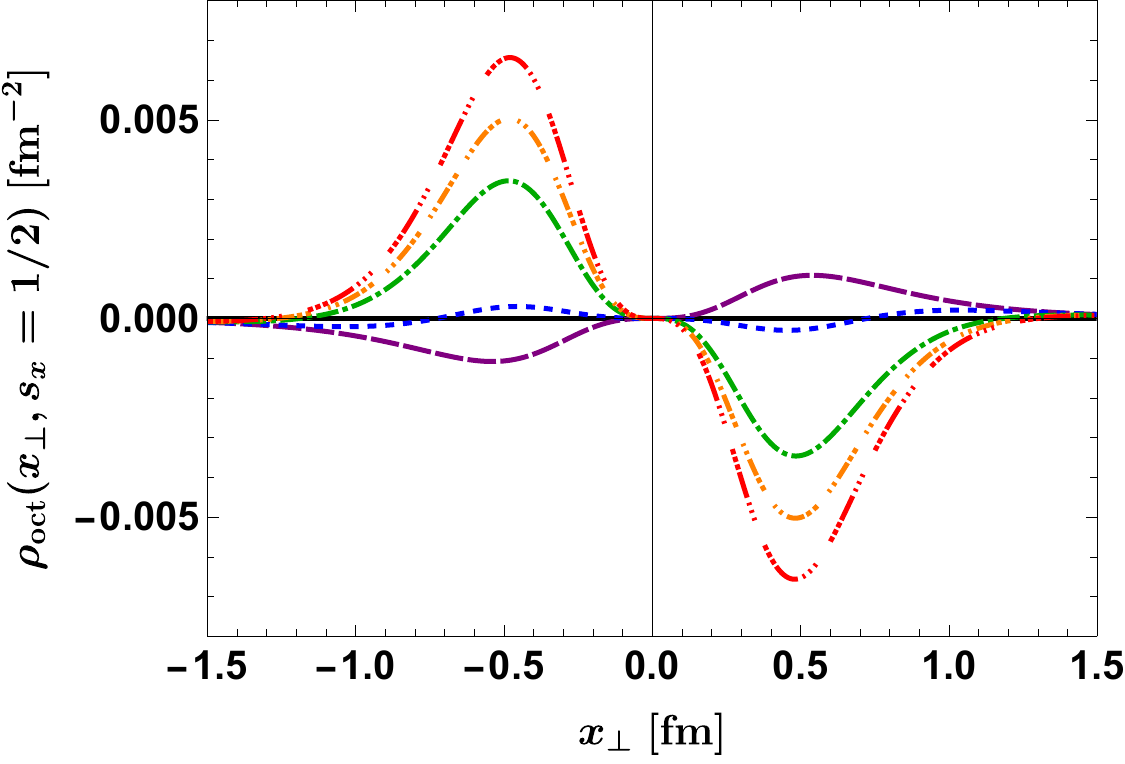}}\\ 
    \noindent
    \subfigure[$s_x=1/2$]{\includegraphics[width=0.55\linewidth]{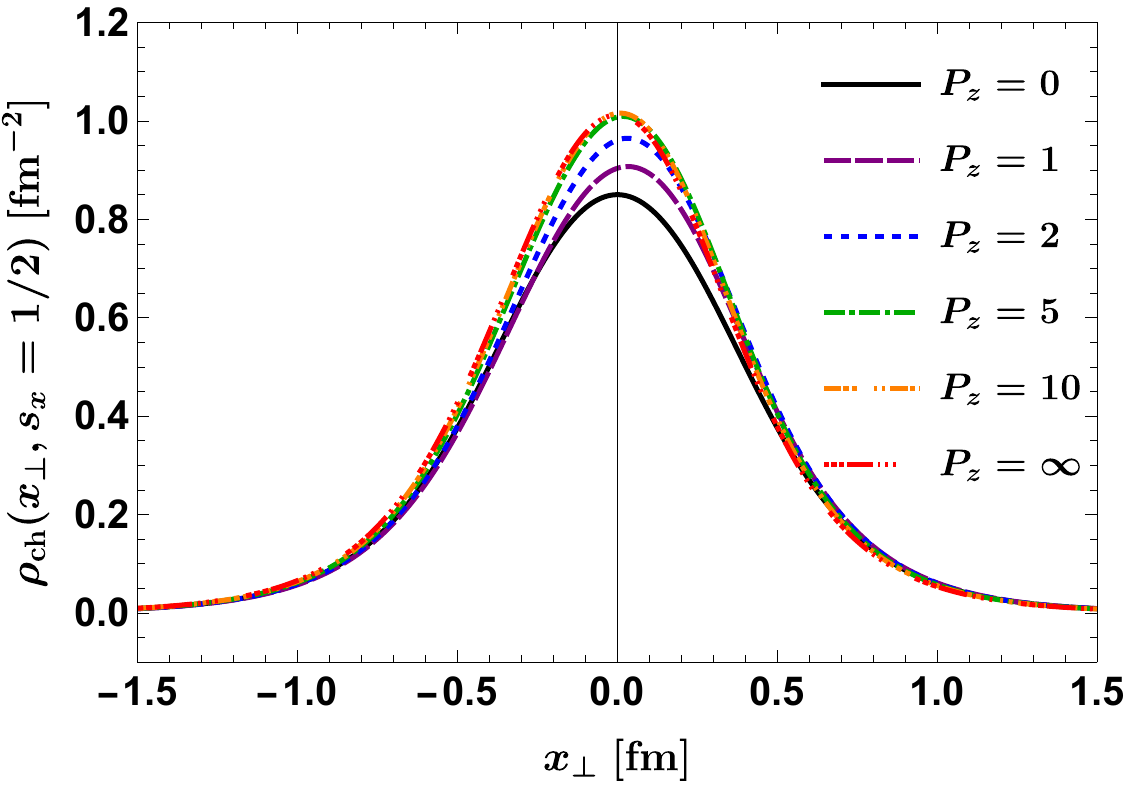}}
    \caption{(a) monopole, (b) dipole, (c) quadrupole, and (d)
      octupole contributions to the $y$-axis profiles of the (e) 2D
      charge distributions of the $\Delta^+$ baryon when its spin is
      polarized along the $x$-axis with $s_{x}=1/2$.} 
    \label{fig:4}
\end{figure}

In Figs.~\ref{fig:4}(a)-(d) we present the numerical results for the
monopole, dipole, quadrupole, and octupole patterns of the transverse
$\Delta^{+}$ charge distribution when it is polarized along the
$x$-axis with $s_{x}=1/2$. The sum of the total contributions is drawn
in Fig.~\ref{fig:4}(e). They show a tendency similar to the  
$s_{x}=3/2$ case. However, the strength of the dipole contribution is
almost a half of that with $s_{x}=3/2$. So, the shape of the charge 
distribution is almost kept to be symmetric, and in the IMF they are
shifted to the negative $x_\perp$-direction with respect to $x_\perp=0$,
which is also consistent with the results from
Ref.~\cite{Alexandrou:2009hs}.  

\begin{figure}[htpb]
    \centering  
    \subfigure[$s_x=3/2$]{\includegraphics[width=0.23\linewidth]{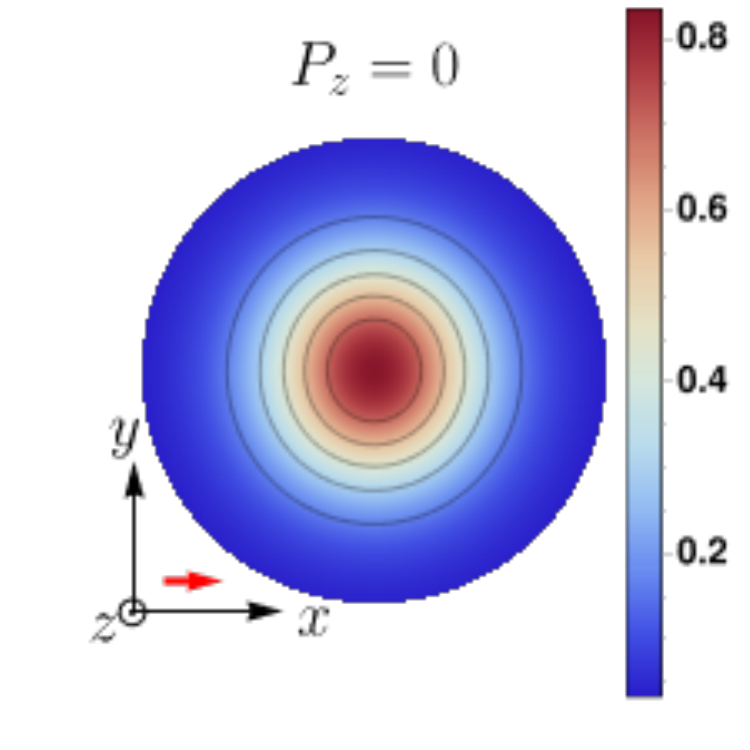}}
    \subfigure[$s_x=3/2$]{\includegraphics[width=0.23\linewidth]{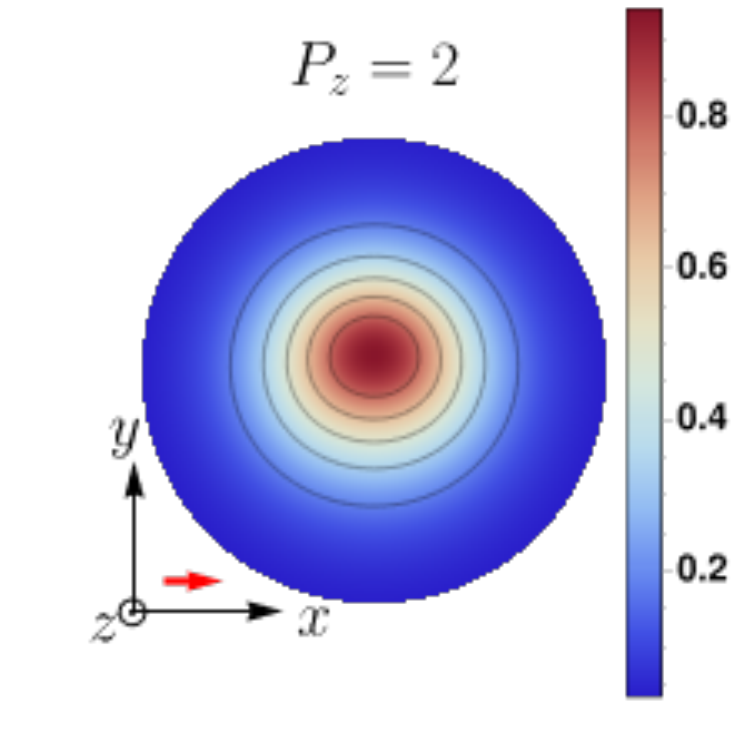}}
    \subfigure[$s_x=3/2$]{\includegraphics[width=0.23\linewidth]{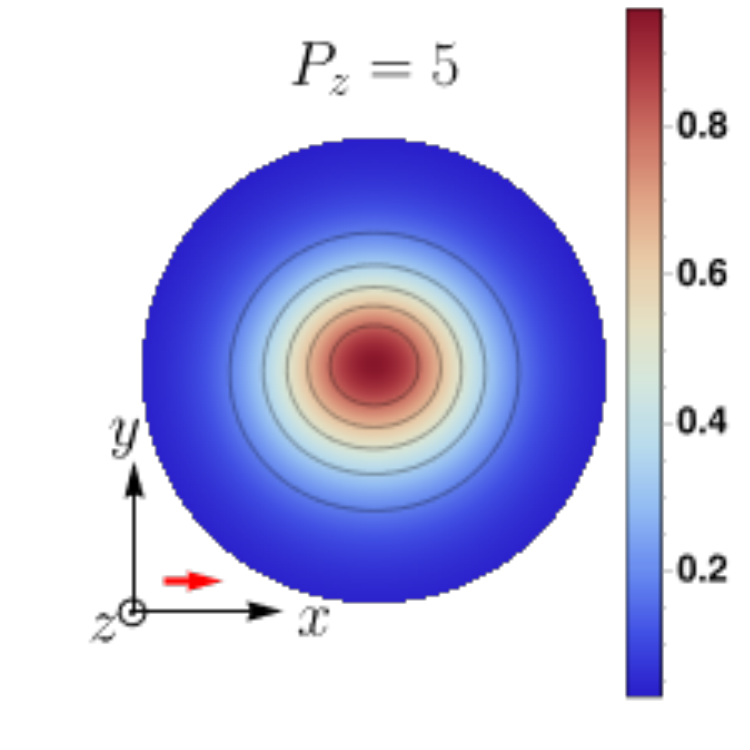}}
    \subfigure[$s_x=3/2$]{\includegraphics[width=0.23\linewidth]{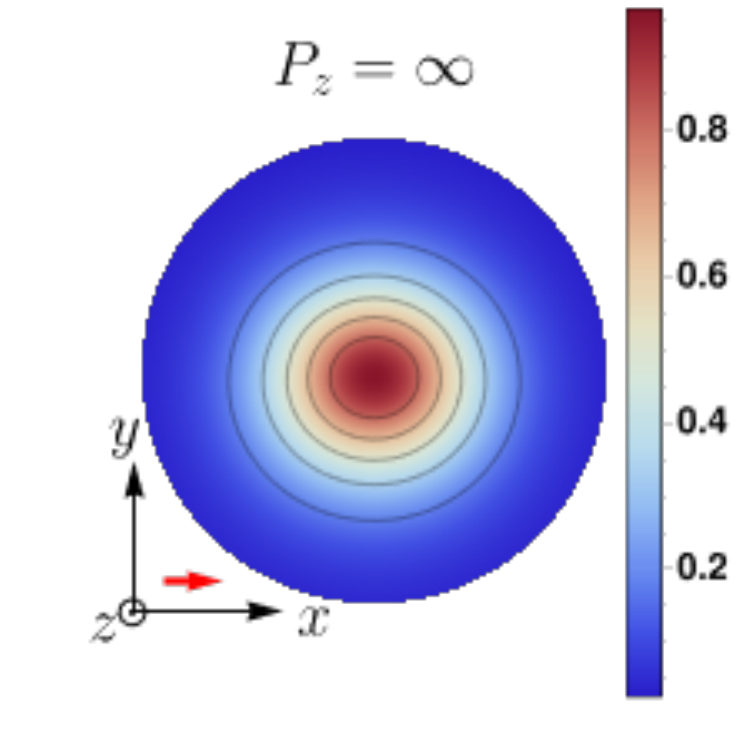}}
    \subfigure[$s_x=1/2$]{\includegraphics[width=0.23\linewidth]{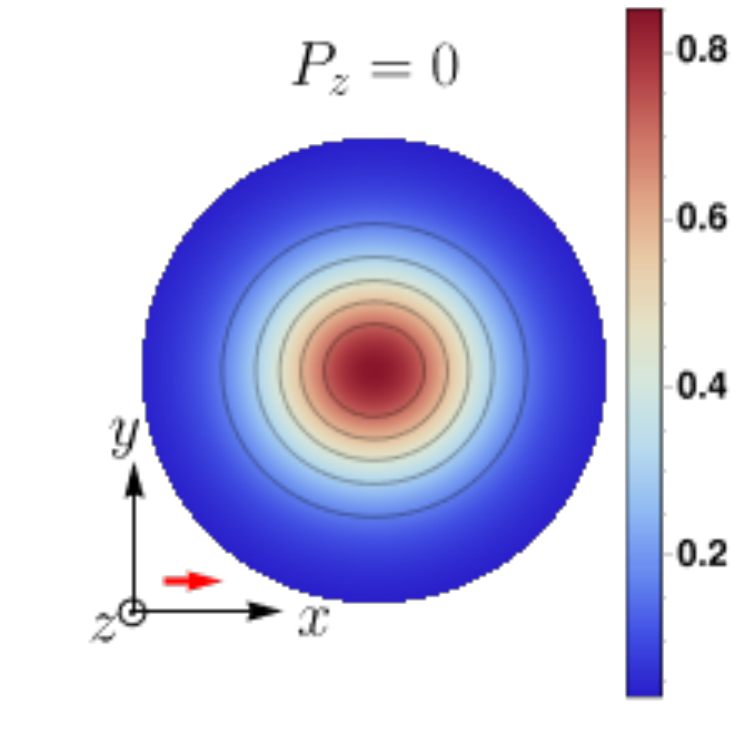}}
    \subfigure[$s_x=1/2$]{\includegraphics[width=0.23\linewidth]{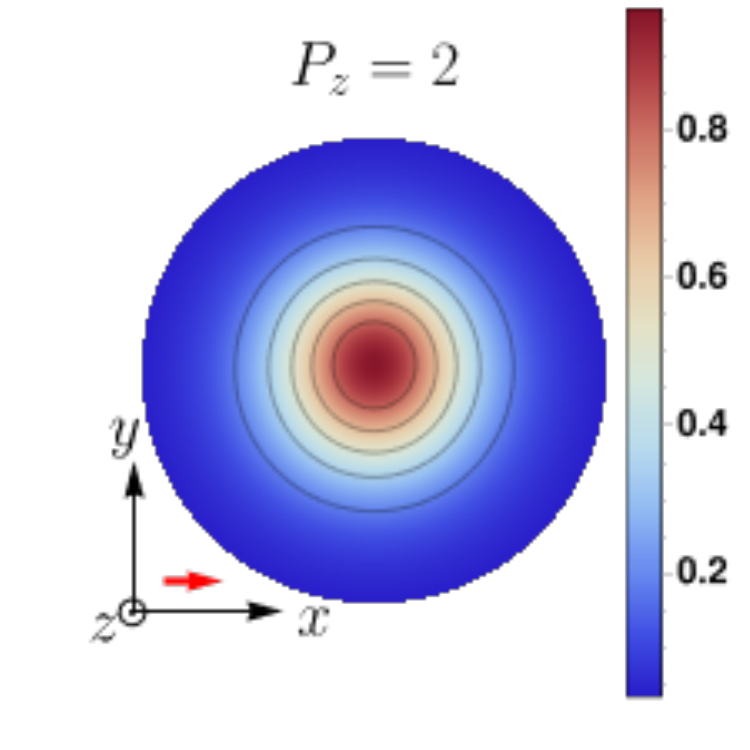}}
    \subfigure[$s_x=1/2$]{\includegraphics[width=0.23\linewidth]{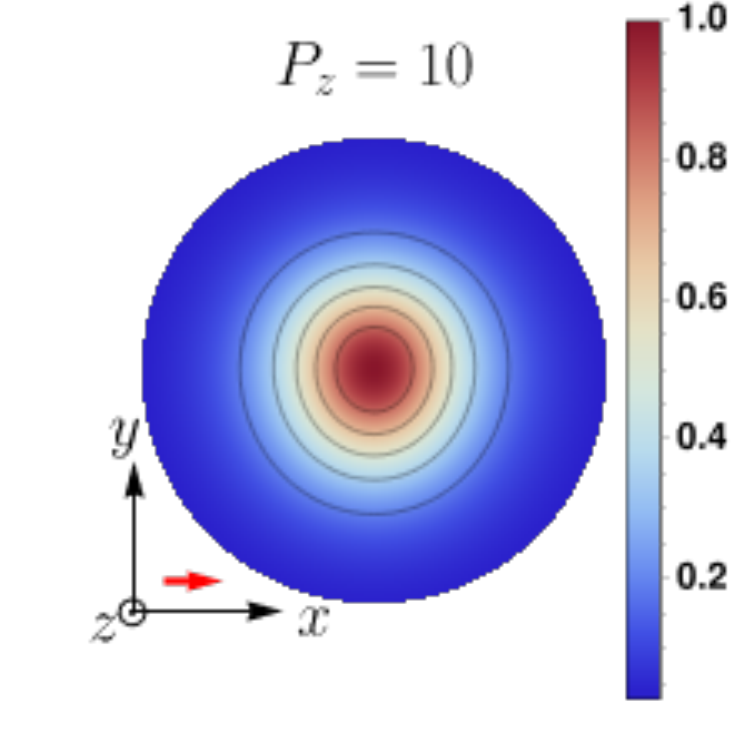}}
    \subfigure[$s_x=1/2$]{\includegraphics[width=0.23\linewidth]{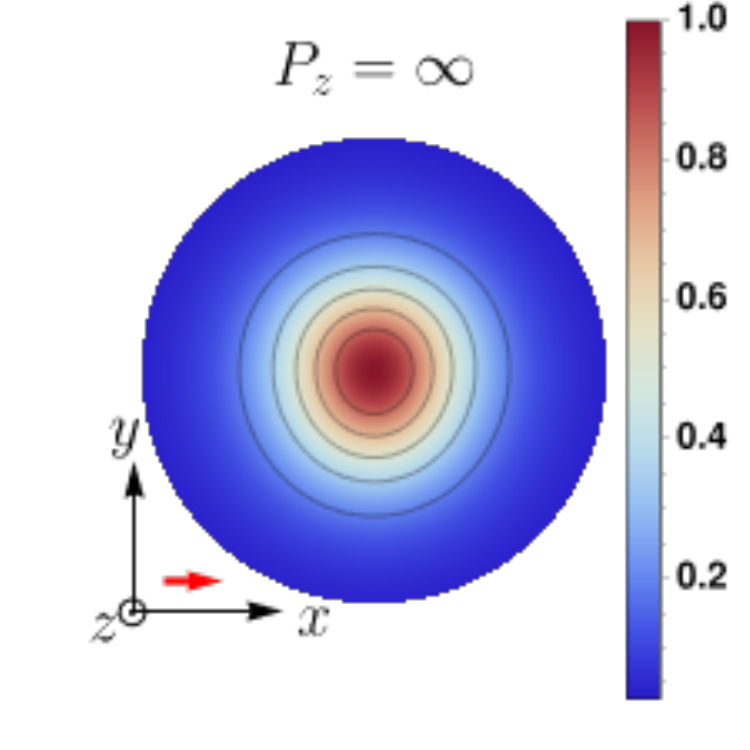}}
    \caption{(a)-(d) 2D charge distributions of the moving $\Delta^+$
      baryon transversely polarized along $x$-axis with $s_x=3/2$;
      (e)-(h) 2D charge distributions of the moving $\Delta^+$ baryon
      transversely polarized along $x$-axis with $s_x=1/2$} 
    \label{fig:5}
\end{figure}

\begin{figure}[htpb]
    \centering
    \subfigure[$s_x=3/2$,
    Monopole]{\includegraphics[width=0.23\linewidth]{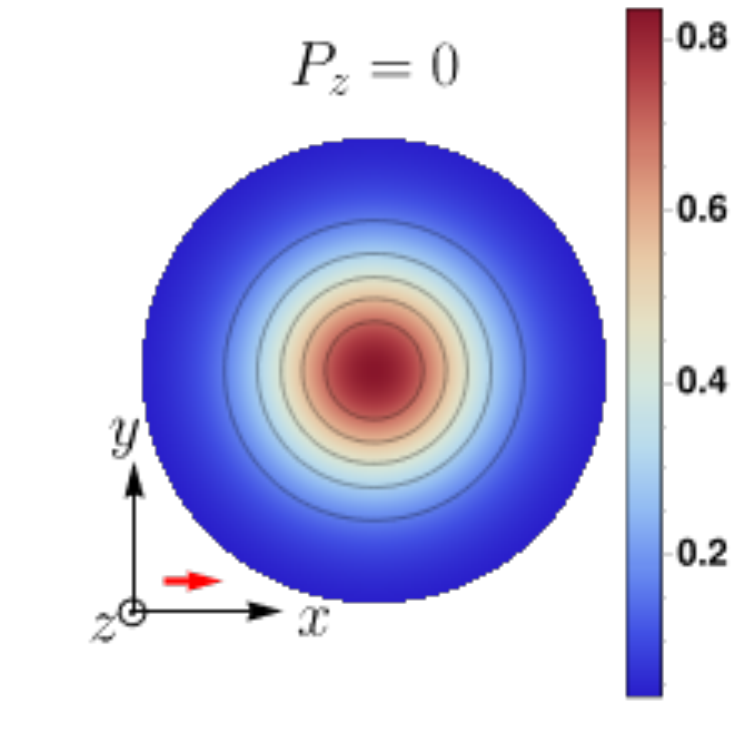}} 
    \subfigure[$s_x=3/2$,
    Monopole]{\includegraphics[width=0.23\linewidth]{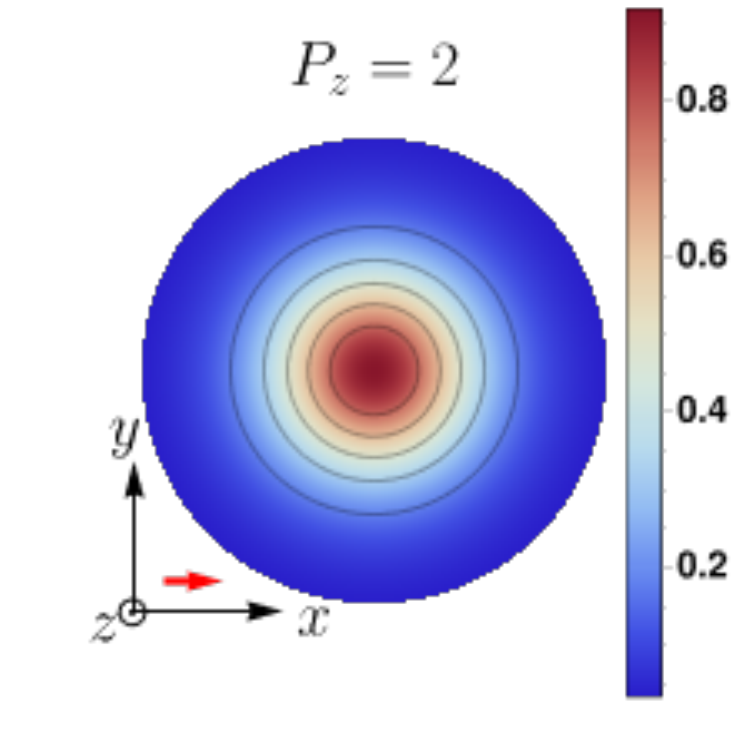}} 
    \subfigure[$s_x=3/2$,
    Monopole]{\includegraphics[width=0.23\linewidth]{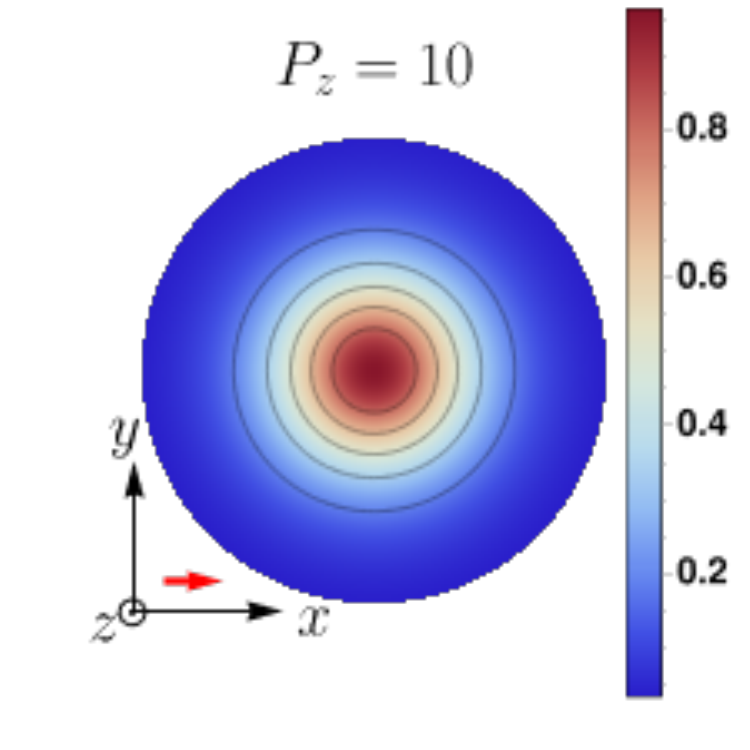}} 
    \subfigure[$s_x=3/2$,
    Monopole]{\includegraphics[width=0.23\linewidth]{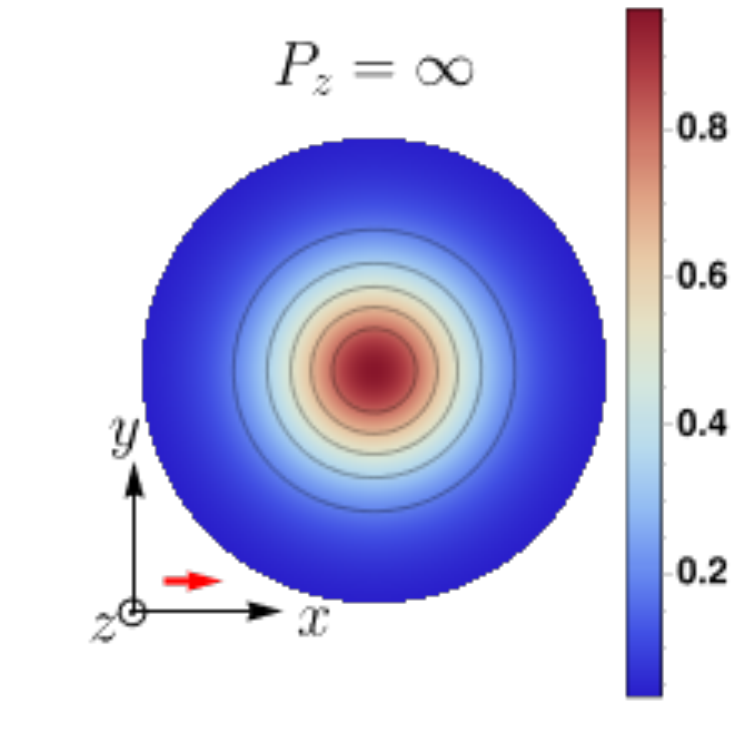}}\\ 
    \noindent
    \subfigure[$s_x=3/2$,
    Dipole]{\includegraphics[width=0.23\linewidth]{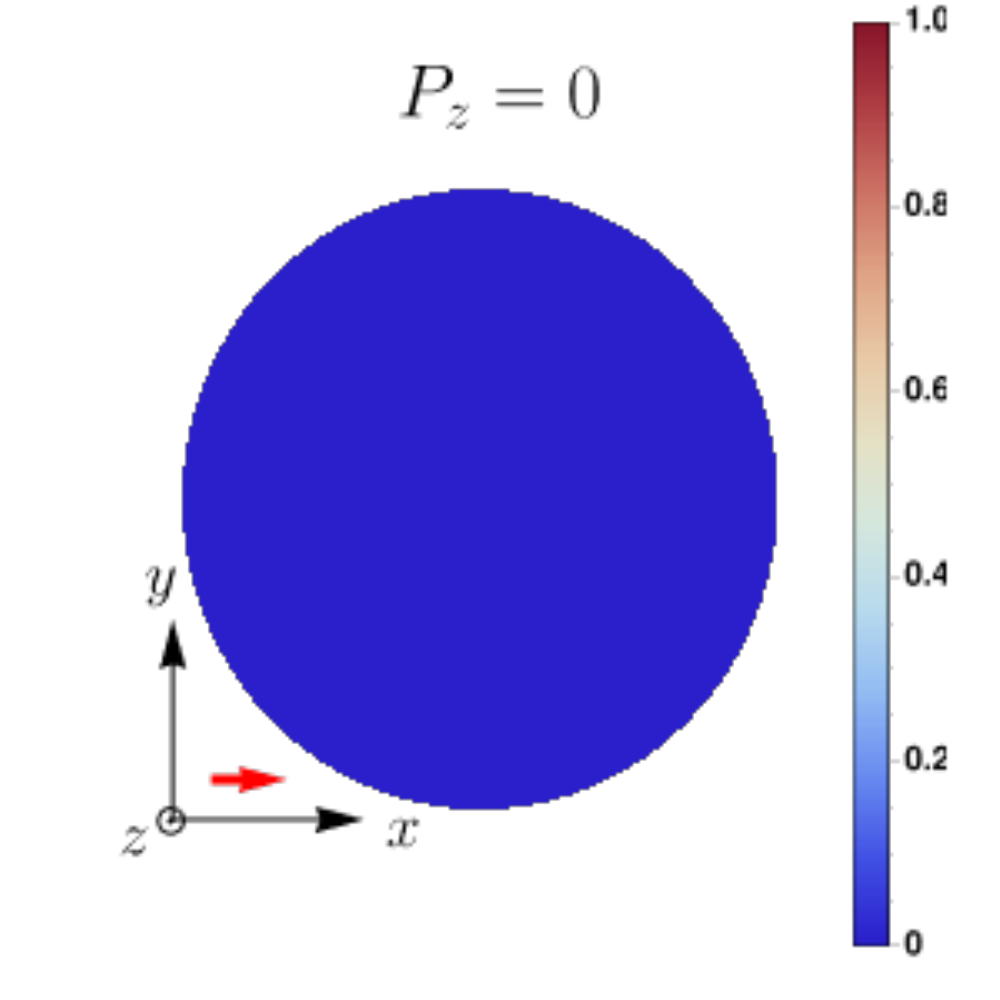}} 
    \subfigure[$s_x=3/2$,
    Dipole]{\includegraphics[width=0.23\linewidth]{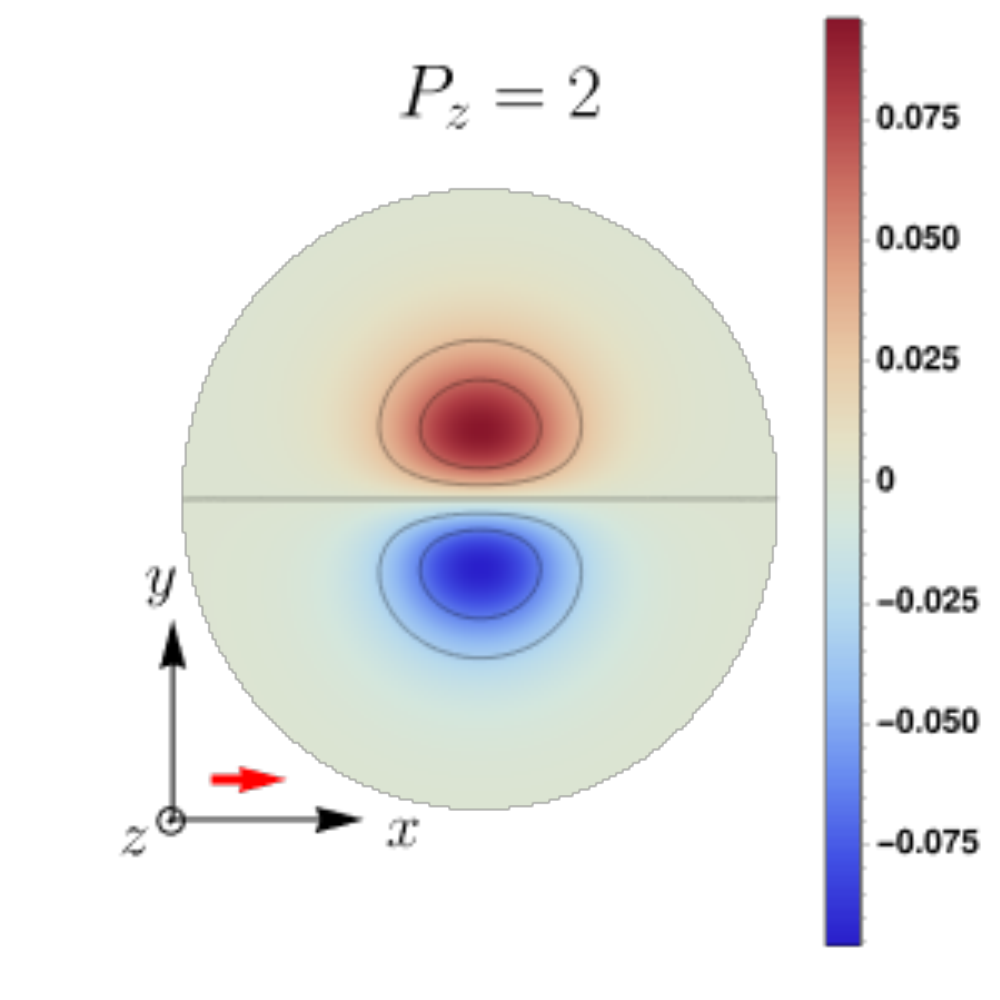}} 
    \subfigure[$s_x=3/2$,
    Dipole]{\includegraphics[width=0.23\linewidth]{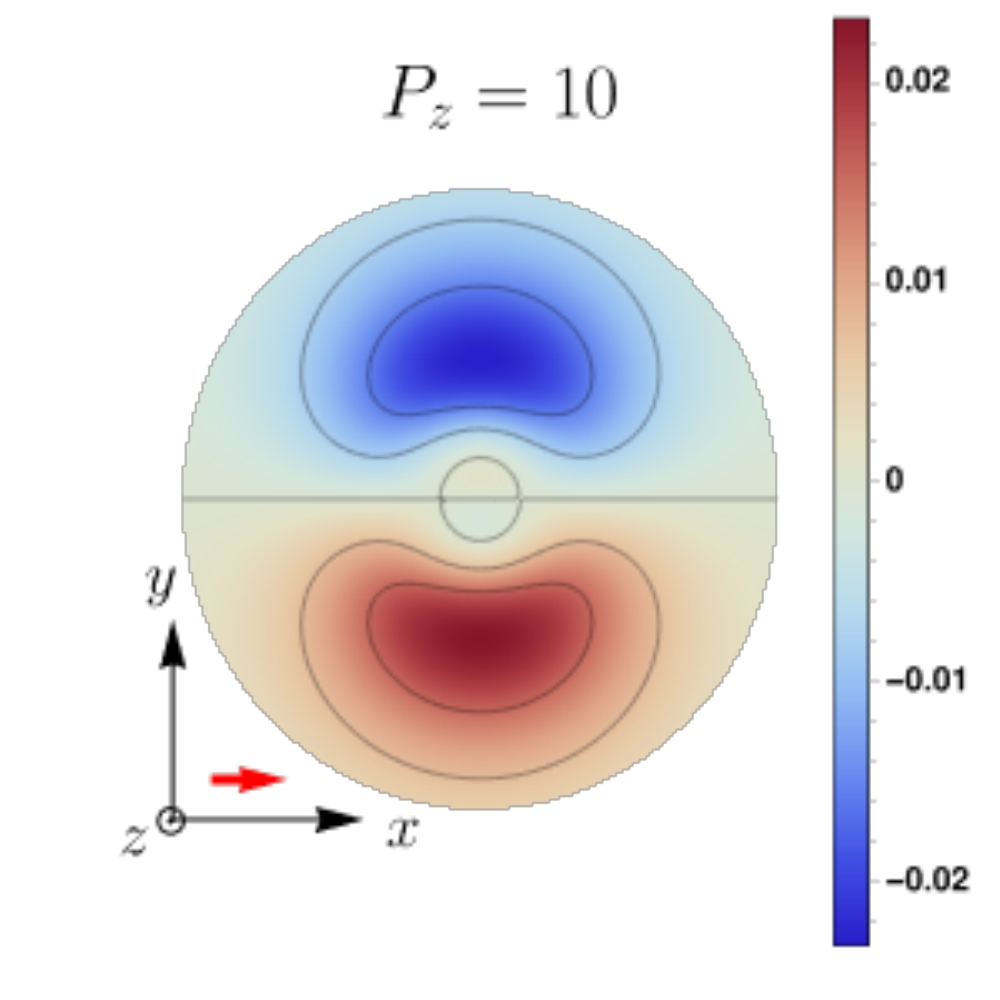}} 
    \subfigure[$s_x=3/2$,
    Dipole]{\includegraphics[width=0.23\linewidth]{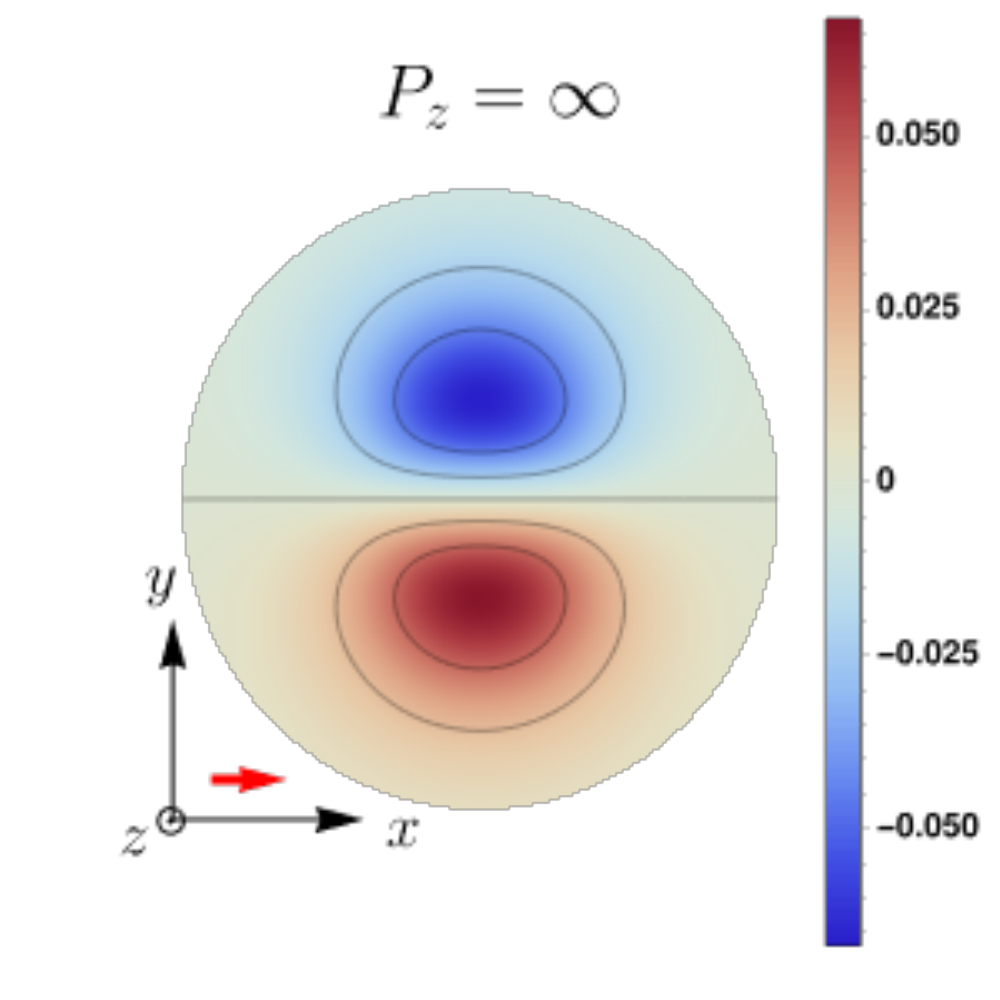}}\\ 
    \noindent
    \subfigure[$s_x=3/2$,
    Quadrupole]{\includegraphics[width=0.23\linewidth]{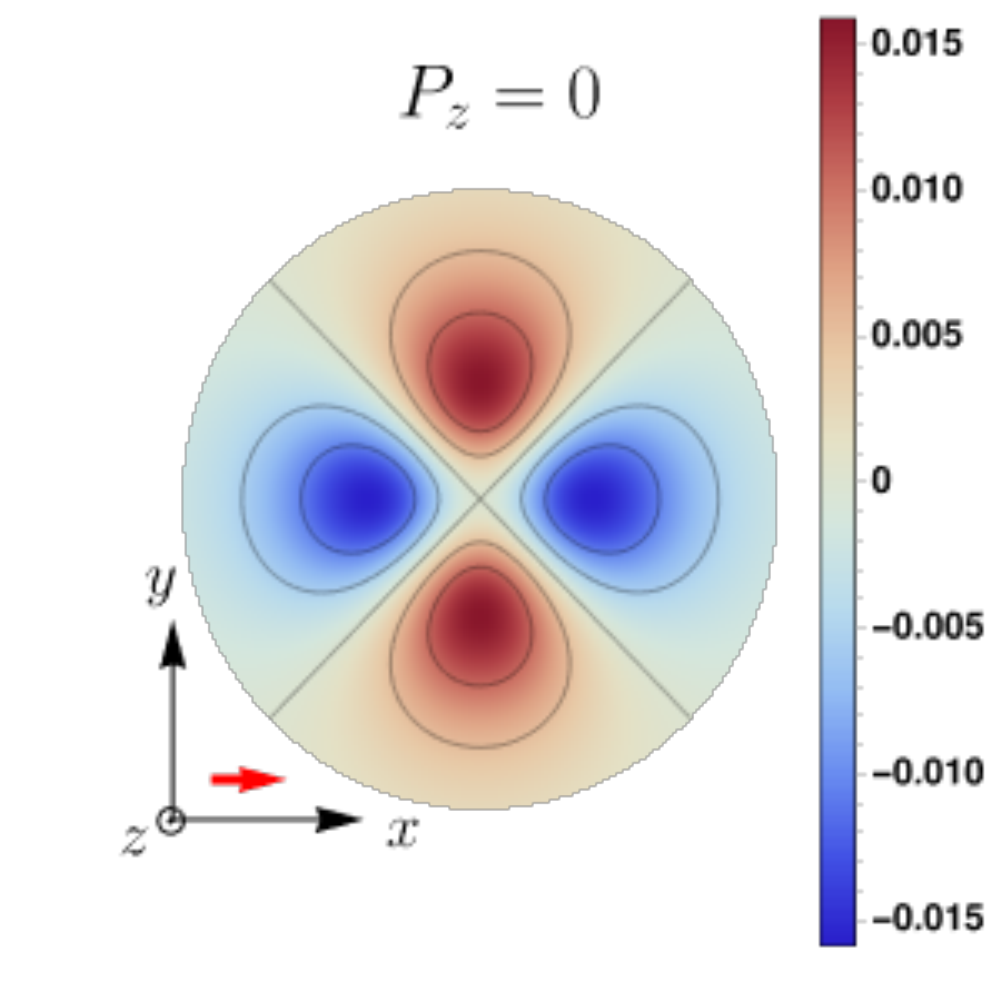}} 
    \subfigure[$s_x=3/2$,
    Quadrupole]{\includegraphics[width=0.23\linewidth]{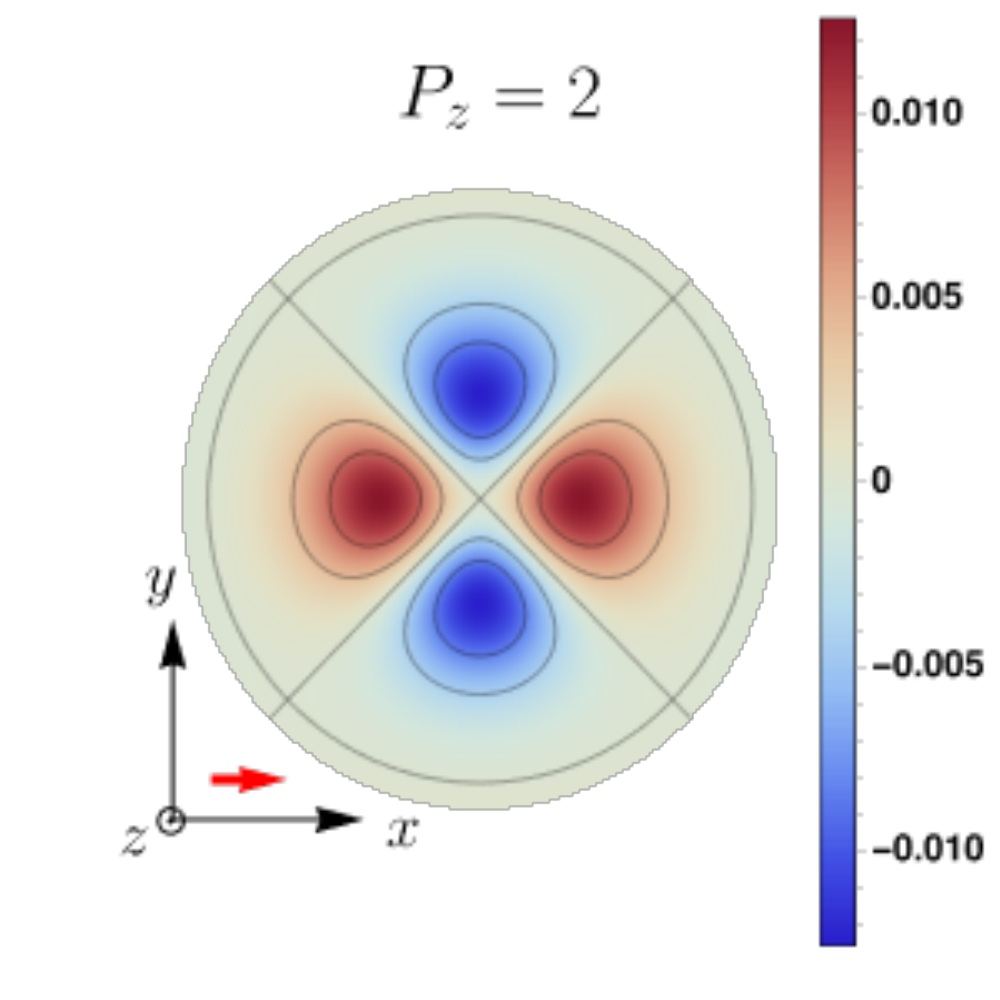}} 
    \subfigure[$s_x=3/2$,
    Quadrupole]{\includegraphics[width=0.23\linewidth]{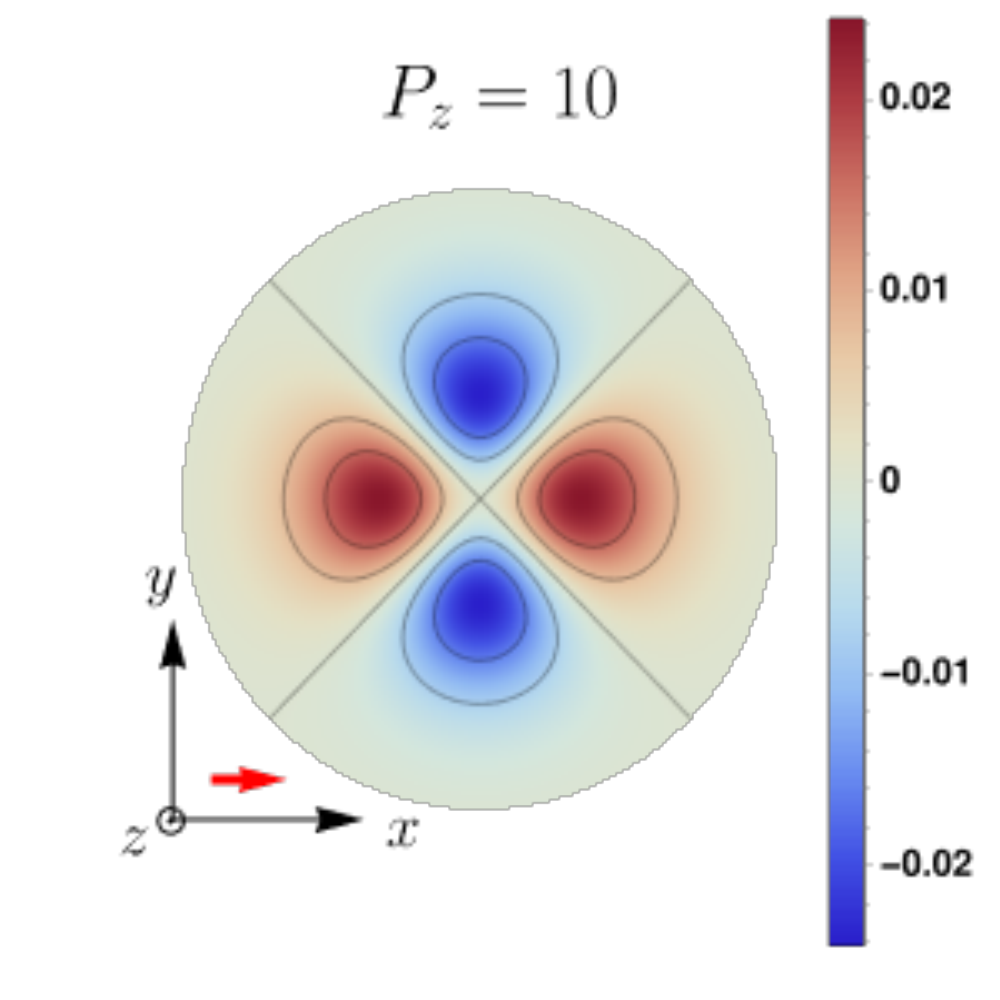}} 
    \subfigure[$s_x=3/2$,
    Quadrupole]{\includegraphics[width=0.23\linewidth]{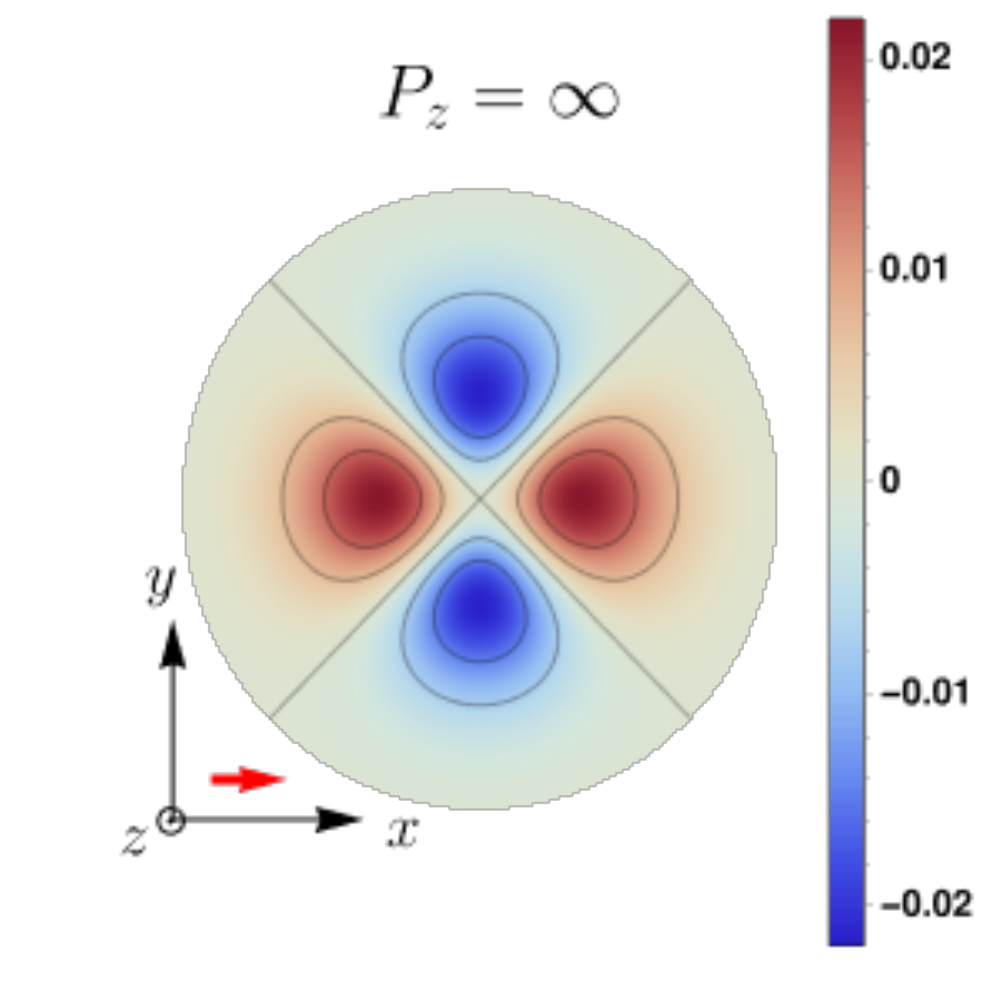}}\\ 
    \noindent
    \subfigure[$s_x=3/2$,
    Octupole]{\includegraphics[width=0.23\linewidth]{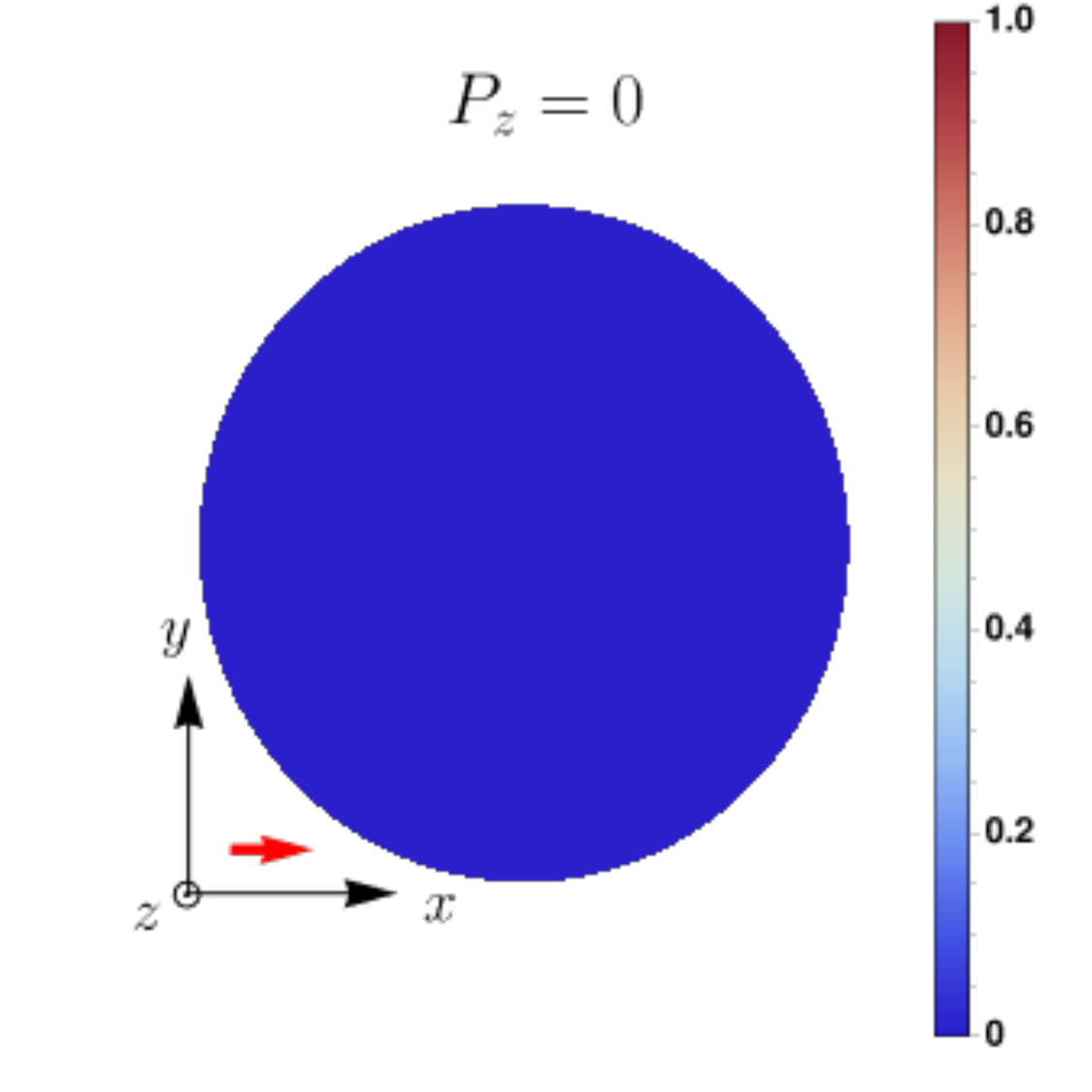}} 
    \subfigure[$s_x=3/2$,
    Octupole]{\includegraphics[width=0.23\linewidth]{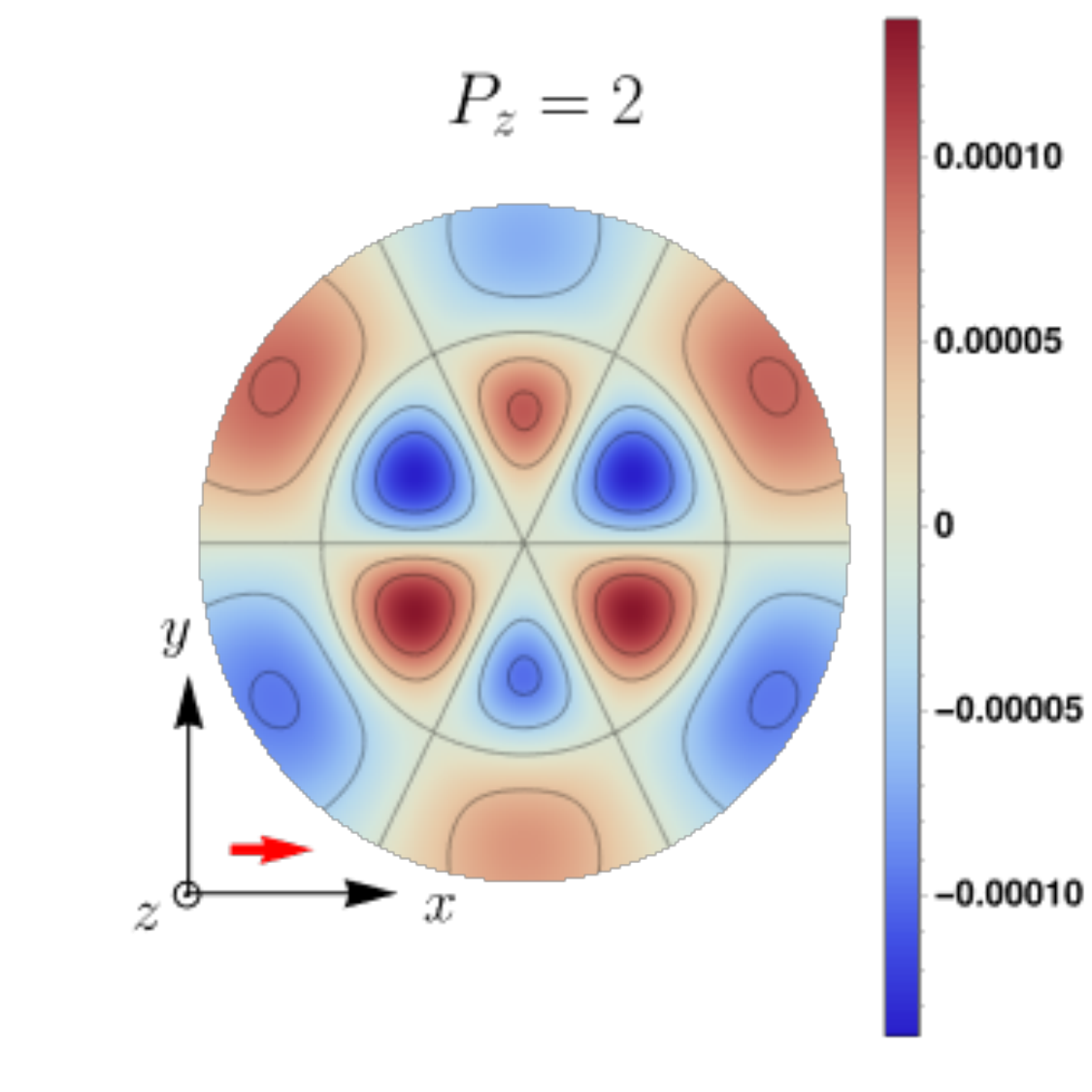}} 
    \subfigure[$s_x=3/2$,
    Octupole]{\includegraphics[width=0.23\linewidth]{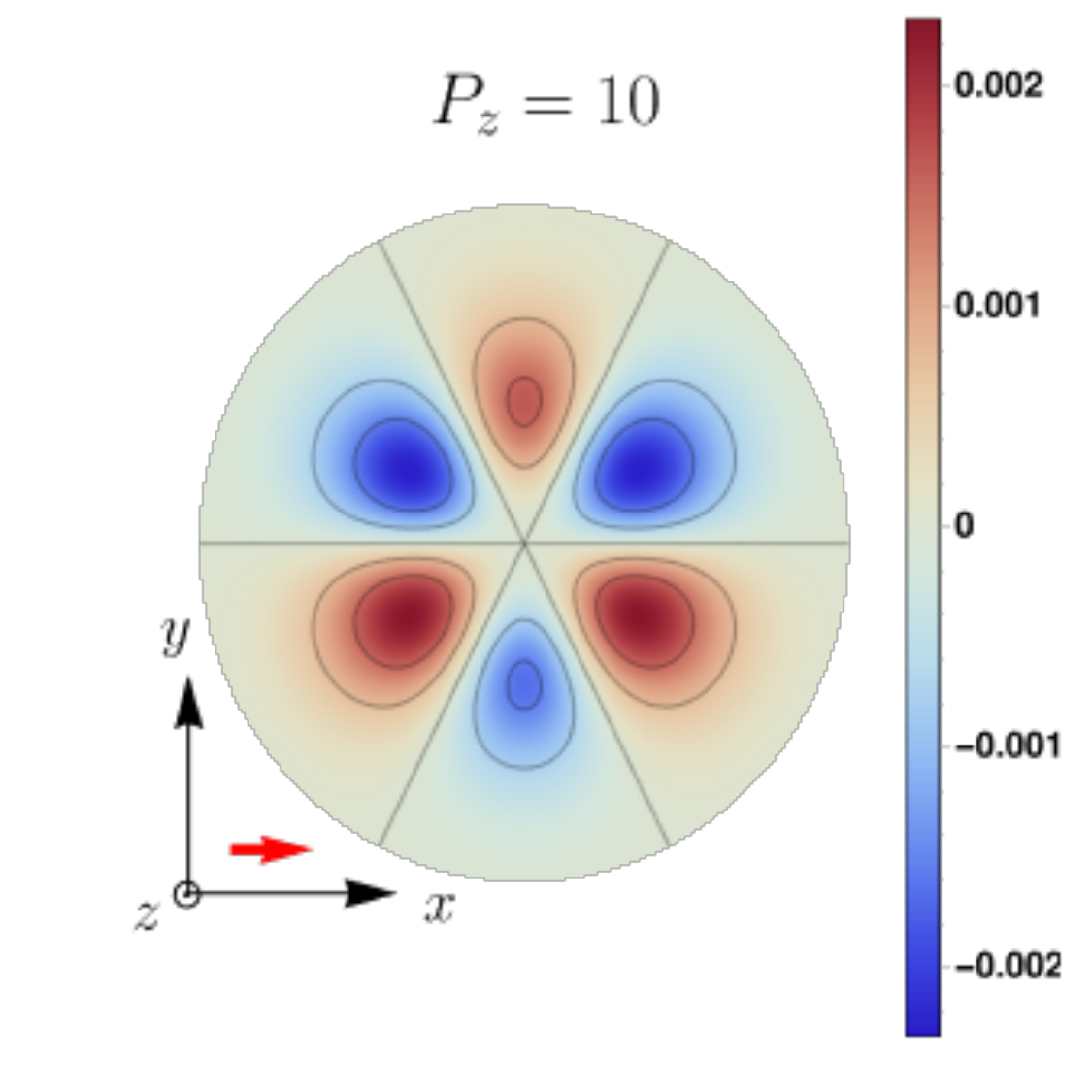}} 
    \subfigure[$s_x=3/2$,
    Octupole]{\includegraphics[width=0.23\linewidth]{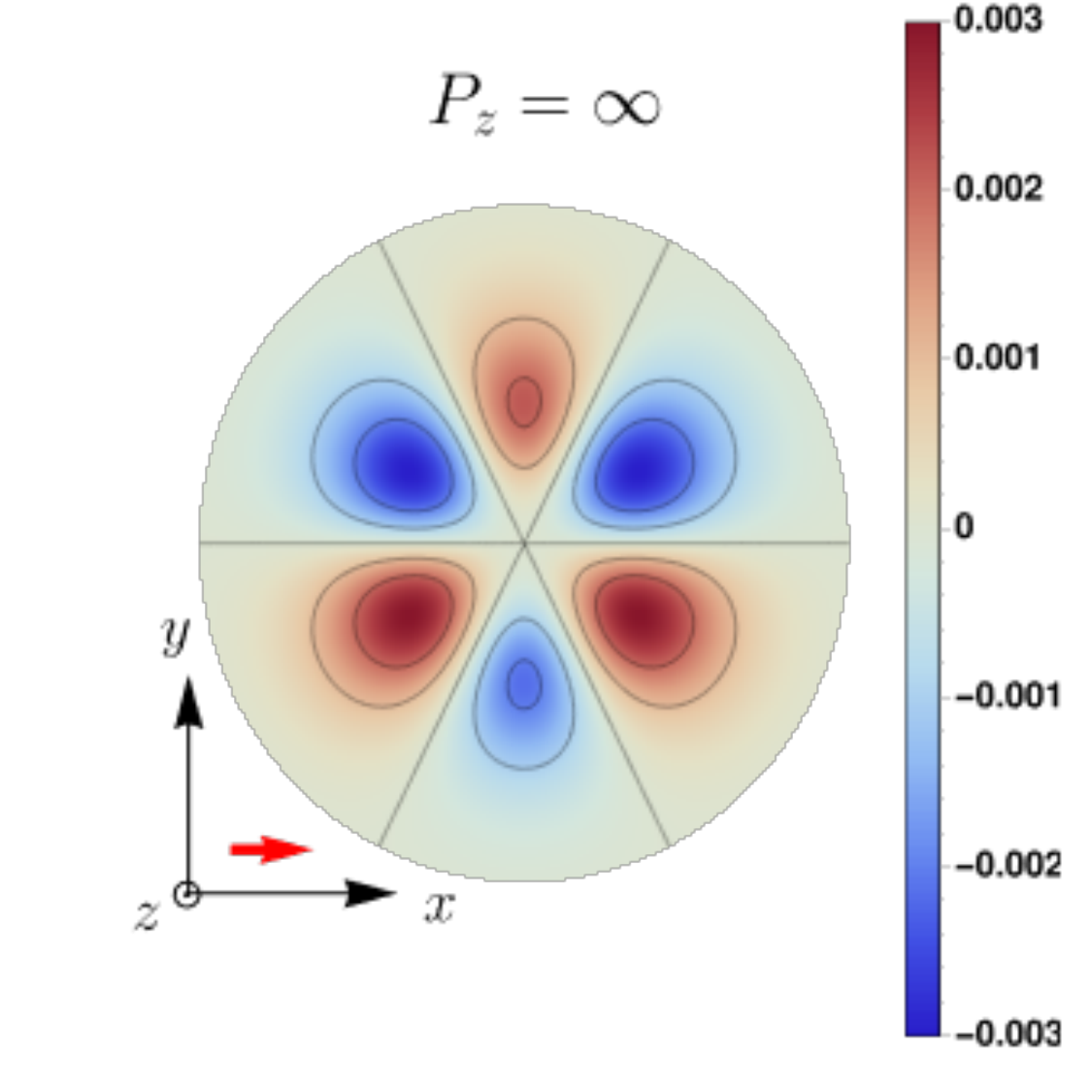}} 
    \caption{(a)-(d) monopole, (e)-(h) dipole, (i)-(l) quadrupole, and
      (m)-(p) octupole contributions of $\Delta^+$ with $s_x=3/2$ to
      the 2D charge distribution.} 
    \label{fig:6}
\end{figure}

\begin{figure}[htpb]
    \centering
    \subfigure[$s_x=1/2$,
    Monopole]{\includegraphics[width=0.23\linewidth]{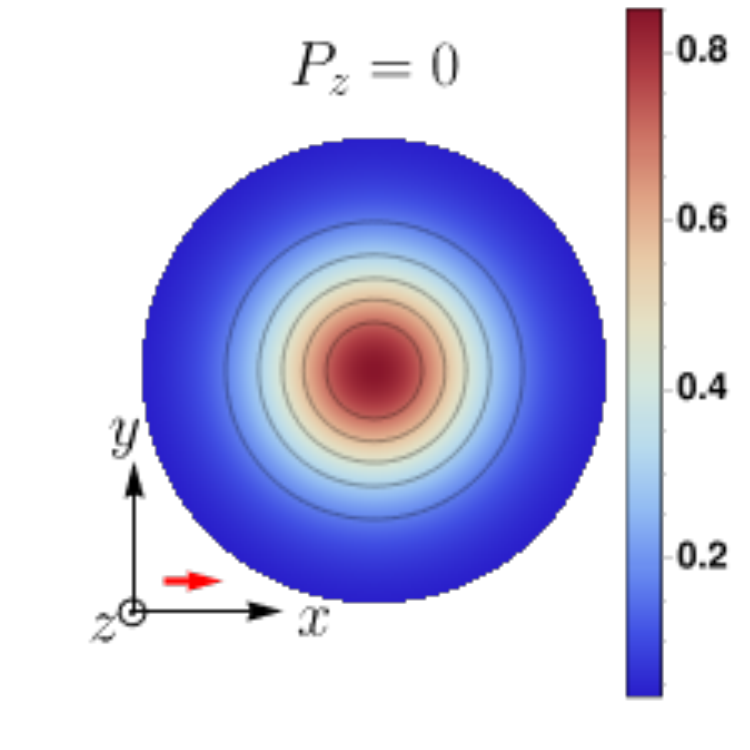}} 
    \subfigure[$s_x=1/2$,
    Monopole]{\includegraphics[width=0.23\linewidth]{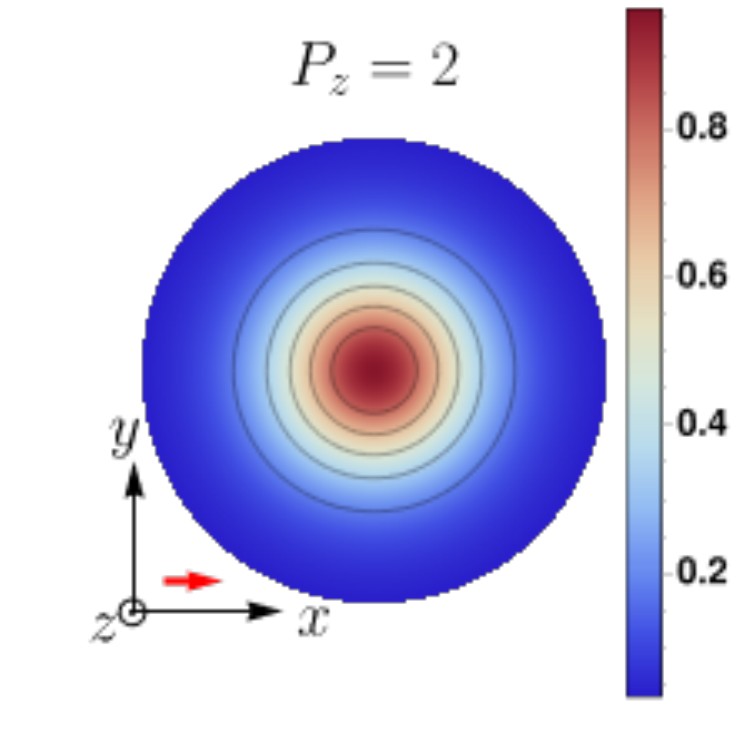}} 
    \subfigure[$s_x=1/2$,
    Monopole]{\includegraphics[width=0.23\linewidth]{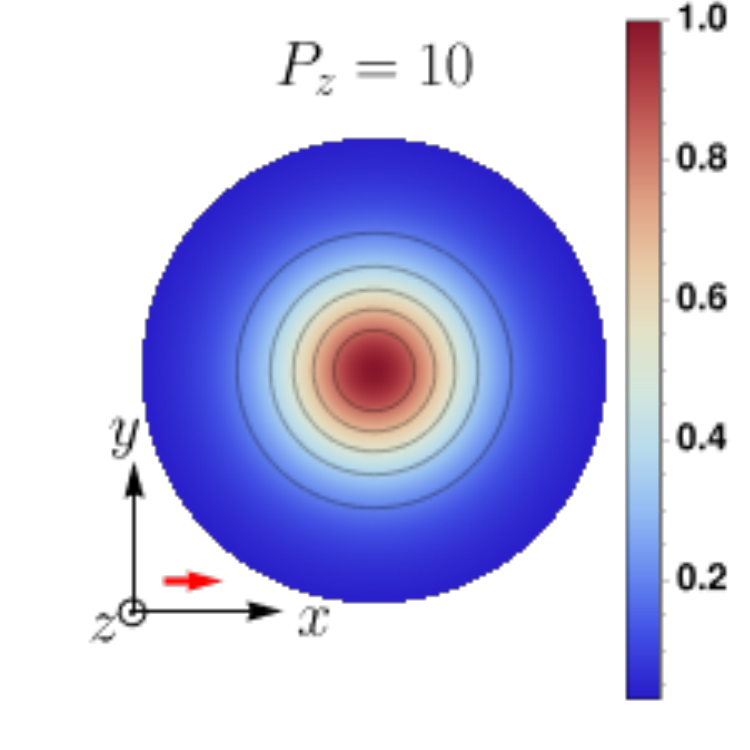}} 
    \subfigure[$s_x=1/2$,
    Monopole]{\includegraphics[width=0.23\linewidth]{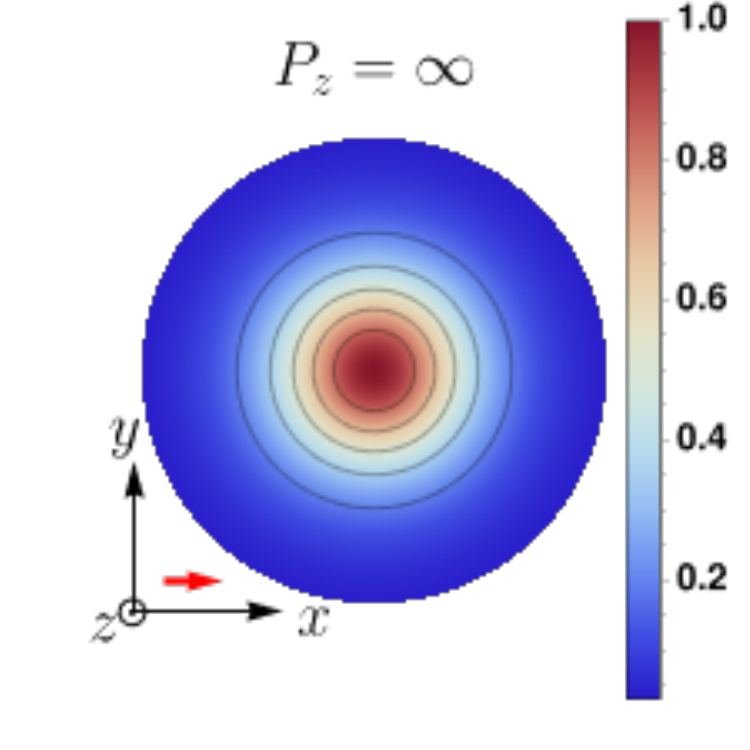}}\\ 
    \noindent
    \subfigure[$s_x=1/2$,
    Dipole]{\includegraphics[width=0.23\linewidth]{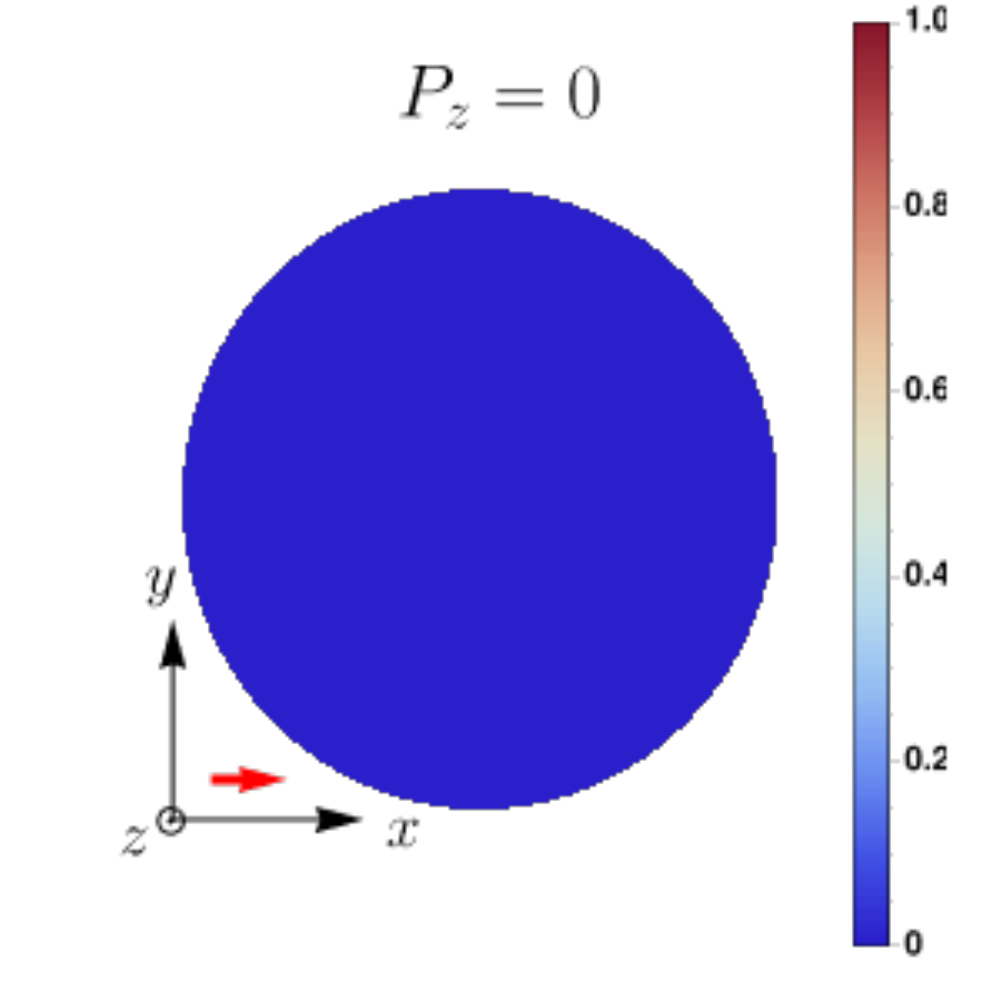}} 
    \subfigure[$s_x=1/2$,
    Dipole]{\includegraphics[width=0.23\linewidth]{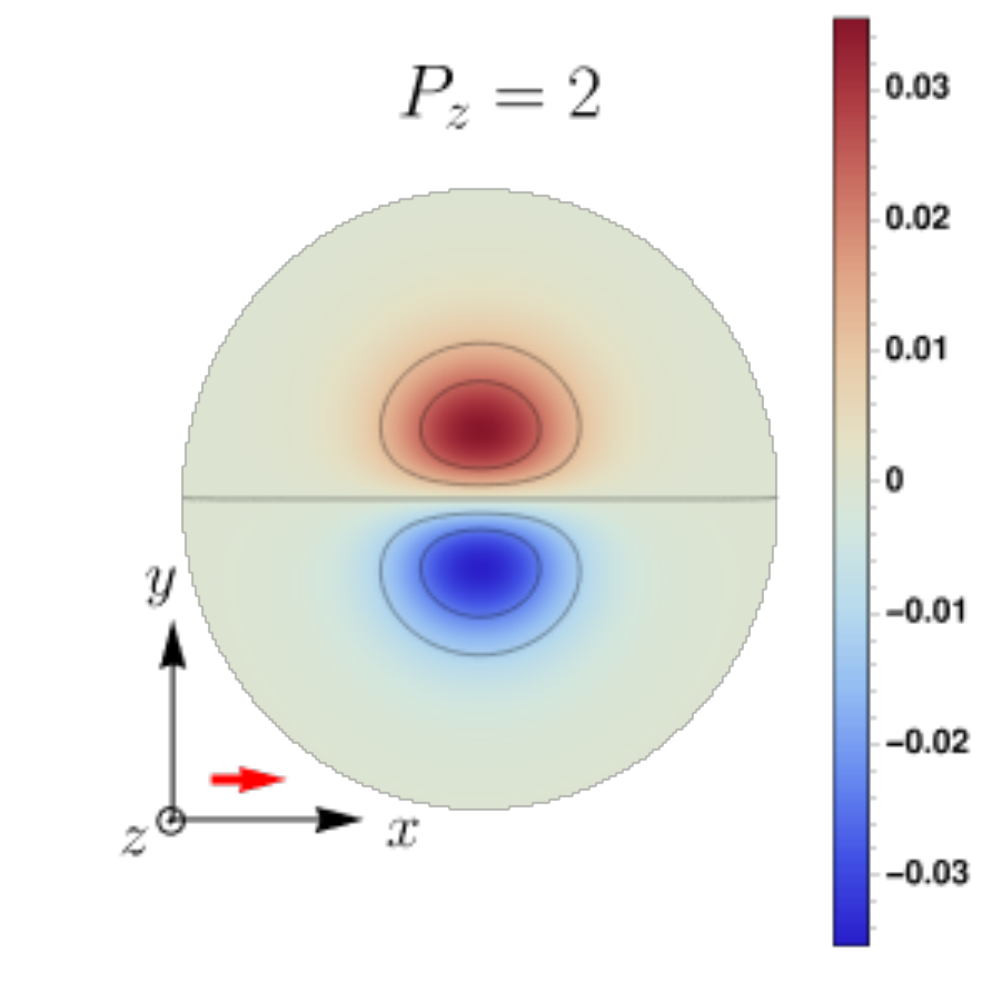}} 
    \subfigure[$s_x=1/2$,
    Dipole]{\includegraphics[width=0.23\linewidth]{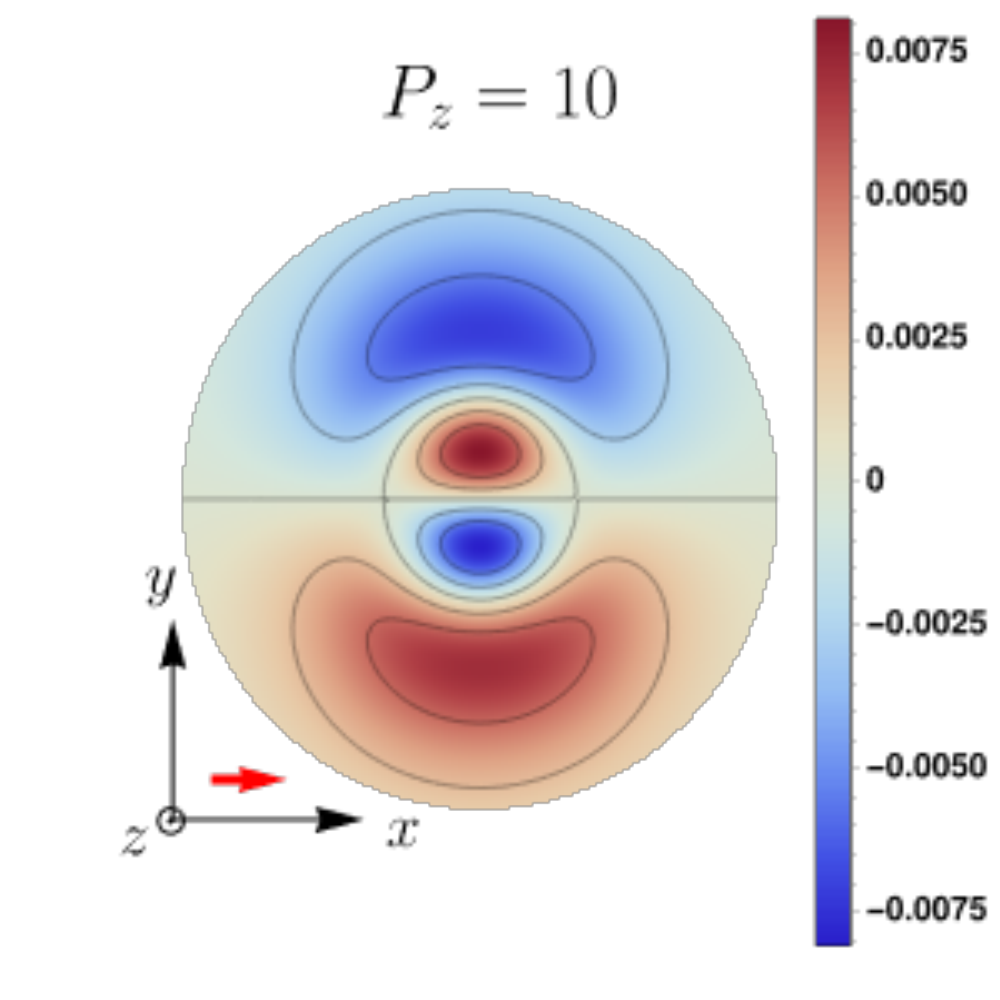}} 
    \subfigure[$s_x=1/2$,
    Dipole]{\includegraphics[width=0.23\linewidth]{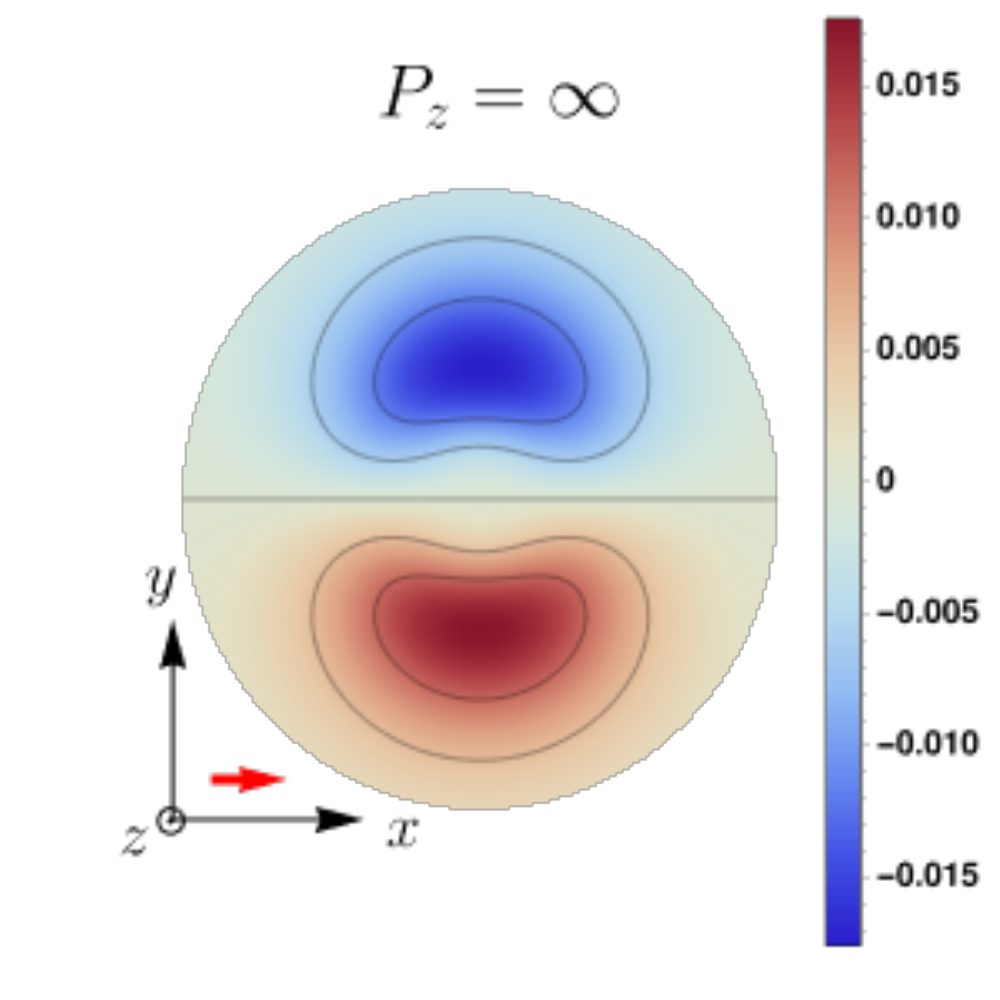}}\\ 
    \noindent
    \subfigure[$s_x=1/2$,
    Quadrupole]{\includegraphics[width=0.23\linewidth]{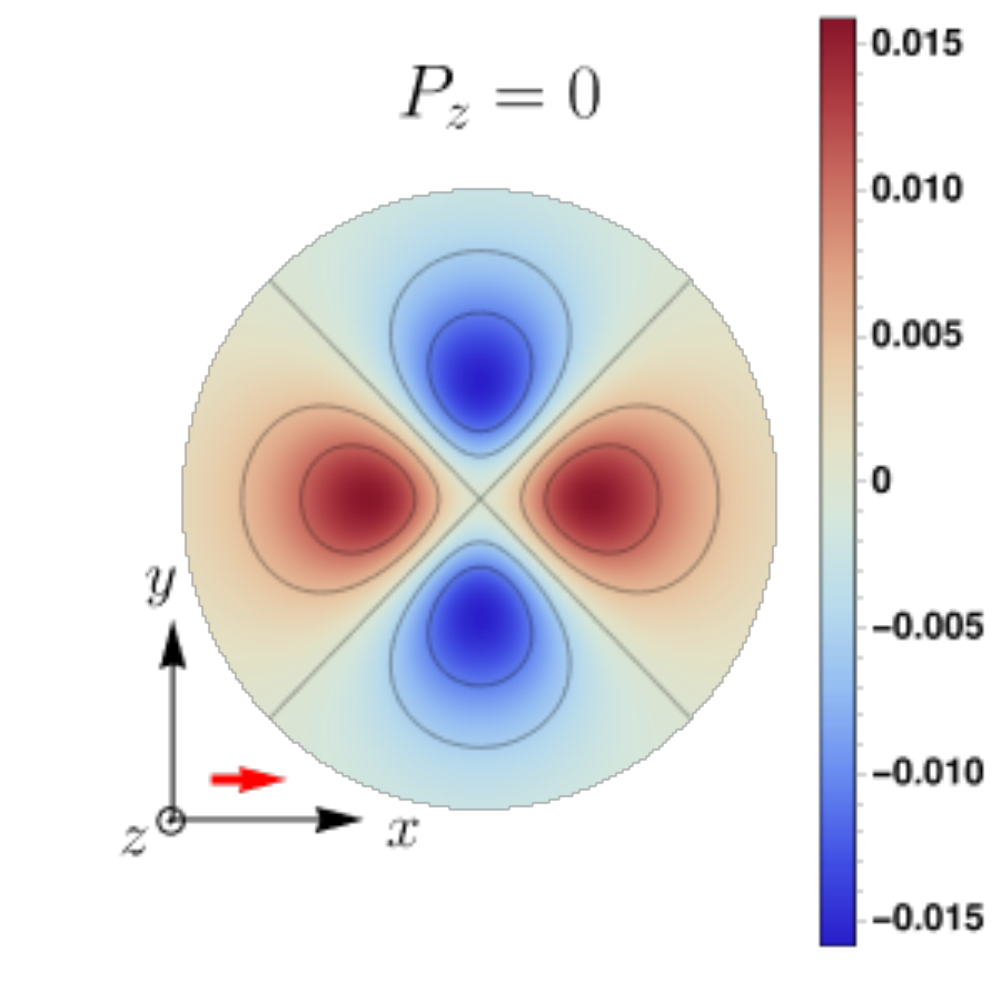}} 
    \subfigure[$s_x=1/2$,
    Quadrupole]{\includegraphics[width=0.23\linewidth]{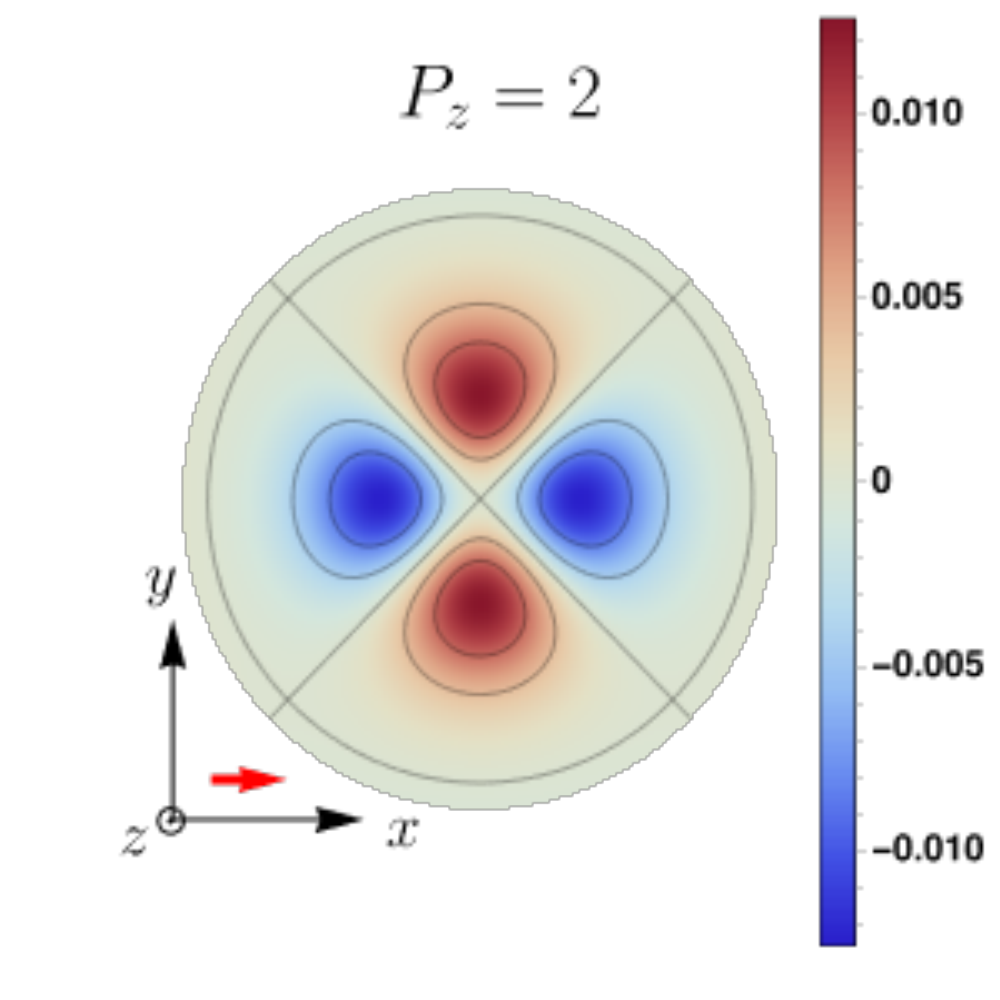}} 
    \subfigure[$s_x=1/2$,
    Quadrupole]{\includegraphics[width=0.23\linewidth]{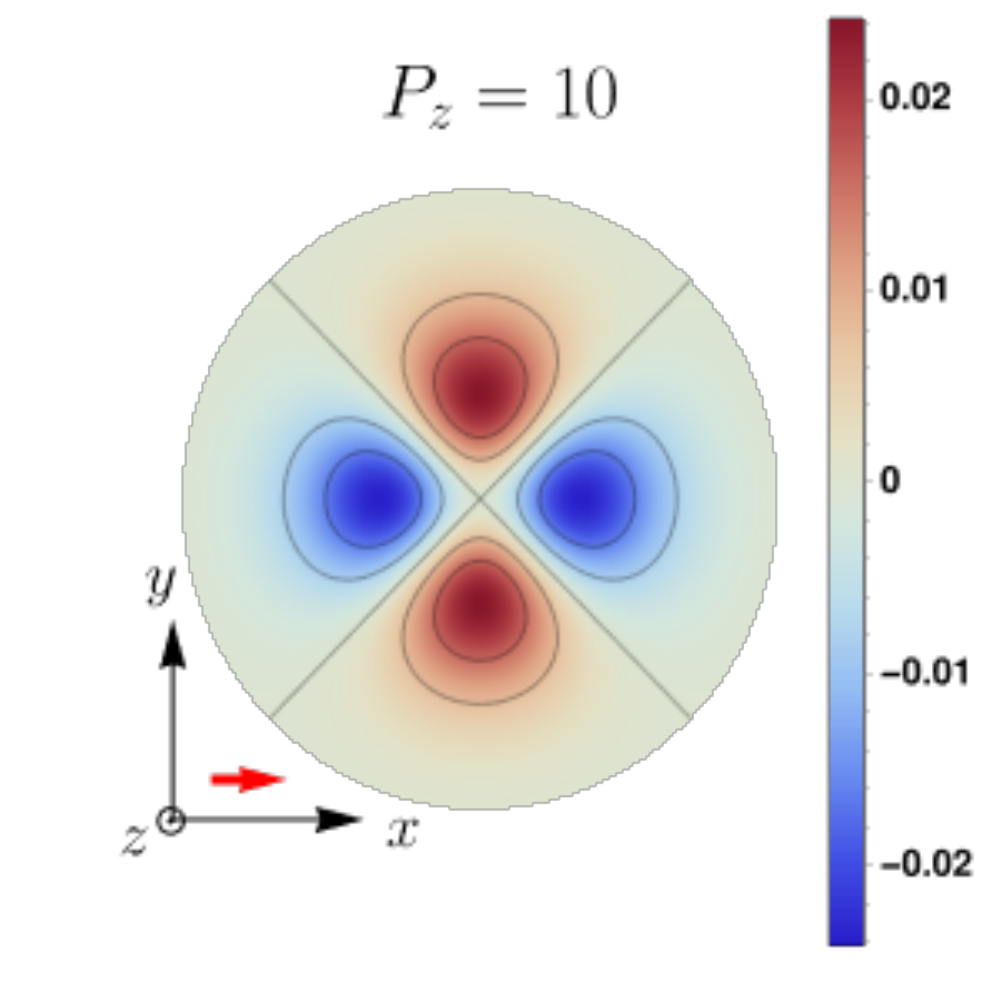}} 
    \subfigure[$s_x=1/2$,
    Quadrupole]{\includegraphics[width=0.23\linewidth]{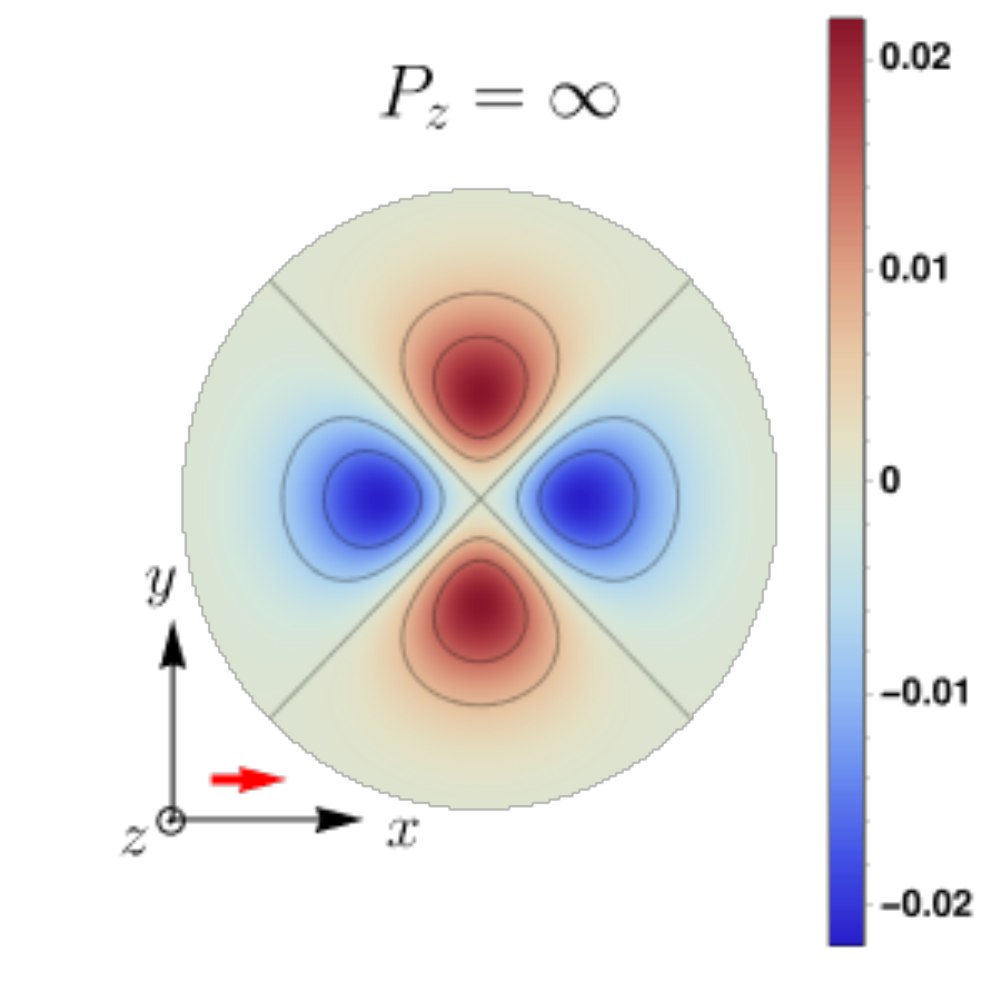}}\\ 
    \noindent
    \subfigure[$s_x=1/2$,
    Octupole]{\includegraphics[width=0.23\linewidth]{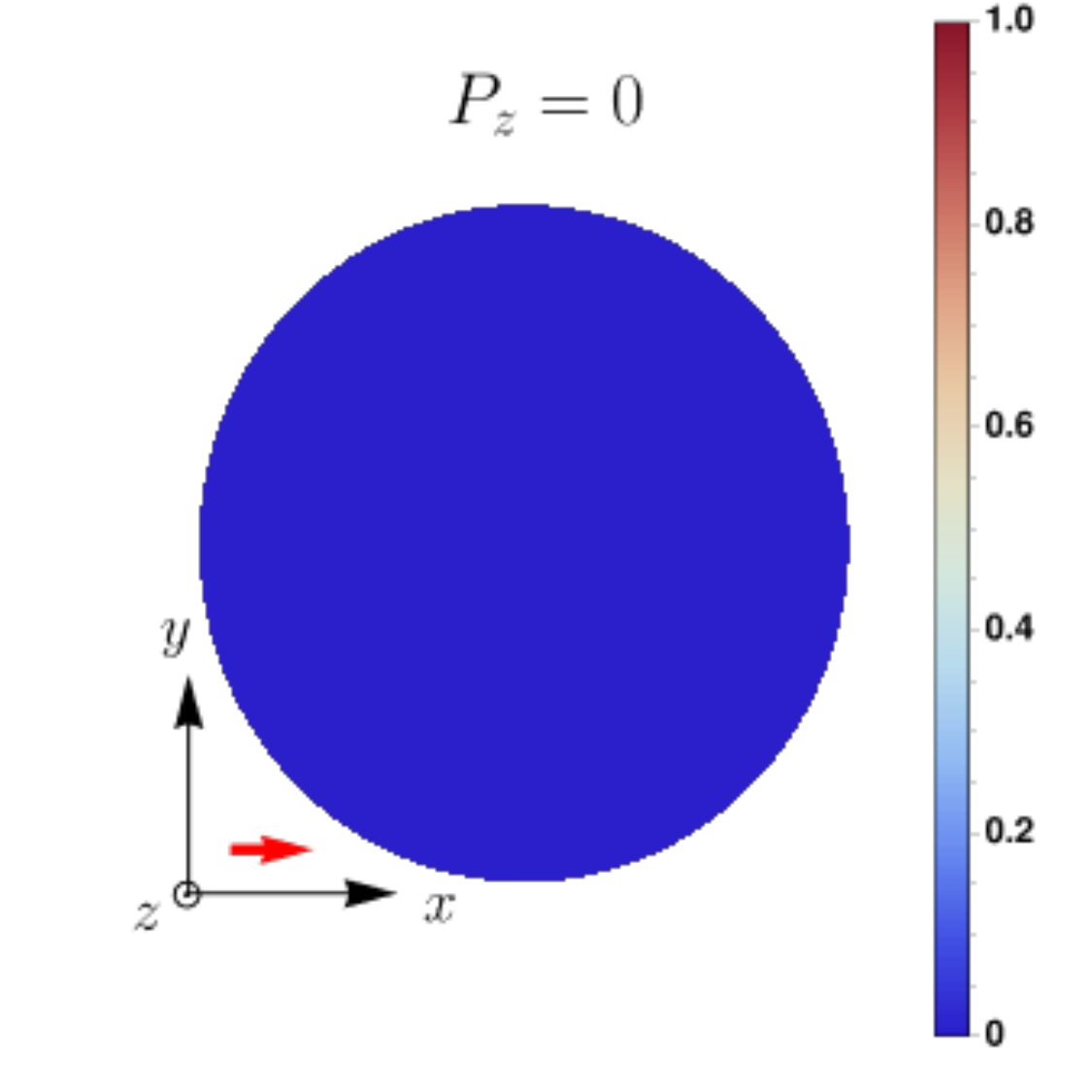}} 
    \subfigure[$s_x=1/2$,
    Octupole]{\includegraphics[width=0.23\linewidth]{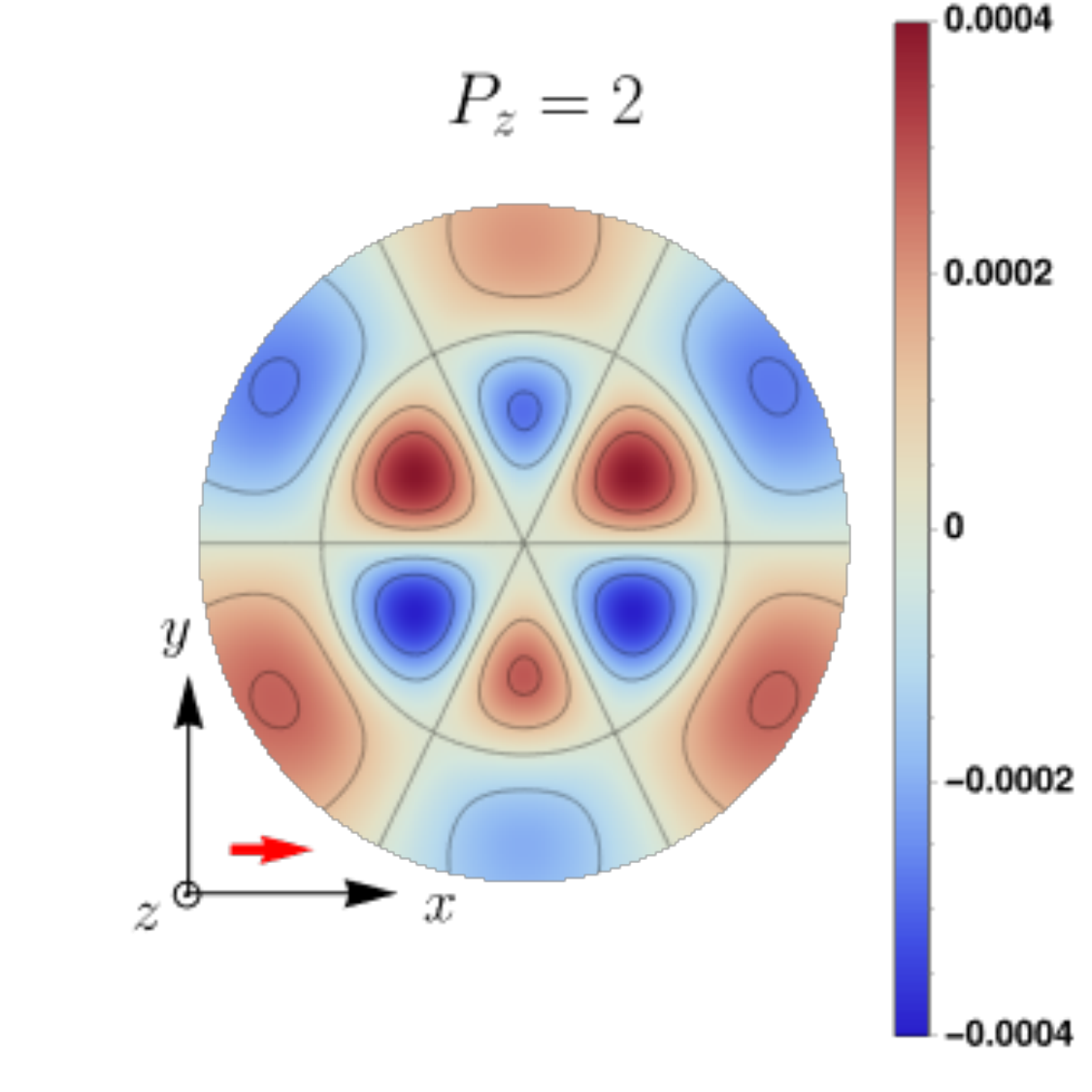}} 
    \subfigure[$s_x=1/2$,
    Octupole]{\includegraphics[width=0.23\linewidth]{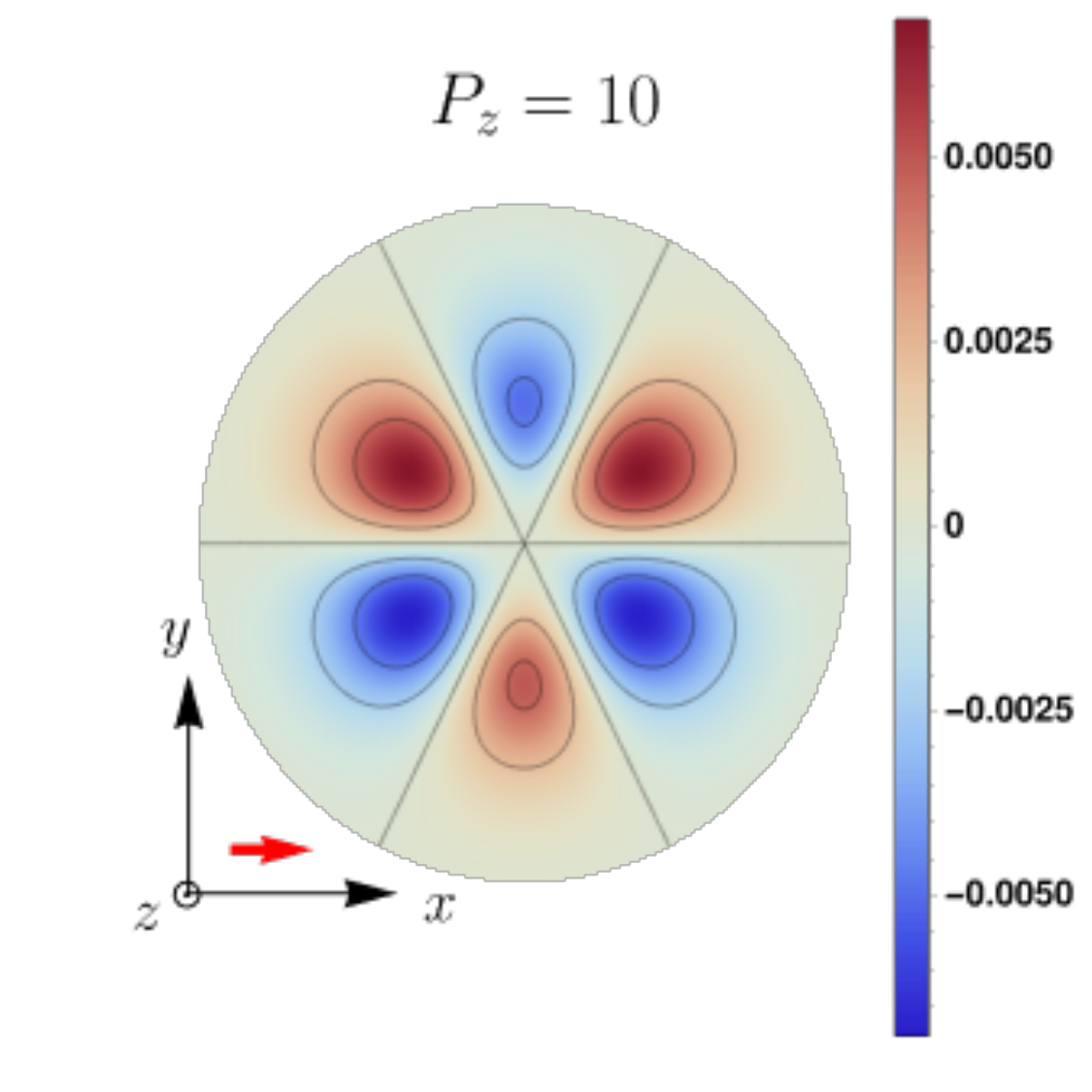}} 
    \subfigure[$s_x=1/2$,
    Octupole]{\includegraphics[width=0.23\linewidth]{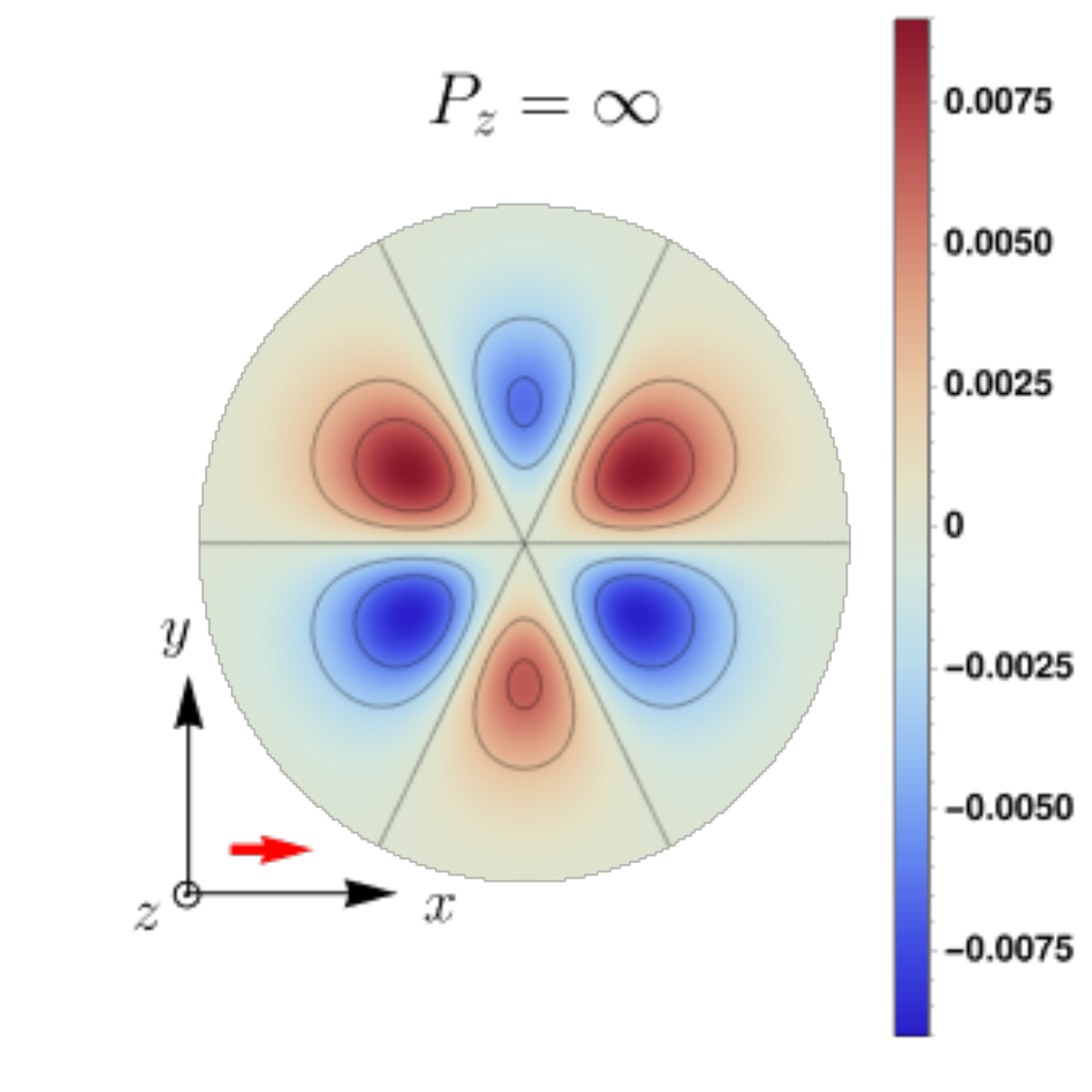}} 
    \caption{(a)-(d) monopole, (e)-(h) dipole, (i)-(l) quadrupole, and
      (m)-(p) octupole contributions of $\Delta^+$ with $s_x=1/2$ to
      the 2D charge distribution.} 
    \label{fig:7}
\end{figure}

In the upper panel of Fig.~\ref{fig:5}, we illustrate the transverse
charge distributions of the moving $\Delta^+$ baryon transversely
polarized along the $x$-axis with $s_x=3/2$. As shown in
Fig.~\ref{fig:3} and Fig.~\ref{fig:4}, the charge distribution is
deformed along the $y$-axis due to the presence of the quadrupole
contribution, so that it is not spherically symmetric. Of course,
there are no induced electric dipole and octupole contributions. The
first column in Fig.~\ref{fig:6} (Figs.~\ref{fig:6}(a), \ref{fig:6}(e),
\ref{fig:6}(i), and \ref{fig:6}(m)) shows the separate multipole
contributions when the $\Delta^+$ is at rest. Since the electric
dipole moment is induced as the $\Delta^{+}$ baryon is boosted along
the $z$-axis, the transverse charge distribution starts to get
deformed. At around $P_{z}=2\,\mathrm{GeV}$, the charge distribution
is shifted to the positive $x_\perp$-axis due to the induced dipole
contribution. On the other hand, the quadrupole contribution is
relatively small in comparison with the dipole one. One of the
remarkable features is that the sign of the quadrupole contribution is
reversed at around $P_{z}\sim1.4\,\mathrm{GeV}$. See the 
second column in Fig.~\ref{fig:6} (Figs.~\ref{fig:6}(b),
\ref{fig:6}(f), \ref{fig:6}(j), and \ref{fig:6}(n)). However, when
the system is boosted larger than $P_z \sim 10\,\mathrm{GeV}$, the
sign of the induced dipole contribution is reversed, so the charge
distribution is moved to the opposite direction, while the higher
multipoles contribute marginally to the charge distribution. In the
IMF, however, the quadrupole contribution dominates over the dipole 
contribution. See the last column in Fig.~\ref{fig:6} (Figs.~\ref{fig:6}(d),
\ref{fig:6}(h), \ref{fig:6}(l), and \ref{fig:6}(p)). When it comes
to the $s_x=1/2$, the tendency is almost the same as the case of 
$s_x=3/2$, but the quadrupole contribution has the opposite sign at
the rest frame. So, the charge distribution broadens
along the $x$-axis instead of the $y$-axis. See the first
column in Fig.~\ref{fig:7} (Figs.~\ref{fig:7}(a), \ref{fig:7}(e),
\ref{fig:7}(i), and \ref{fig:7}(m)).  

\begin{figure}[htpb]
    \centering
    \subfigure[Monopole,
    $s_x=3/2$]{\includegraphics[width=0.48\linewidth]{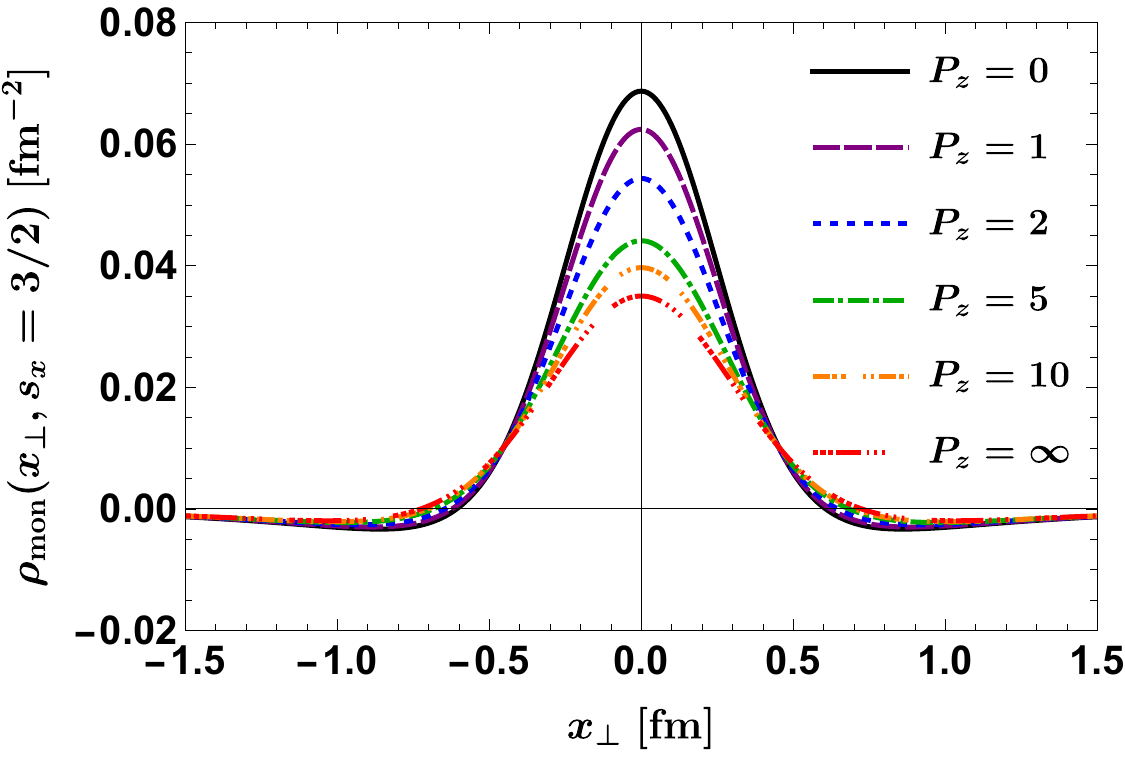}} 
    \subfigure[Dipole,$s_x=3/2$]{\includegraphics[width=0.48\linewidth]{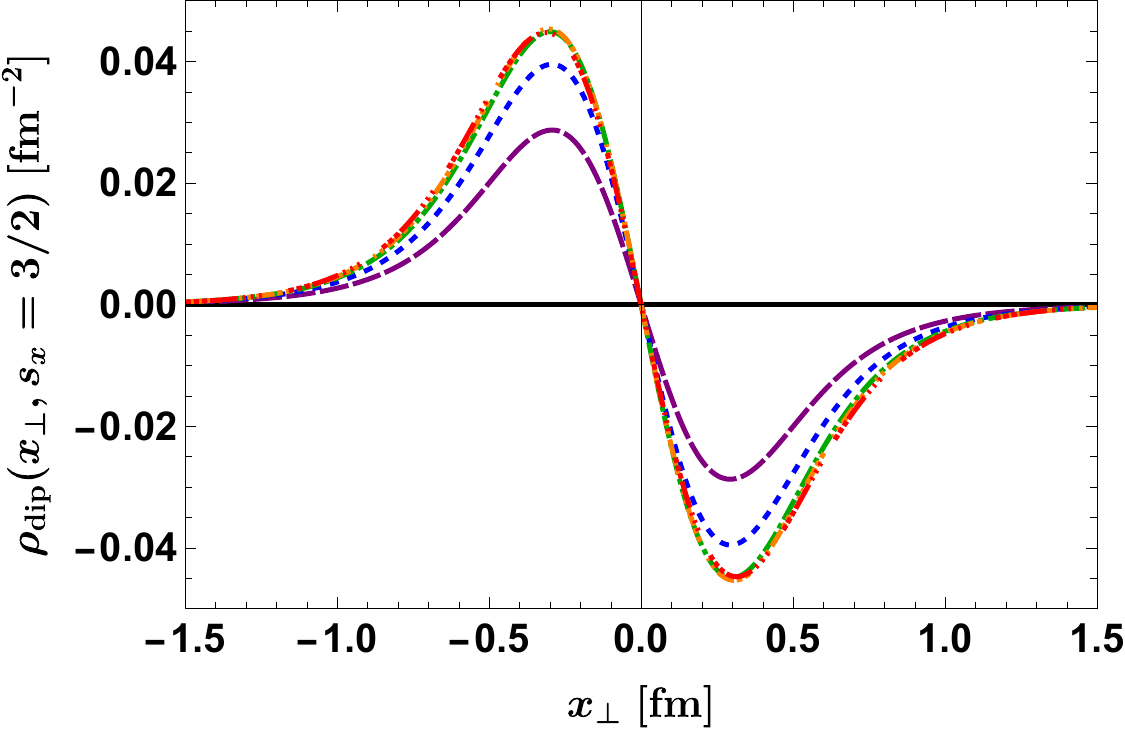}}\\ 
    \noindent
    \subfigure[Quadrupole,$s_x=3/2$]{\includegraphics[width=0.48\linewidth]{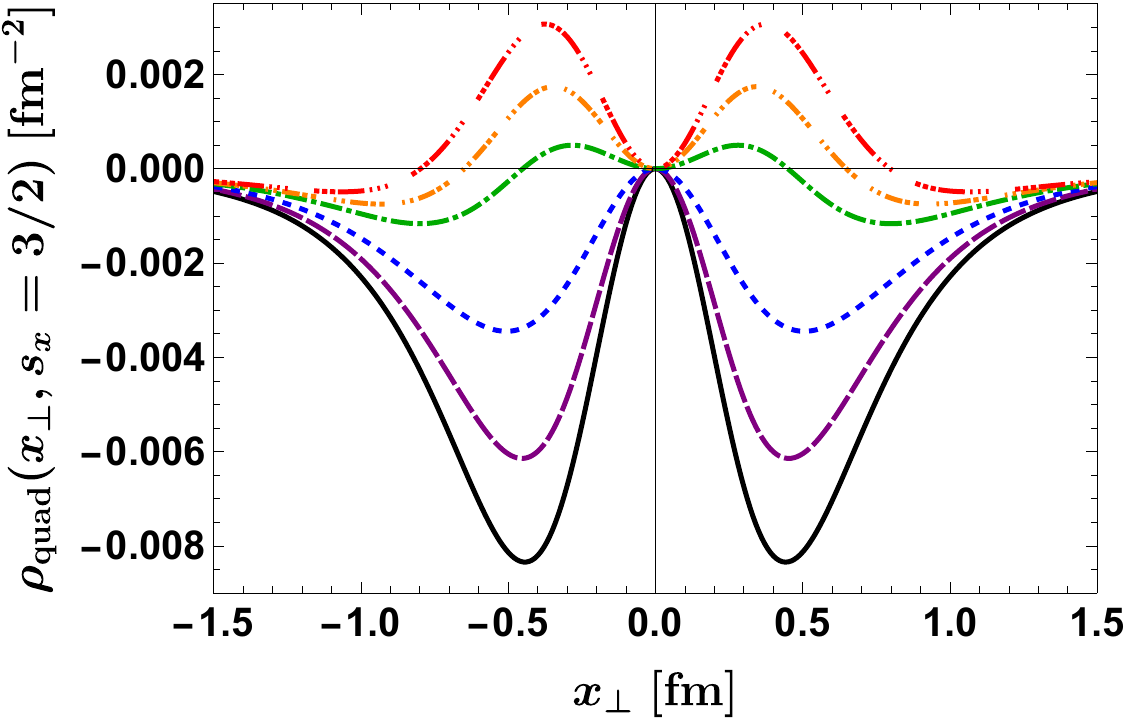}} 
    \subfigure[Octupole,$s_x=3/2$]{\includegraphics[width=0.48\linewidth]{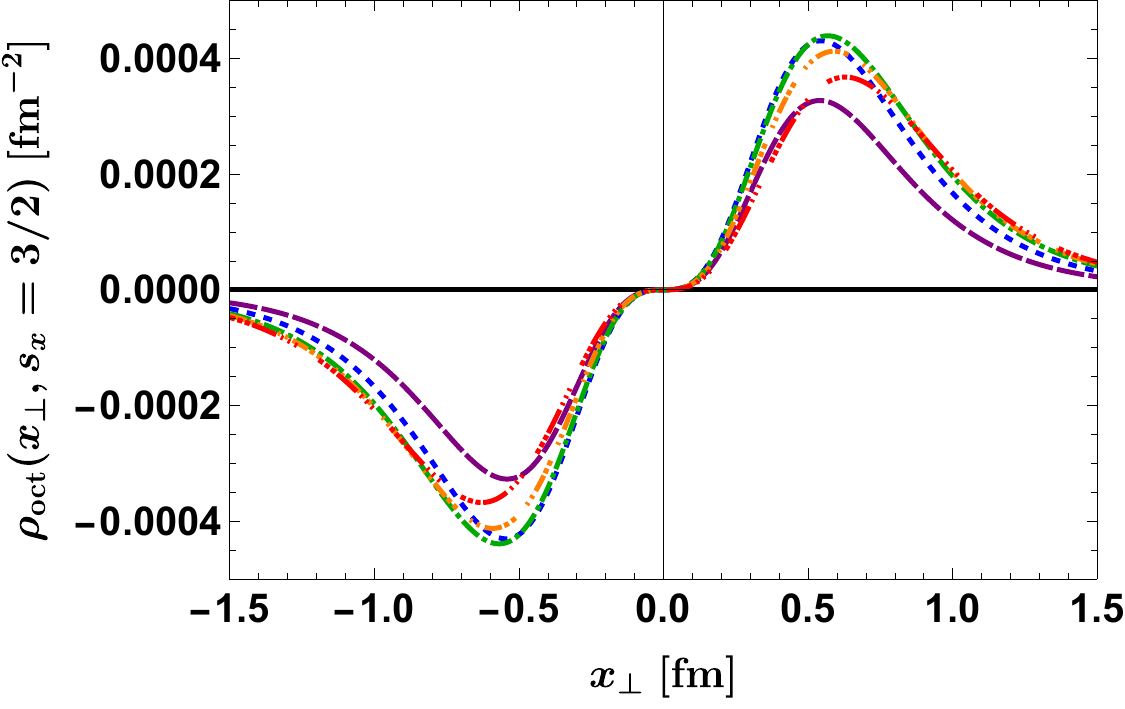}}\\
    \noindent
    \subfigure[$s_x=3/2$]{\includegraphics[width=0.55\linewidth]{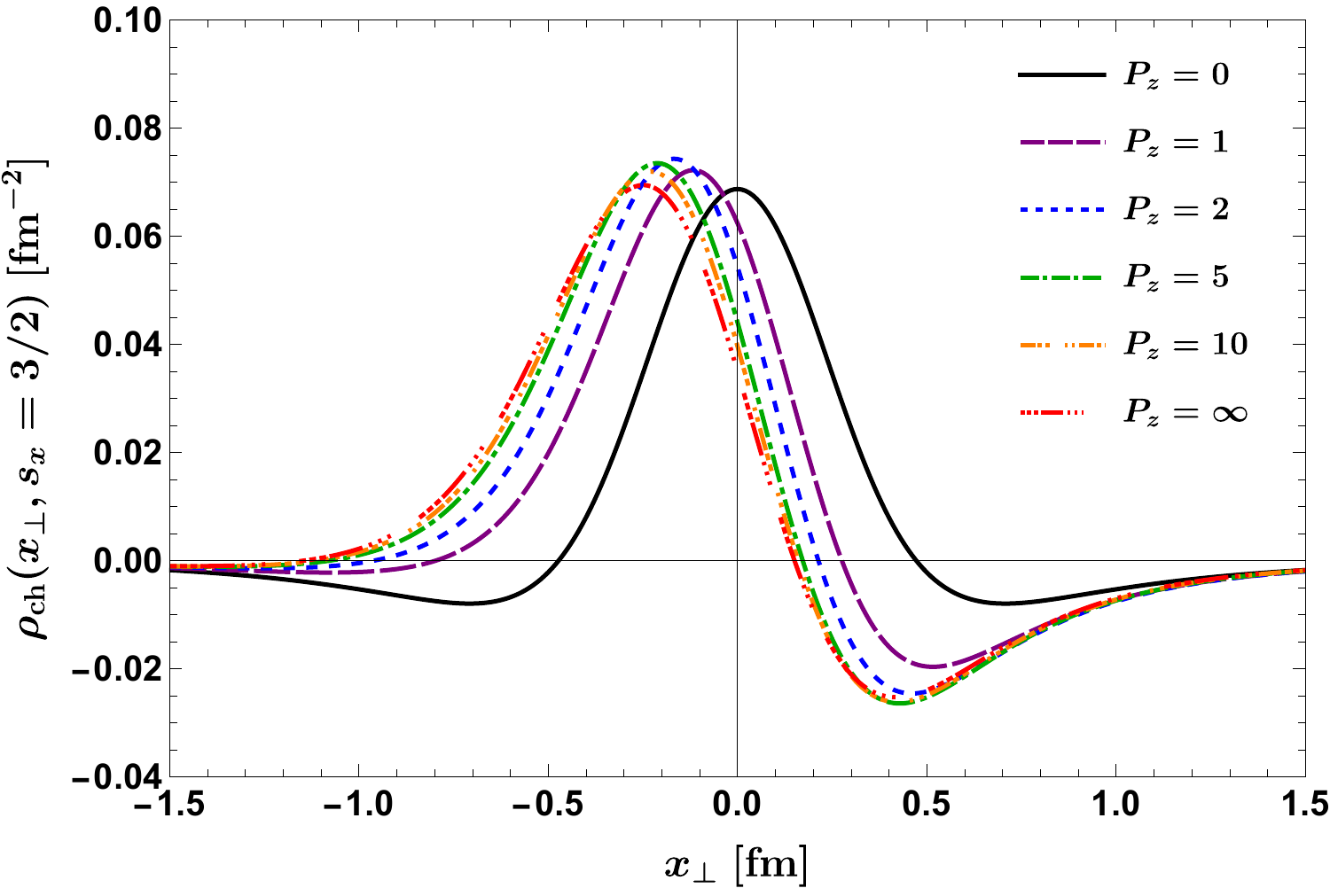}}
    \caption{(a) Monopole, (b) dipole, (c) quadrupole, and (d)
      octupole contributions to the $y$-axis profiles of the (e) 2D
      charge distributions of the $\Delta^0$ baryon when its spin is
      polarized along the $x$-axis with $s_{x}=3/2$.} 
    \label{fig:8}
\end{figure}

We also examine how the transverse charge distribution of the
$\Delta^{0}$ baryon transversely polarized along the $x$-axis varies 
under the Lorentz boost. In Fig.~\ref{fig:8}, we draw the transverse
charge distributions of the $\Delta^0$ baryon when its spin is
polarized along the $x$-axis with $s_{x}=3/2$. See Fig.~\ref{fig:8}
(Figs.~\ref{fig:8}(a)-(e)). We found that the transverse $\Delta^0$
charge distribution is dramatically changed under the Lorentz boost,
in contrast with that of the $\Delta^{+}$ baryon. At the rest frame,
the monopole contribution is positive at the inner part, whereas the
quadrupole contribution is negative over $r$. Obviously, there are no
induced dipole and octupole contributions. So, while the transverse
charge distribution is kept to be positive at the core part, the
quadrupole contribution pulls it down to be negative at the outer
part. As a result, the nodal point of the transverse charge
distribution gets close to the center of the $\Delta^{0}$ baryon due
to the quadrupole contribution. As $P_z$ increases, the charge
distribution starts to be deformed. The dominant contribution to 
$\rho_{\mathrm{ch}}(x_\perp)$ is the monopole one, and it is always
kept to be positive at the core part. As we explained before, the
monopole contribution to the transversely polarized charge
distribution under the Lorentz boost turns out to be always positive
at the core part, though there is a sign flip of the longitudinally
polarized charge distribution. The quadrupole contribution turns  
positive at around $P_z \sim 4.0\,\mathrm{GeV}$ at the core part. At
the same moment, the induced dipole contribution pushes the charge
distribution to the negative $y$-direction, which dominates over the
higher multipole contributions. The value of the
$G^{\Delta^{0}}_{M1}(0)\sim -0.3\,\mu_N$~\cite{Kim:2019gka} is solely
governed by the induced dipole moment because of the
$G^{\Delta^{0}}_{E0}(0)=0$. This is the reason why the transverse charge
distribution of the $\Delta^{0}$ is deformed as that of the neutron, i.e,
$G^{n}_{M1}(0) = -1.91\,\mu_N$~\cite{Workman:2022ynf}, as
shown in Fig.~\ref{fig:8}(e).

\begin{figure}[htpb]
    \subfigure[Monopole,$s_x=1/2$]{\includegraphics[width=0.48\linewidth]{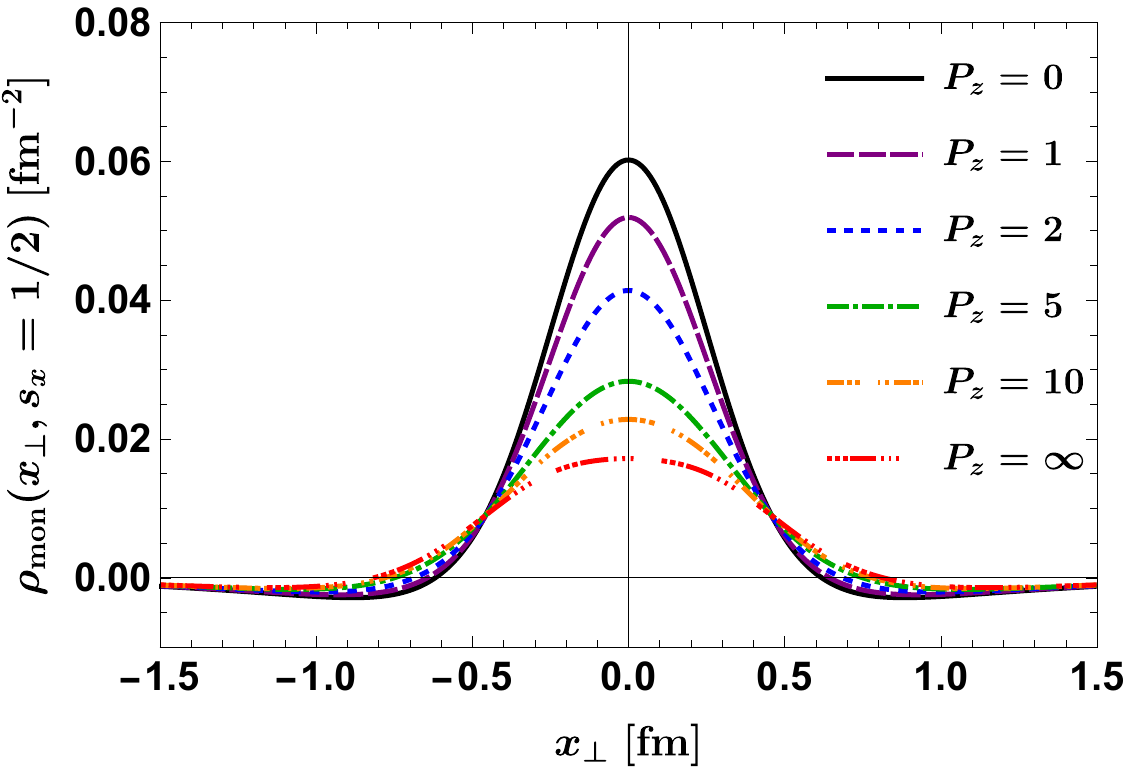}}
    \subfigure[Dipole,$s_x=1/2$]{\includegraphics[width=0.48\linewidth]{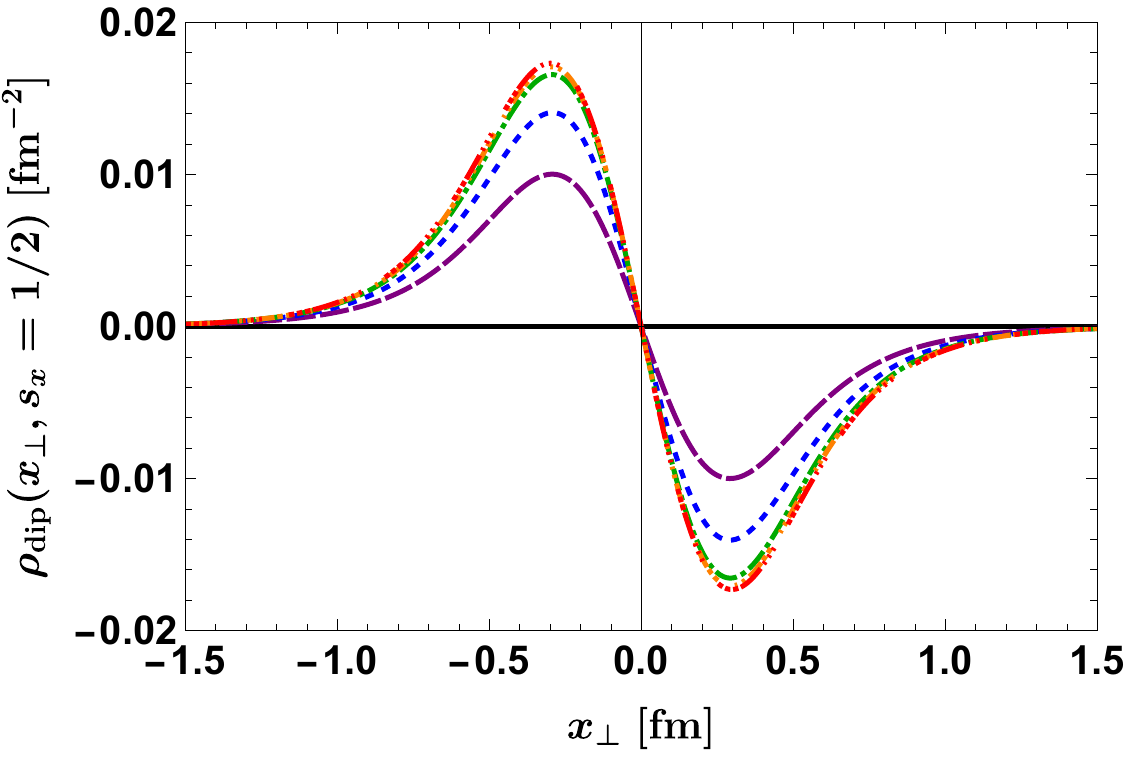}}\\
    \subfigure[Quadrupole,$s_x=1/2$]{\includegraphics[width=0.48\linewidth]{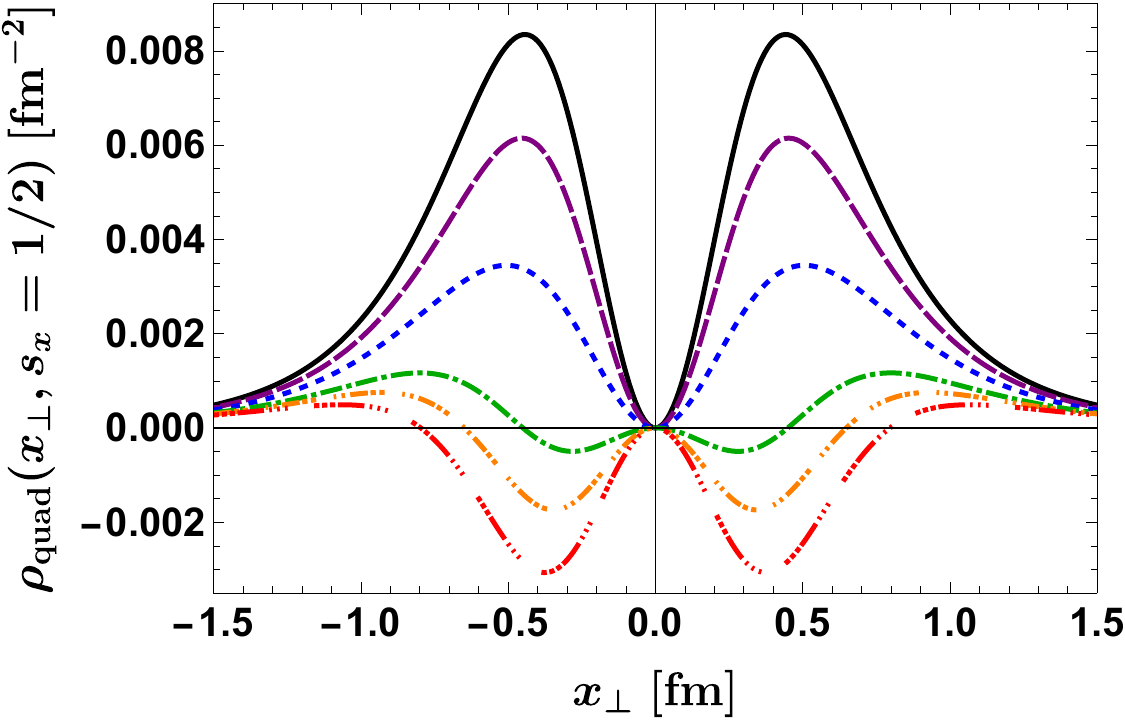}}
    \subfigure[Octupole,$s_x=1/2$]{\includegraphics[width=0.48\linewidth]{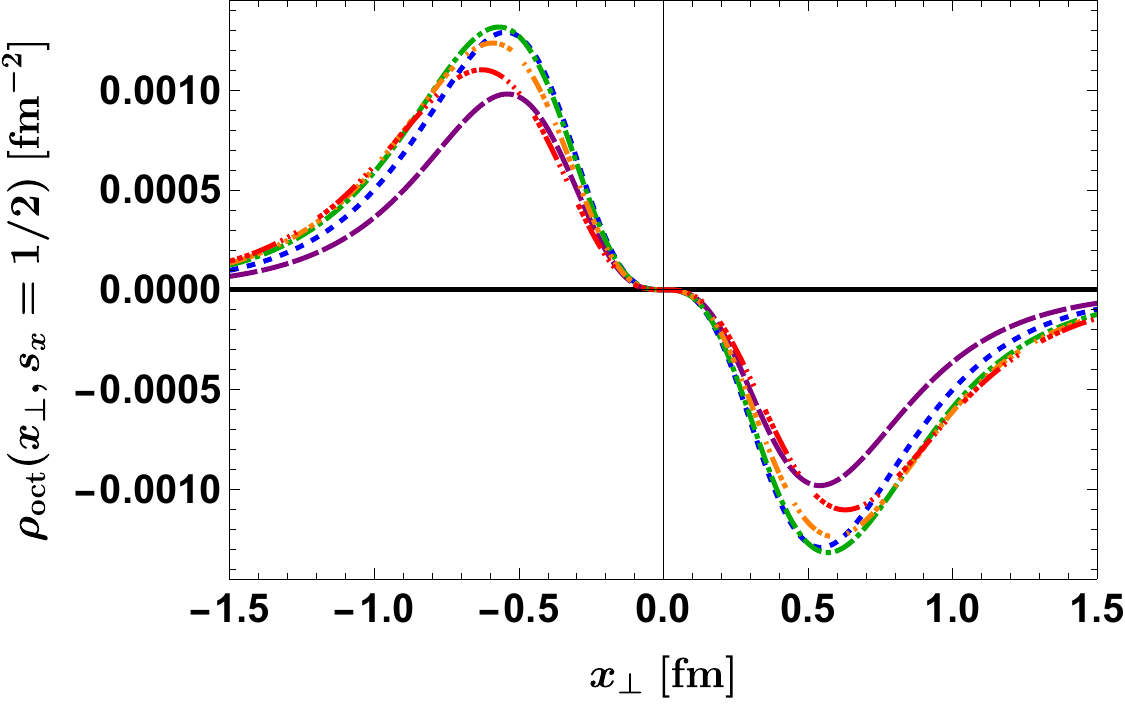}}\\
    \noindent
    \subfigure[$s_x=1/2$]{\includegraphics[width=0.55\linewidth]{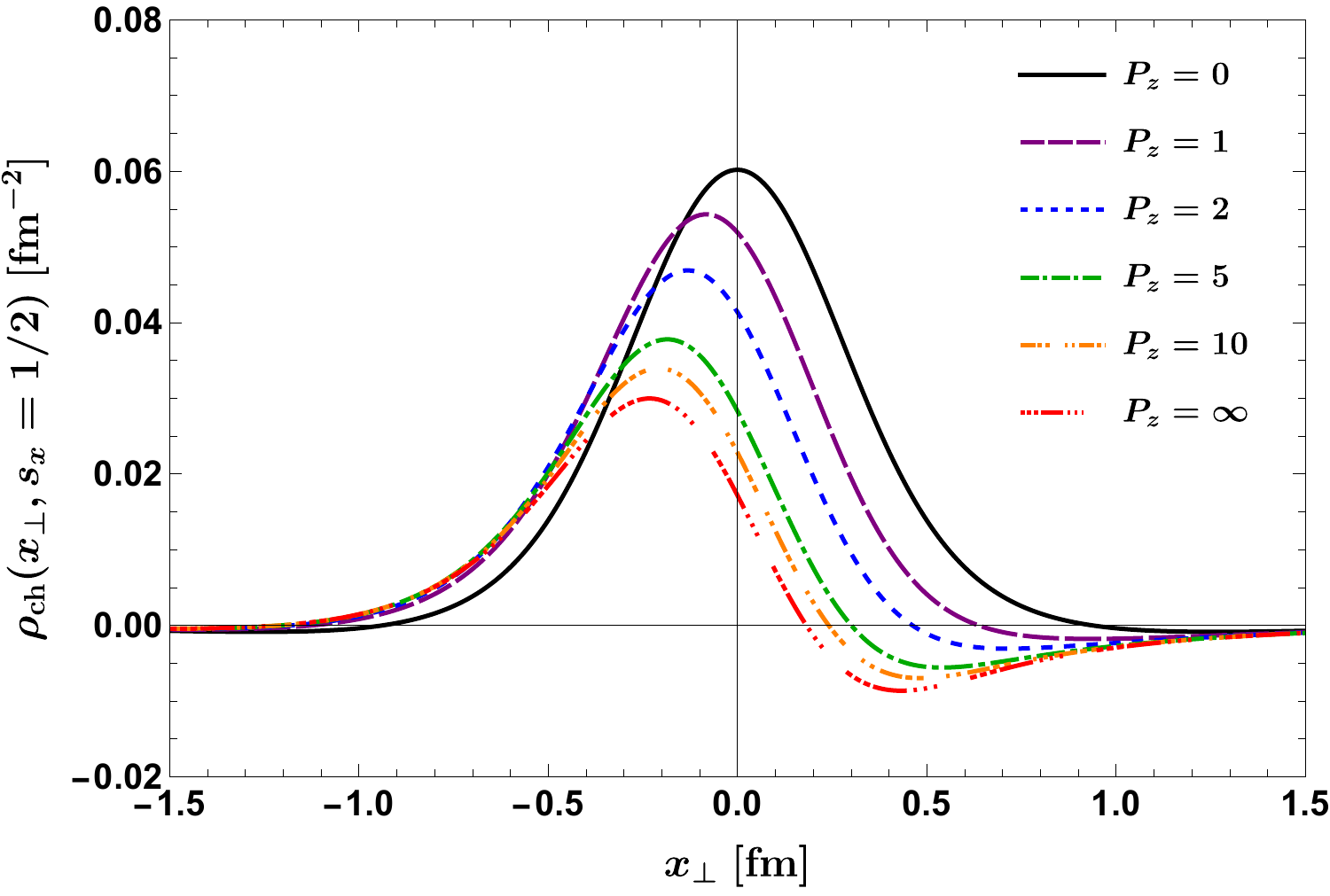}}
    \caption{(a) monopole, (b) dipole, (c) quadrupole, and (d) octupole contributions to the $y$-axis profiles of the (e) 2D charge distributions of the $\Delta^0$ baryon when its spin is polarized along the $x$-axis with $s_{x}=1/2$}
    \label{fig:9}
\end{figure}

In the IMF, we finally obtain the transverse charge distribution
shifted to the negative $y$-axis. When the spin projection is
$s_{x}=1/2$, we are able to see that quadrupole 
contribution is opposite to the $s_{x}=3/2$ case in the rest frame. See Fig.~\ref{fig:9} (Fig.~\ref{fig:9}(a)-(d)). The
quadrupole contribution makes a rather weak plateau at the core part of the charge distribution,
which is a similar feature to the deuteron charge
distribution~\cite{Kim:2022bia,   Lorce:2022jyi}. When the system is
boosted, the quadrupole contribution is relatively suppressed and the
induced dipole contribution dominates over it. In the IMF, we obtained
the $\Delta^{0}$ charge distribution $s_{x}=1/2$, which has a similar 
shape and strength to that with $s_{x}=3/2$. 

\begin{figure}[htpb]
    \centering  
    \subfigure[$s_x=3/2$]{\includegraphics[width=0.23\linewidth]{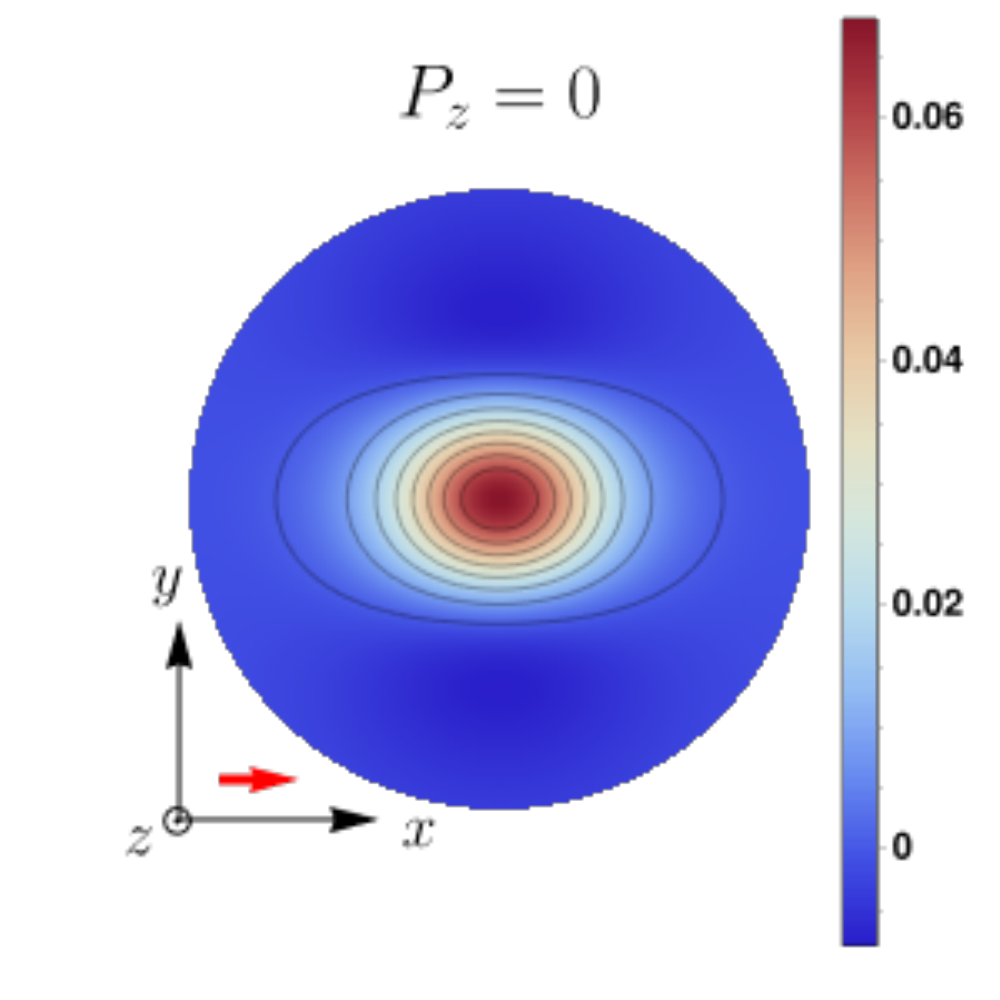}}
    \subfigure[$s_x=3/2$]{\includegraphics[width=0.23\linewidth]{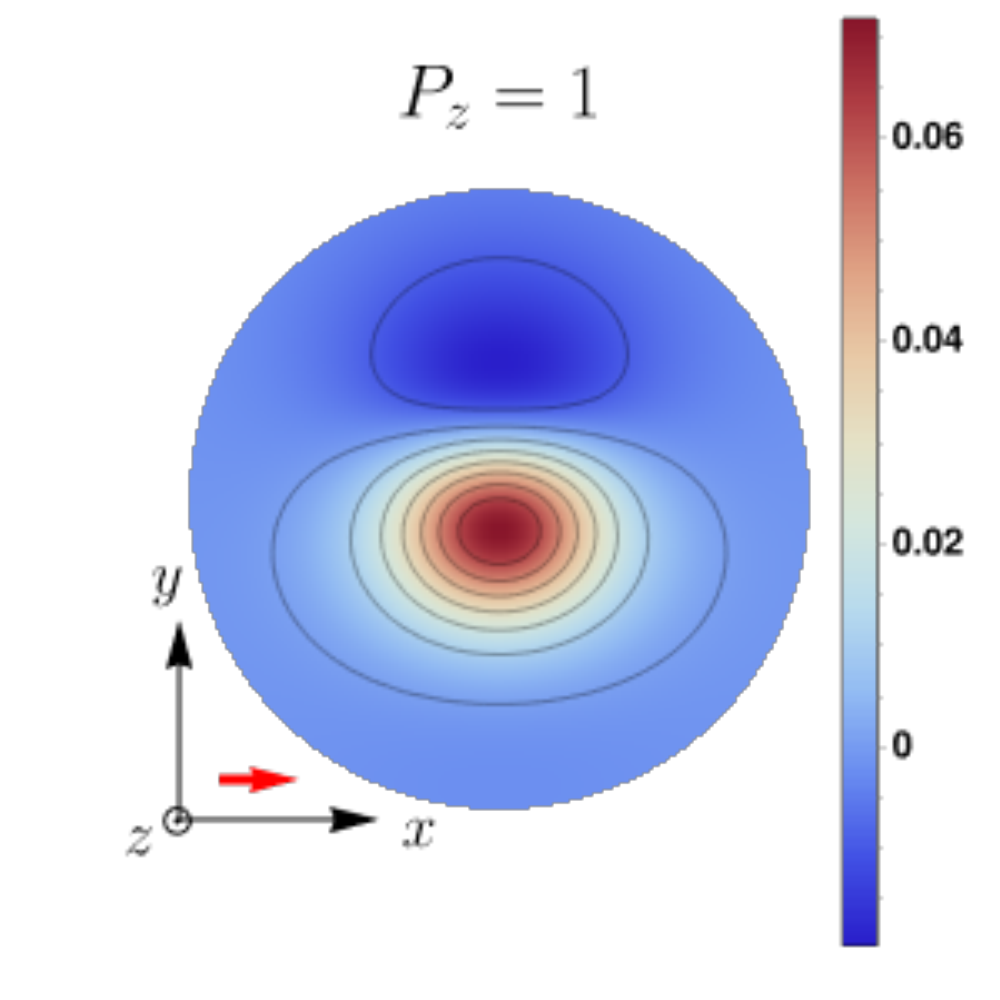}}
    \subfigure[$s_x=3/2$]{\includegraphics[width=0.23\linewidth]{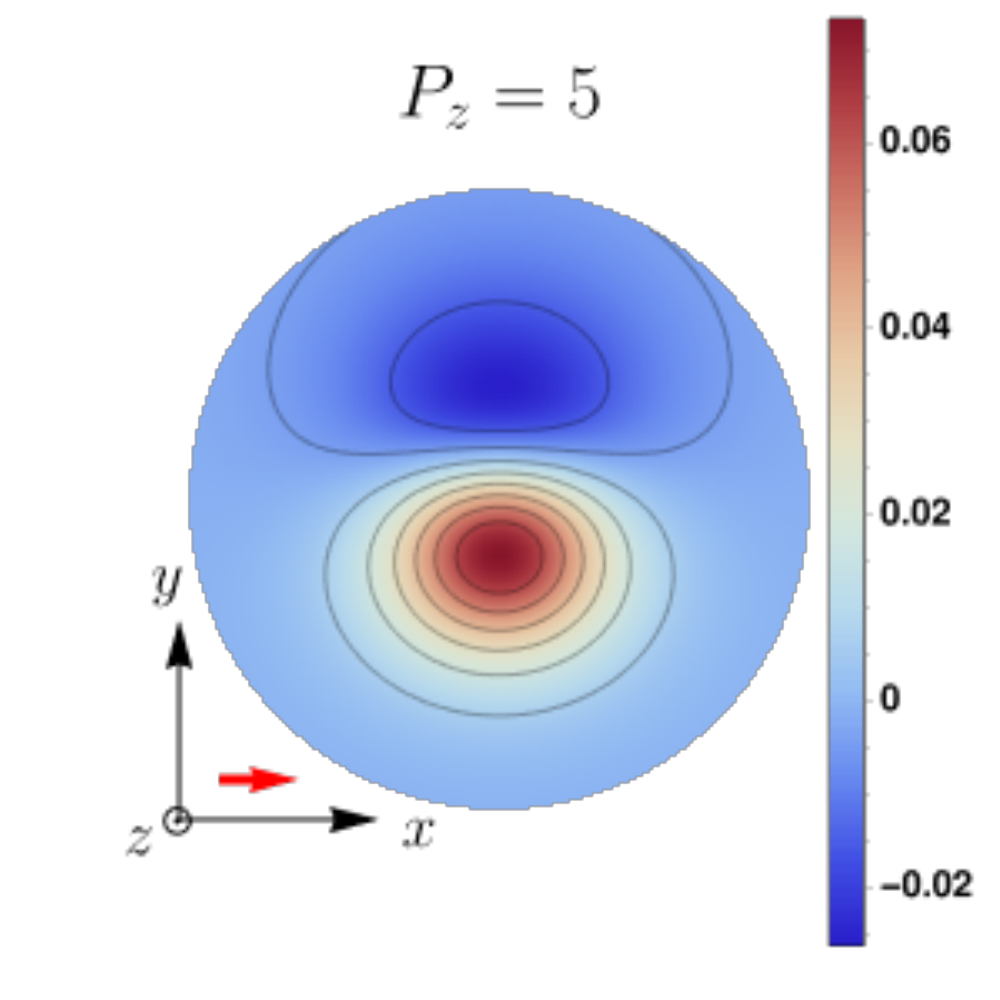}}
    \subfigure[$s_x=3/2$]{\includegraphics[width=0.23\linewidth]{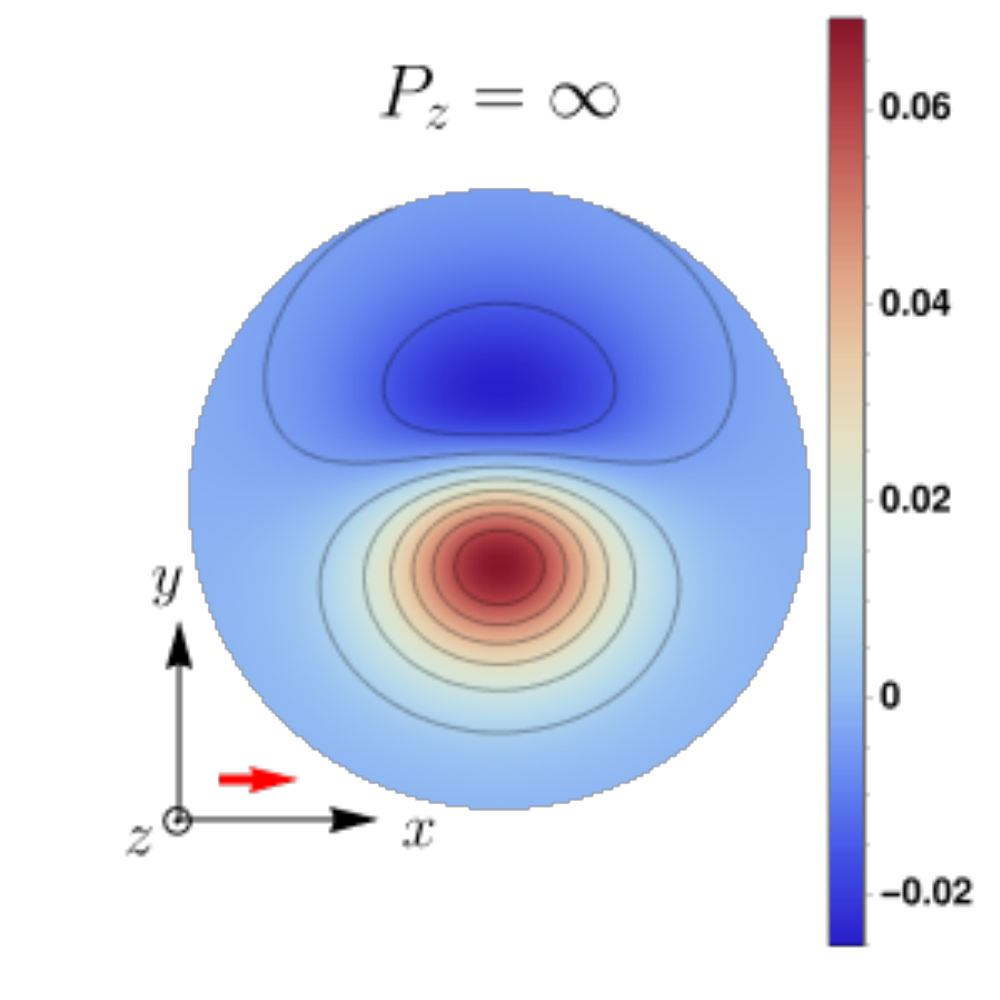}}
    \subfigure[$s_x=1/2$]{\includegraphics[width=0.23\linewidth]{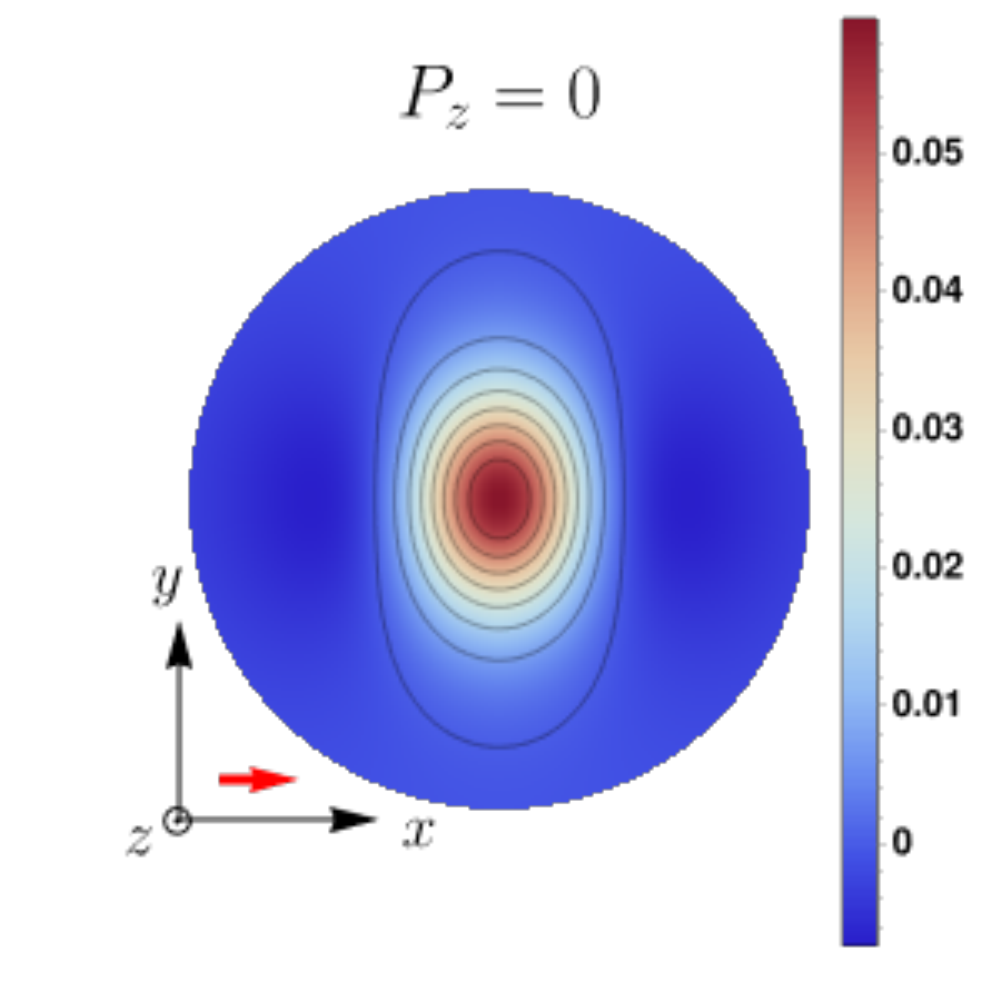}}
    \subfigure[$s_x=1/2$]{\includegraphics[width=0.23\linewidth]{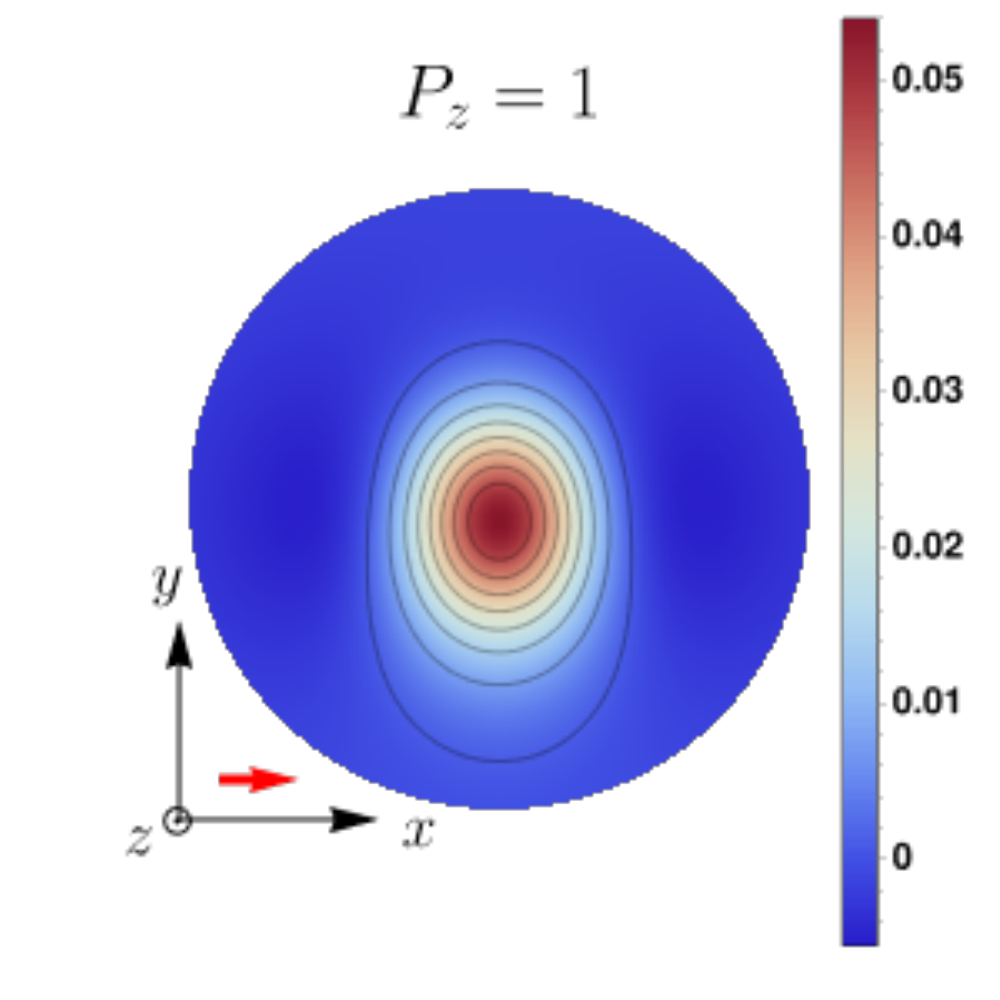}}
    \subfigure[$s_x=1/2$]{\includegraphics[width=0.23\linewidth]{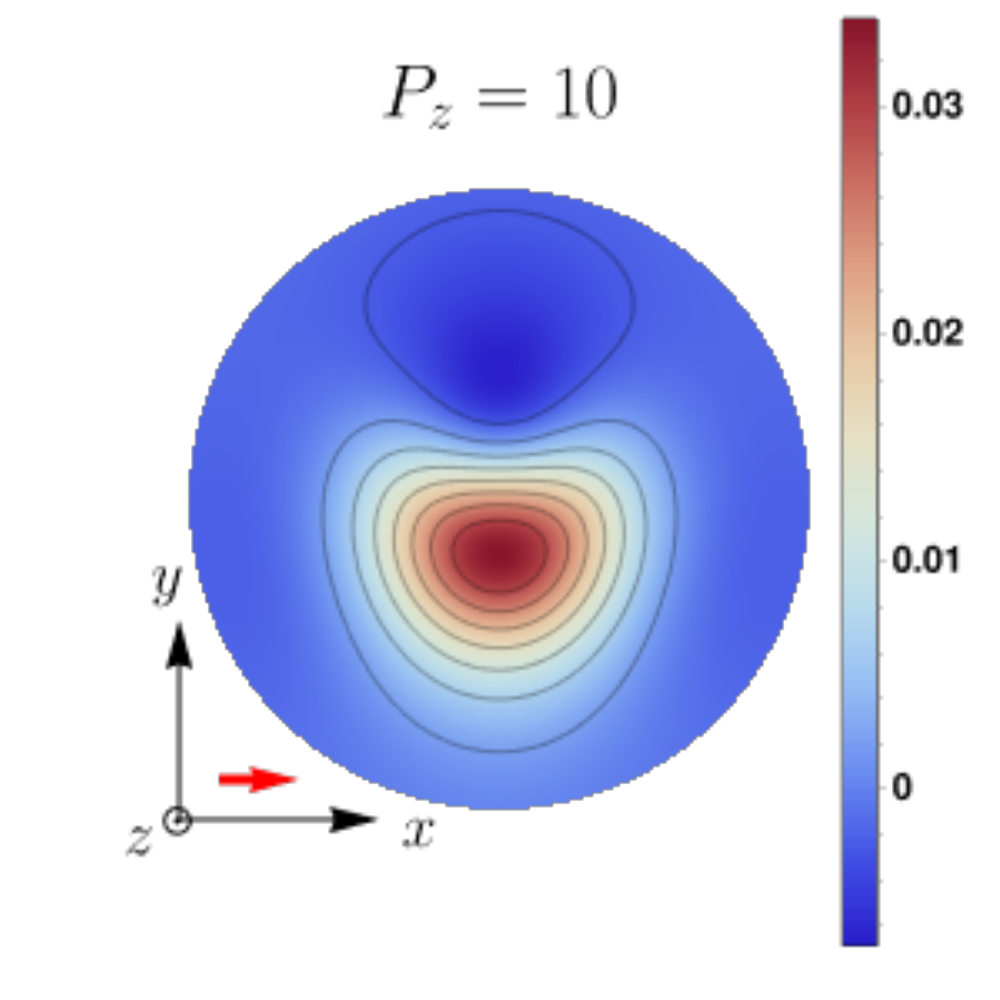}}
    \subfigure[$s_x=1/2$]{\includegraphics[width=0.23\linewidth]{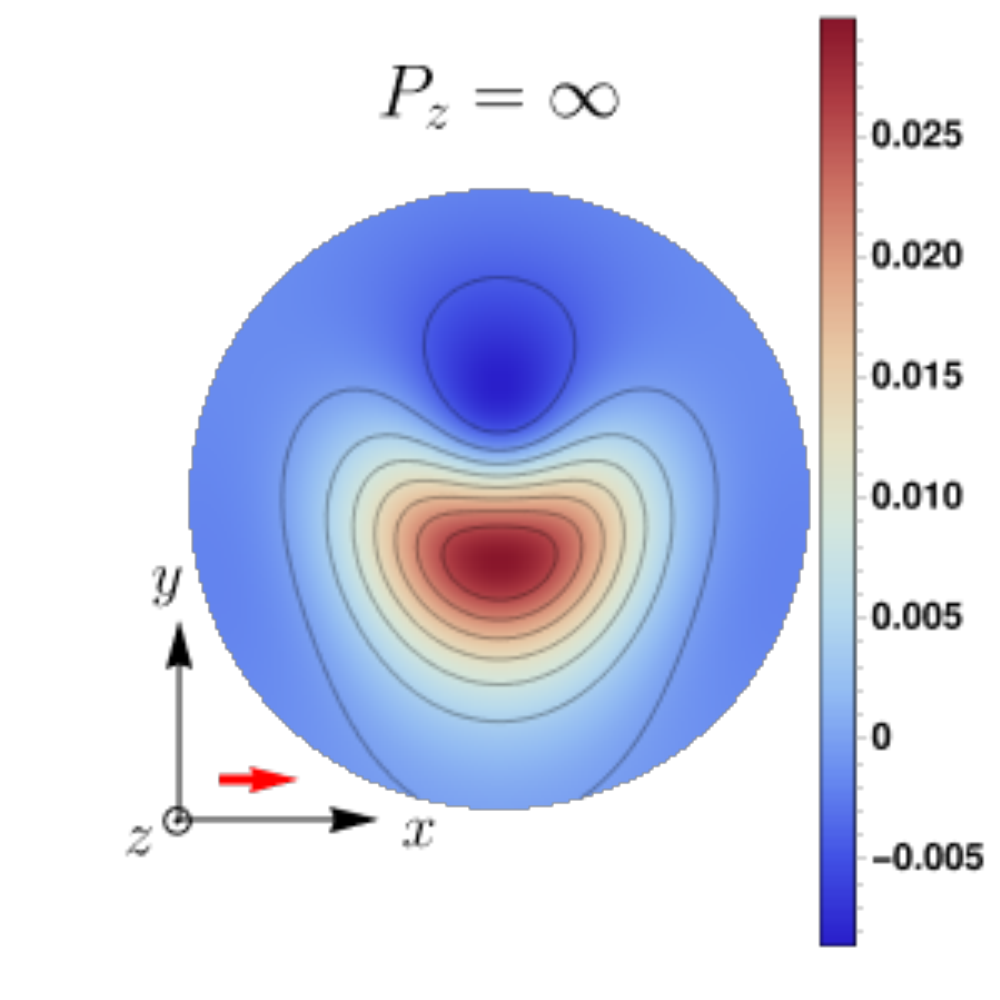}}
    \caption{(a)-(d) 2D charge distributions of the moving $\Delta^0$ baryon transversely polarized along $x$-axis with $s_x=3/2$; (e)-(h) 2D charge distributions of the moving $\Delta^0$ baryon transversely polarized along $x$-axis with $s_x=1/2$}
    \label{fig:10}
\end{figure}

\begin{figure}[htpb]
    \centering
    \subfigure[$s_x=3/2$, Monopole]{\includegraphics[width=0.23\linewidth]{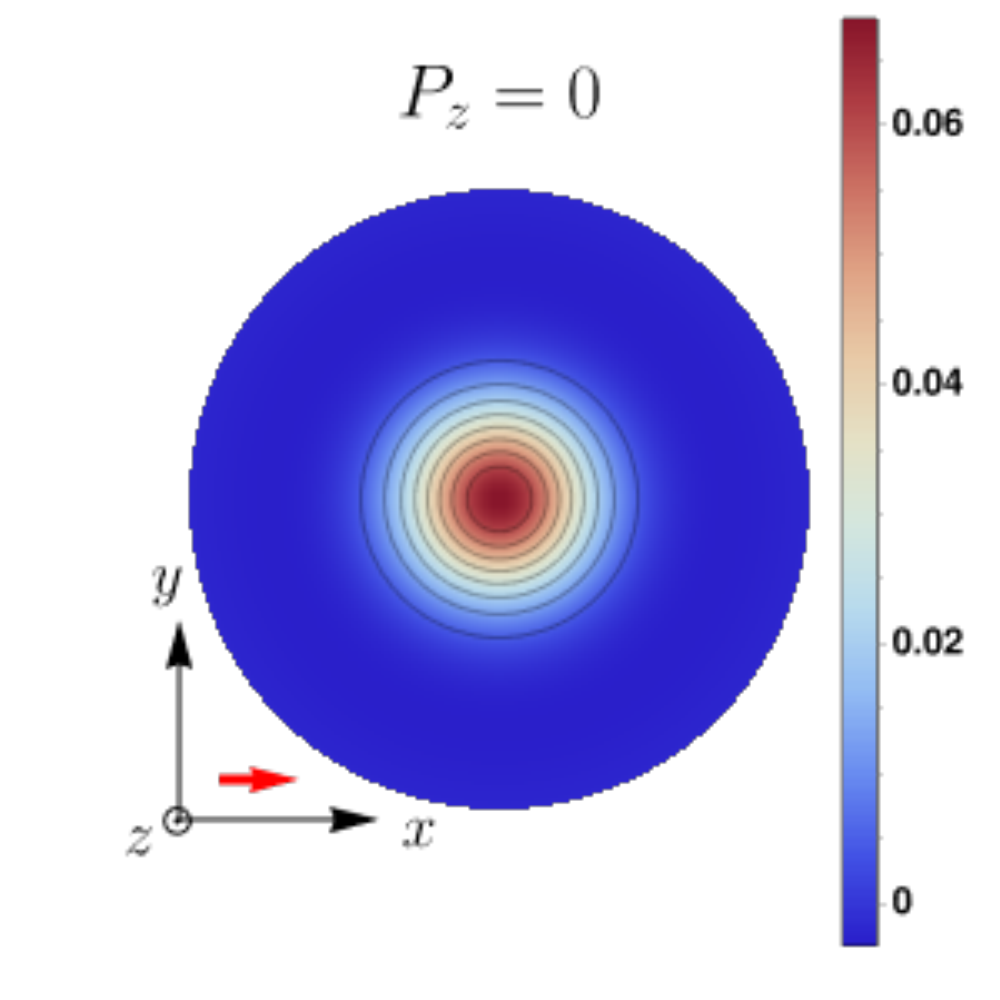}}
    \subfigure[$s_x=3/2$, Monopole]{\includegraphics[width=0.23\linewidth]{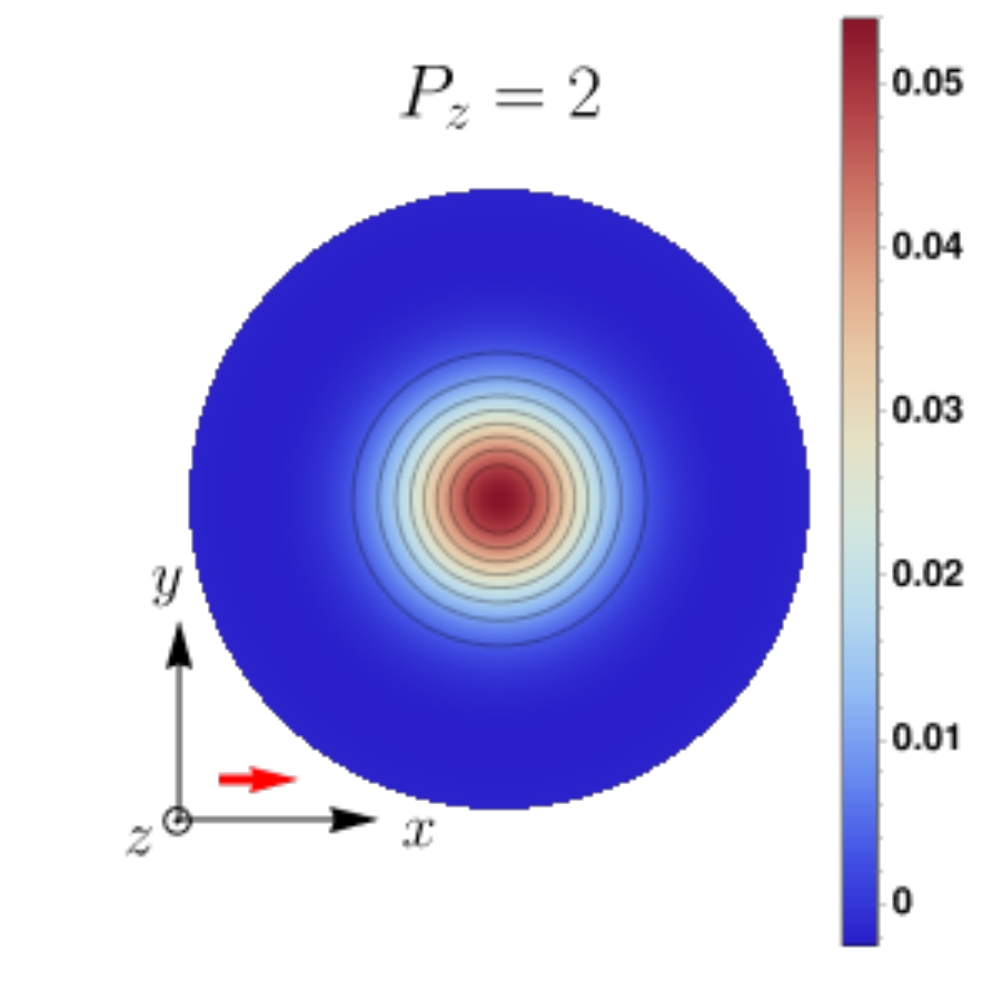}}
    \subfigure[$s_x=3/2$, Monopole]{\includegraphics[width=0.23\linewidth]{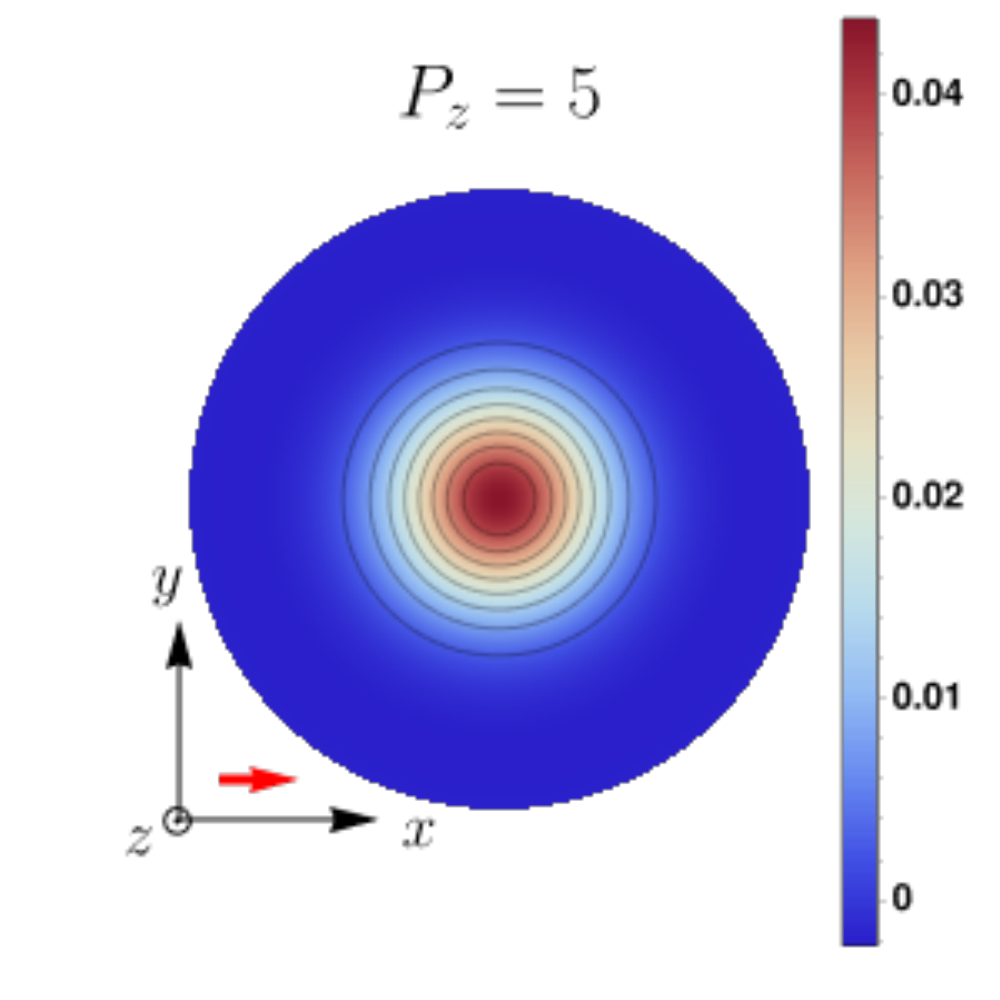}}
    \subfigure[$s_x=3/2$, Monopole]{\includegraphics[width=0.23\linewidth]{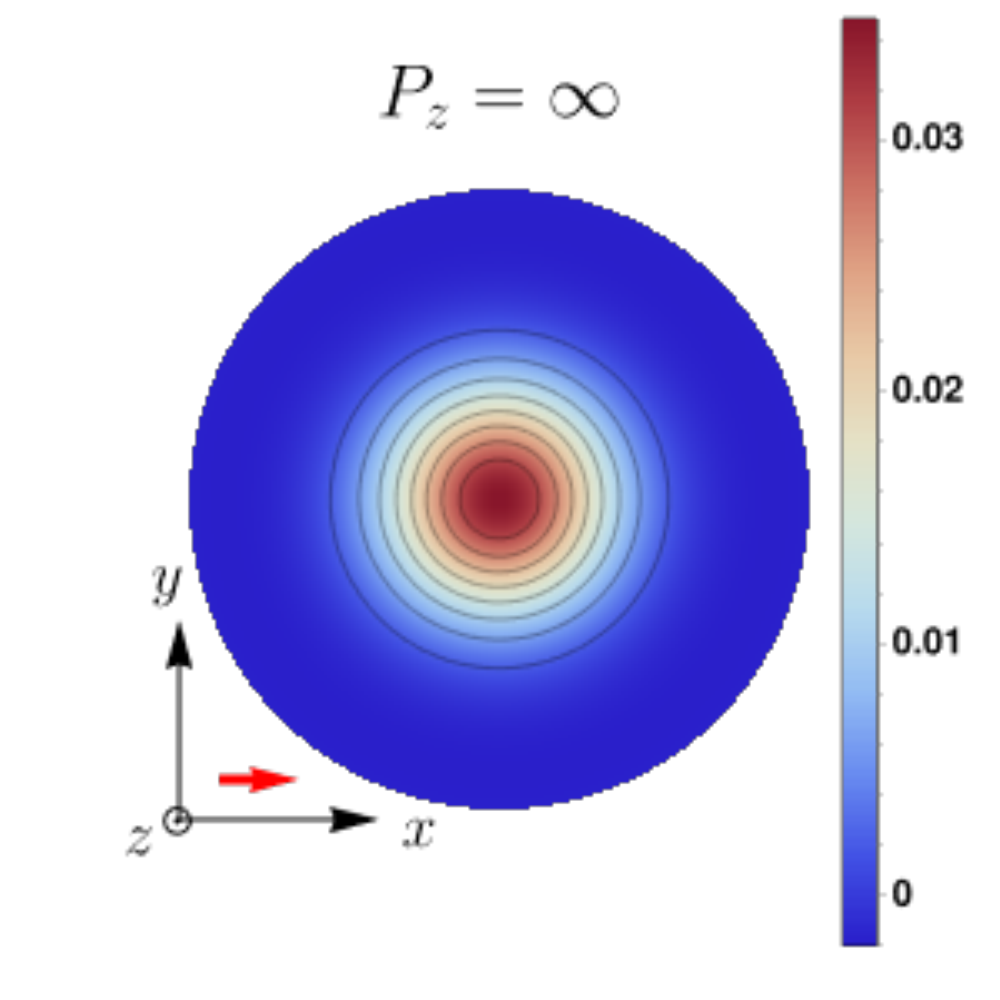}}\\
    \noindent
    \subfigure[$s_x=3/2$, Dipole]{\includegraphics[width=0.23\linewidth]{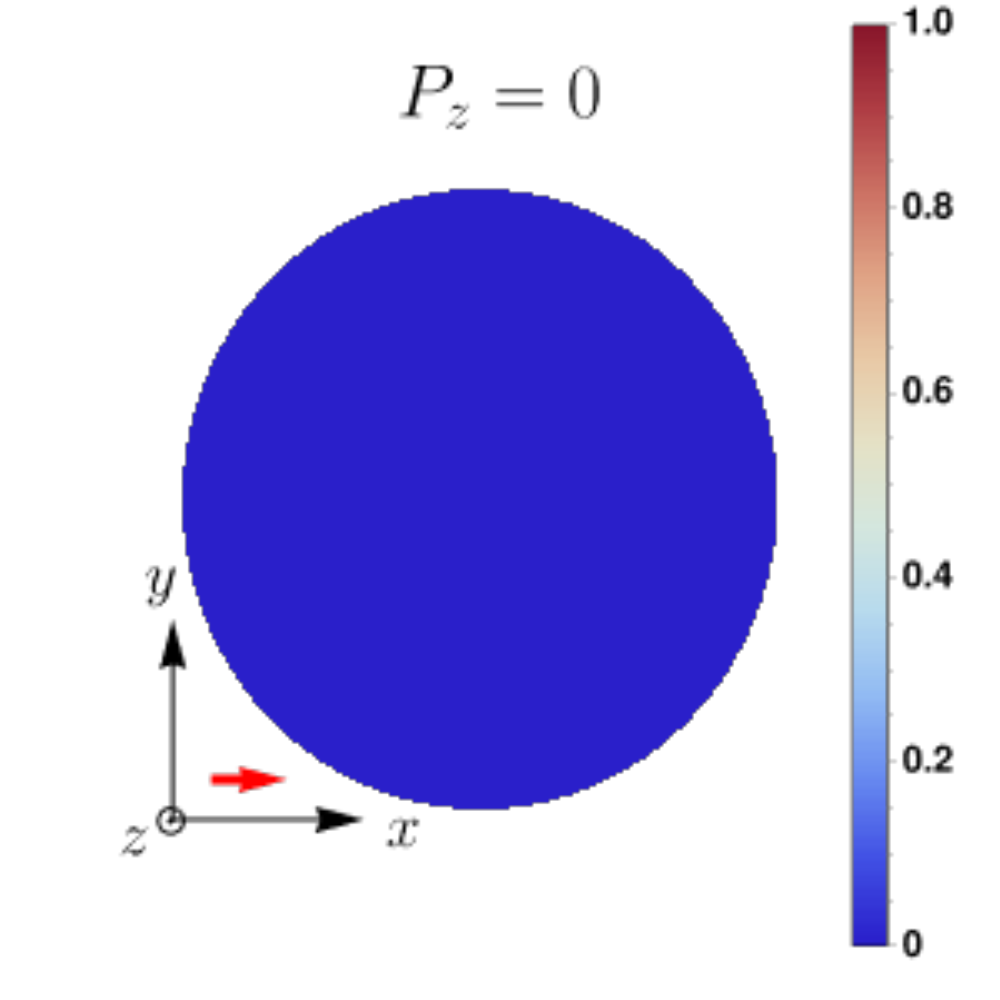}}
    \subfigure[$s_x=3/2$, Dipole]{\includegraphics[width=0.23\linewidth]{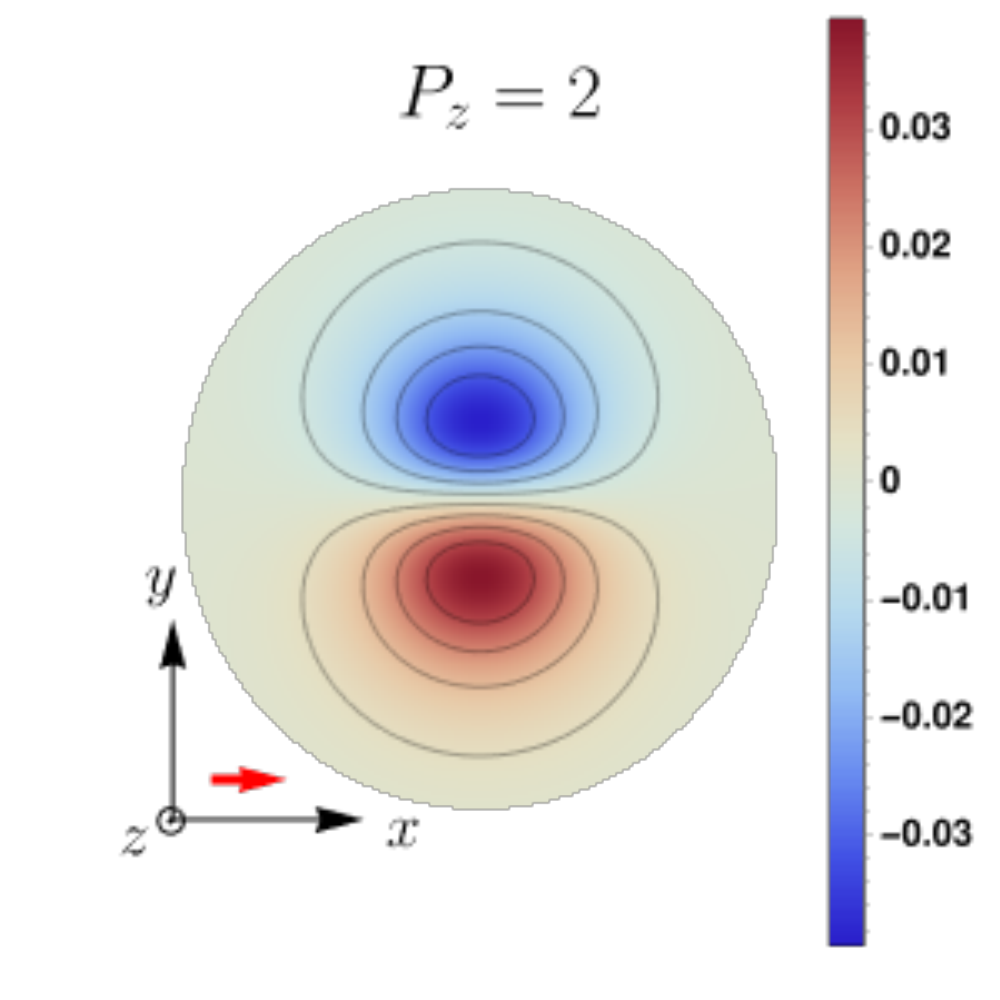}}
    \subfigure[$s_x=3/2$, Dipole]{\includegraphics[width=0.23\linewidth]{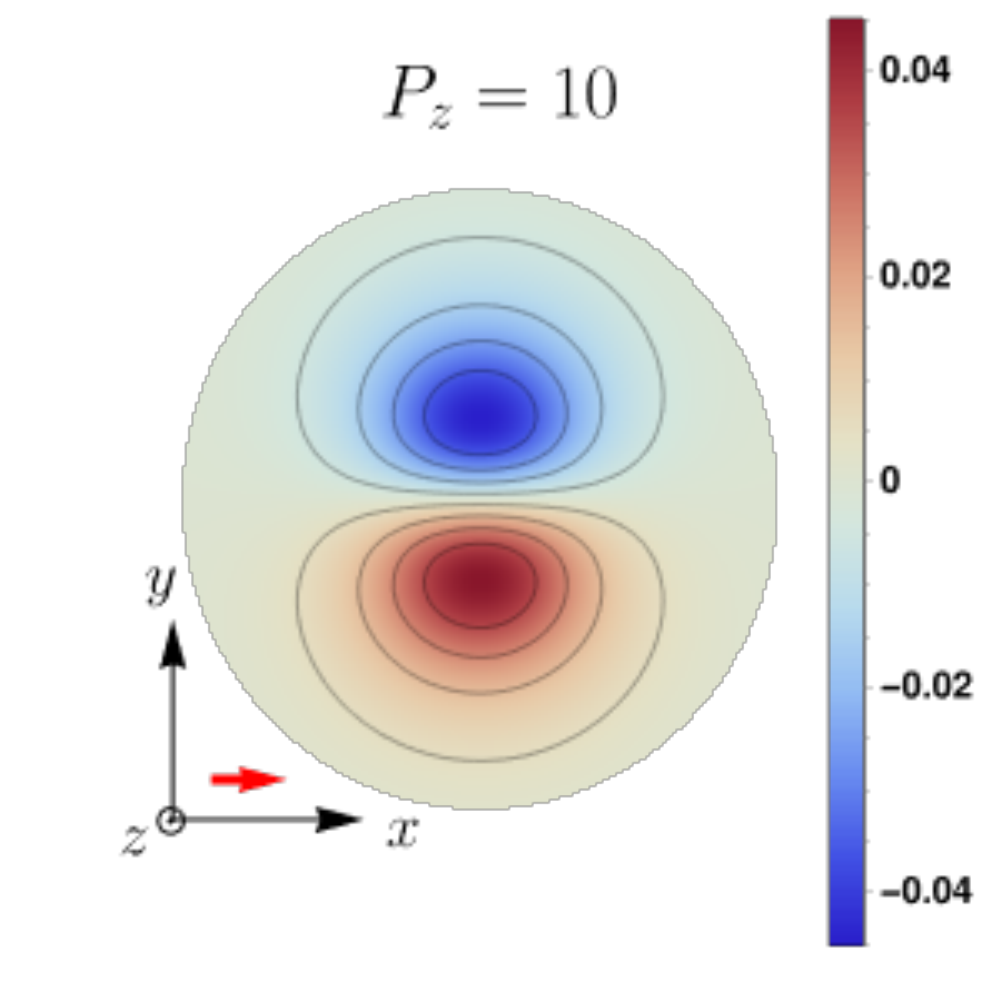}}
    \subfigure[$s_x=3/2$, Dipole]{\includegraphics[width=0.23\linewidth]{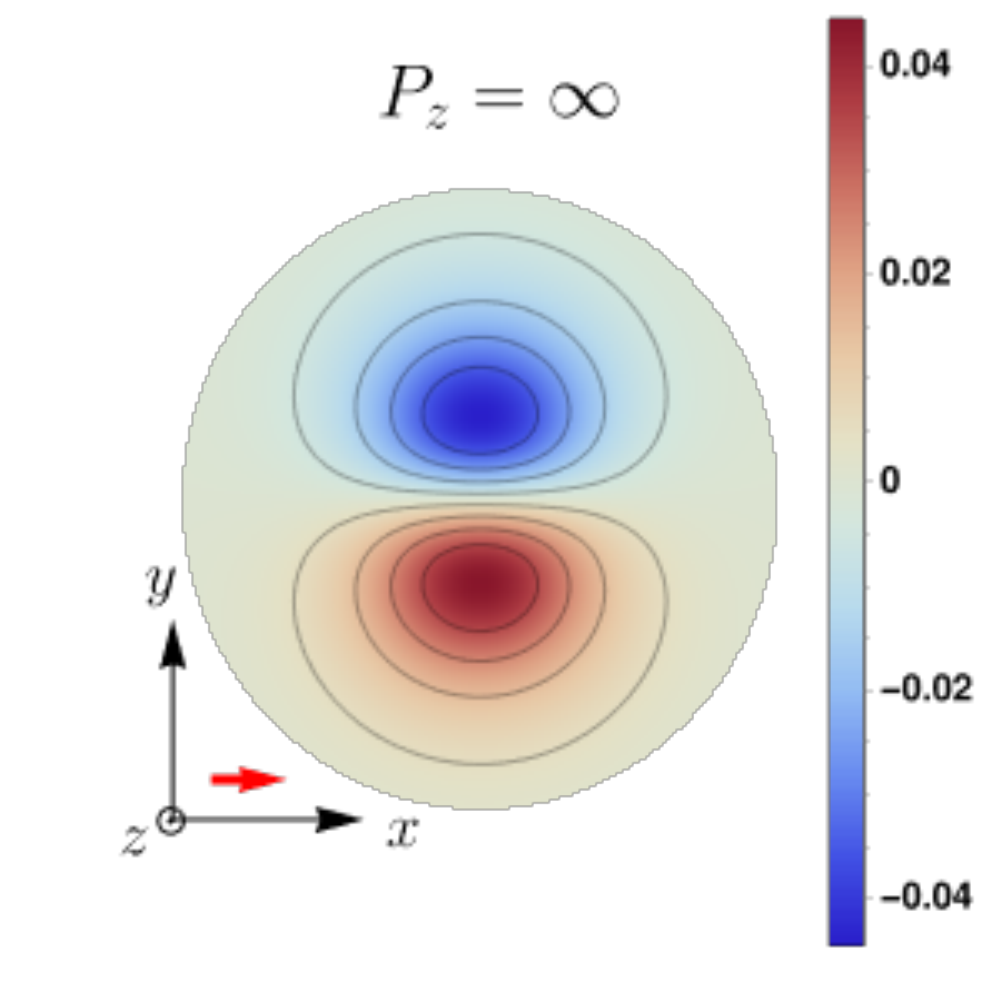}}\\
    \noindent
    \subfigure[$s_x=3/2$, Quadrupole]{\includegraphics[width=0.23\linewidth]{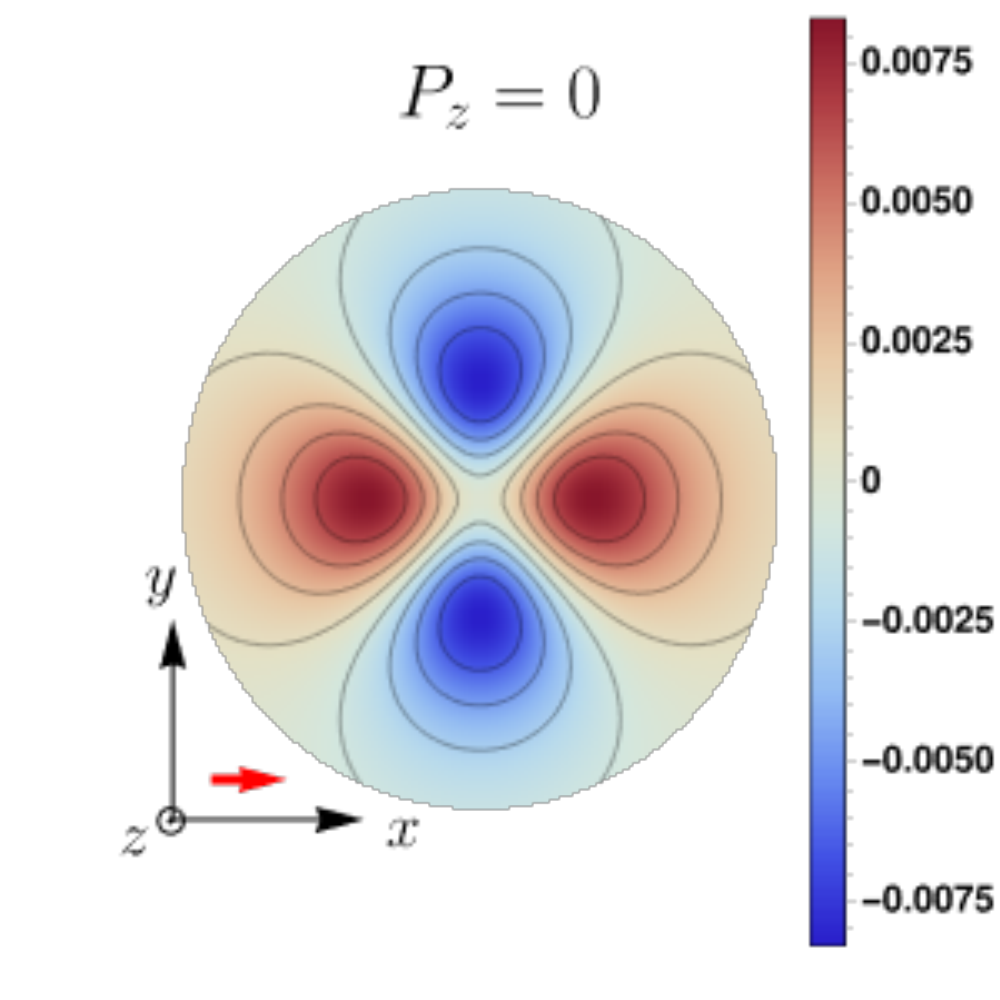}}
    \subfigure[$s_x=3/2$, Quadrupole]{\includegraphics[width=0.23\linewidth]{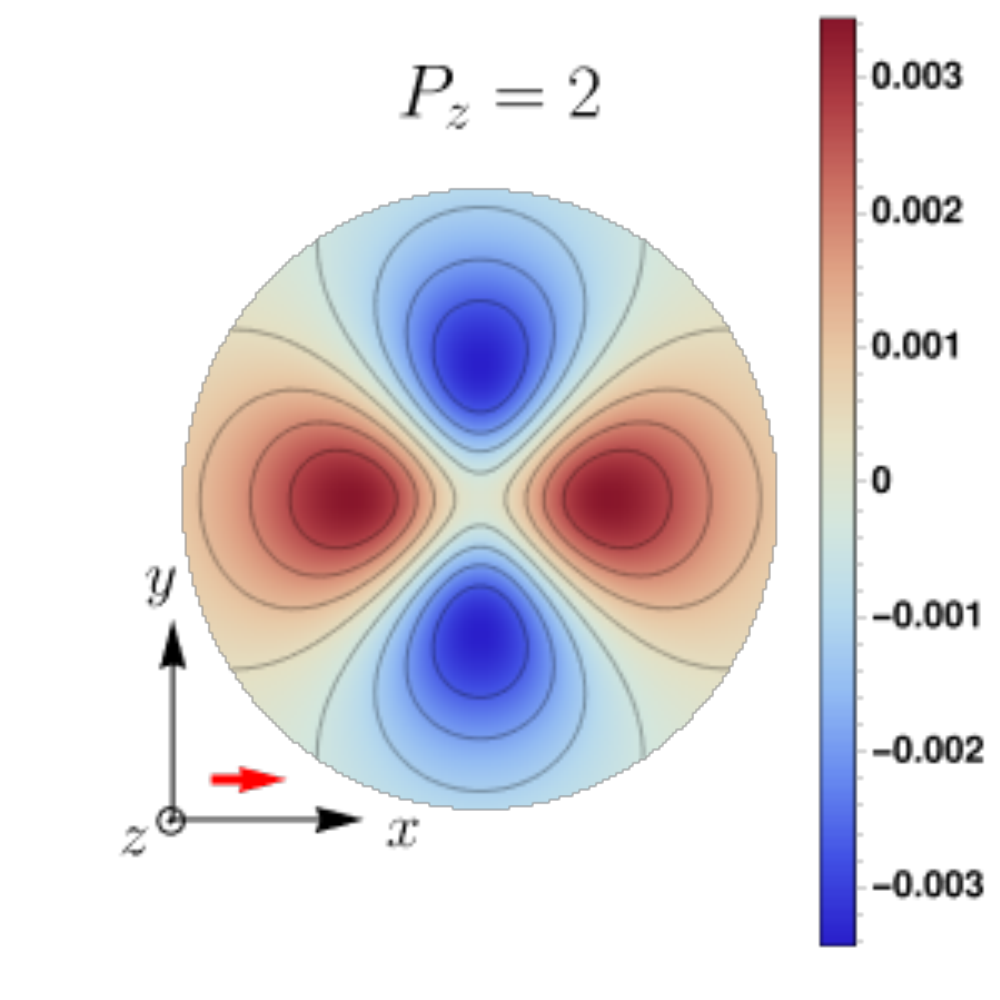}}
    \subfigure[$s_x=3/2$, Quadrupole]{\includegraphics[width=0.23\linewidth]{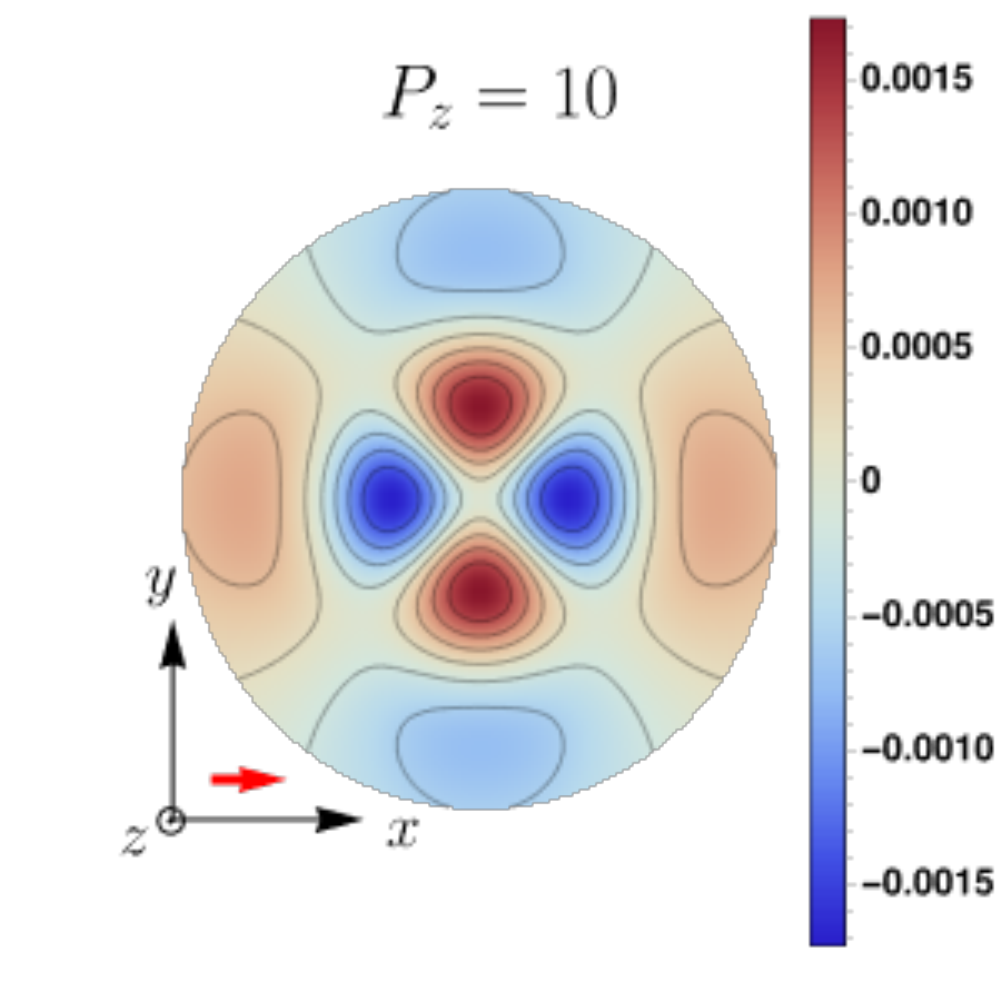}}
    \subfigure[$s_x=3/2$, Quadrupole]{\includegraphics[width=0.23\linewidth]{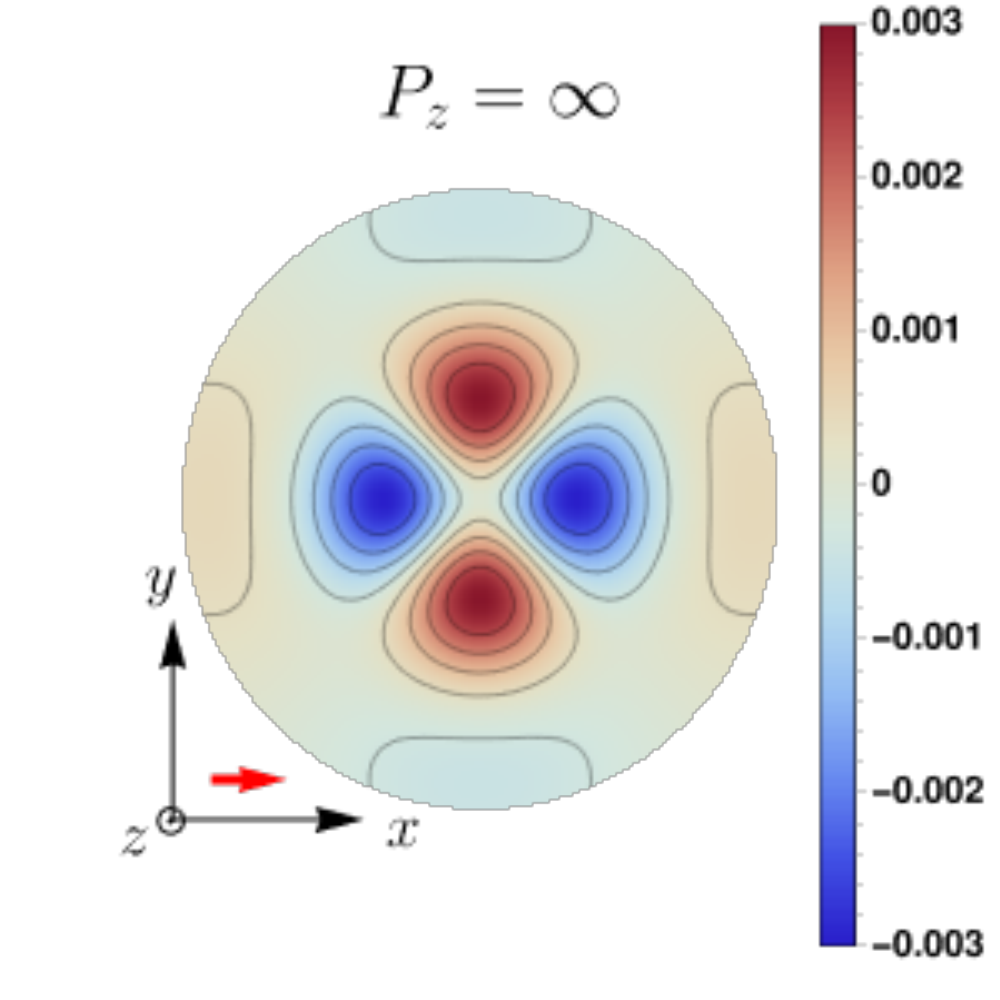}}\\
    \noindent
    \subfigure[$s_x=3/2$, Octupole]{\includegraphics[width=0.23\linewidth]{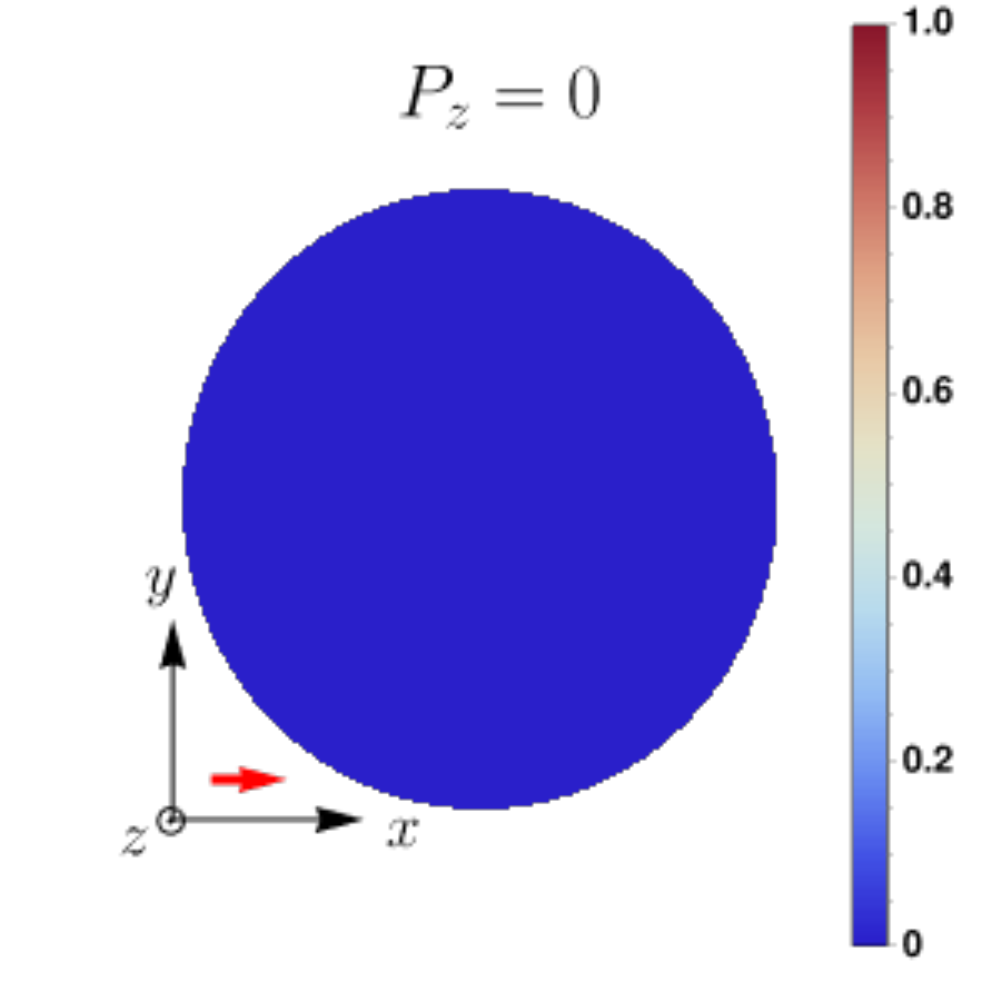}}
    \subfigure[$s_x=3/2$, Octupole]{\includegraphics[width=0.23\linewidth]{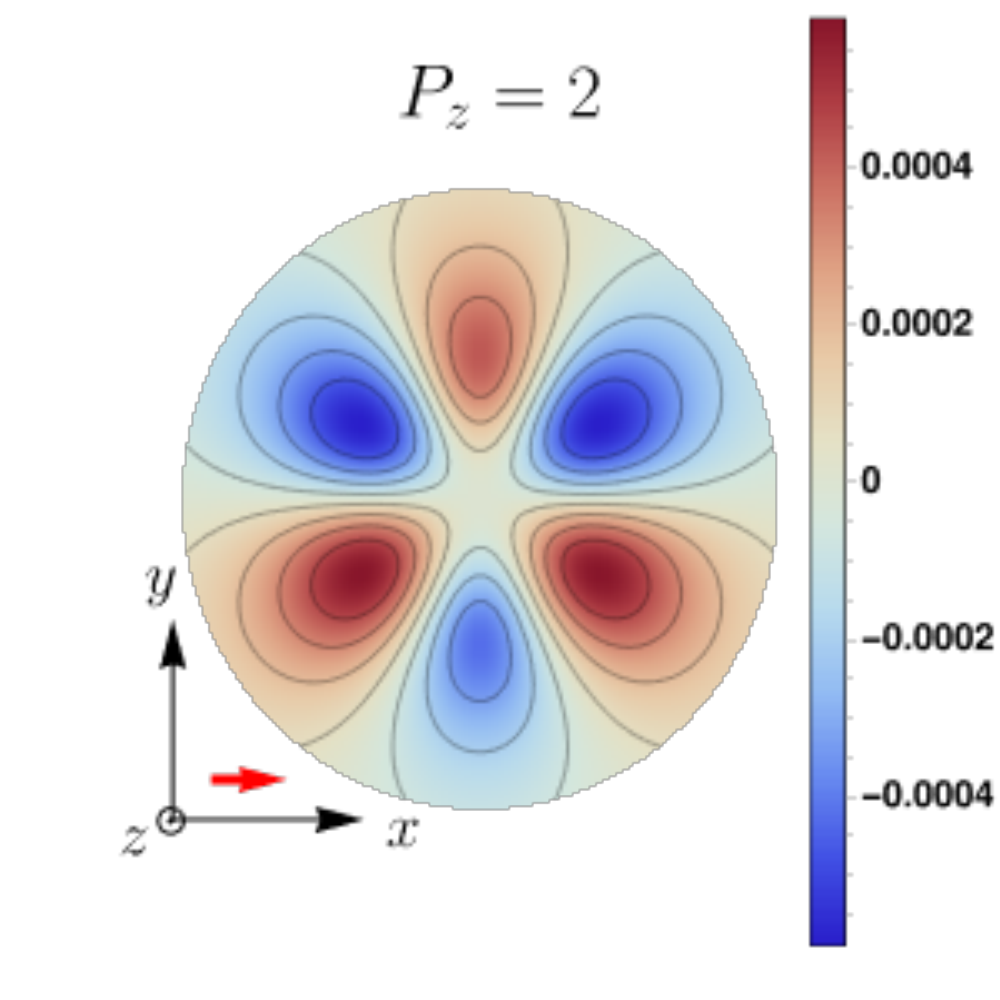}}
    \subfigure[$s_x=3/2$, Octupole]{\includegraphics[width=0.23\linewidth]{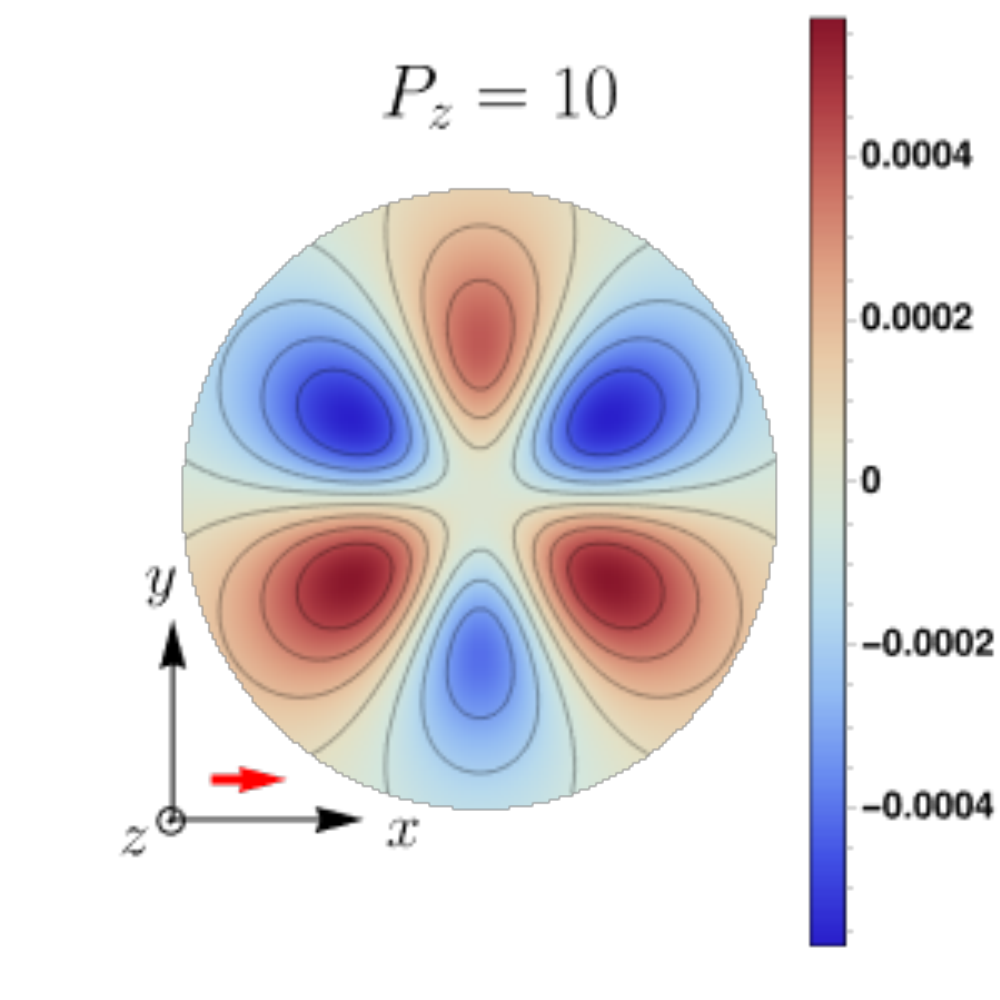}}
    \subfigure[$s_x=3/2$, Octupole]{\includegraphics[width=0.23\linewidth]{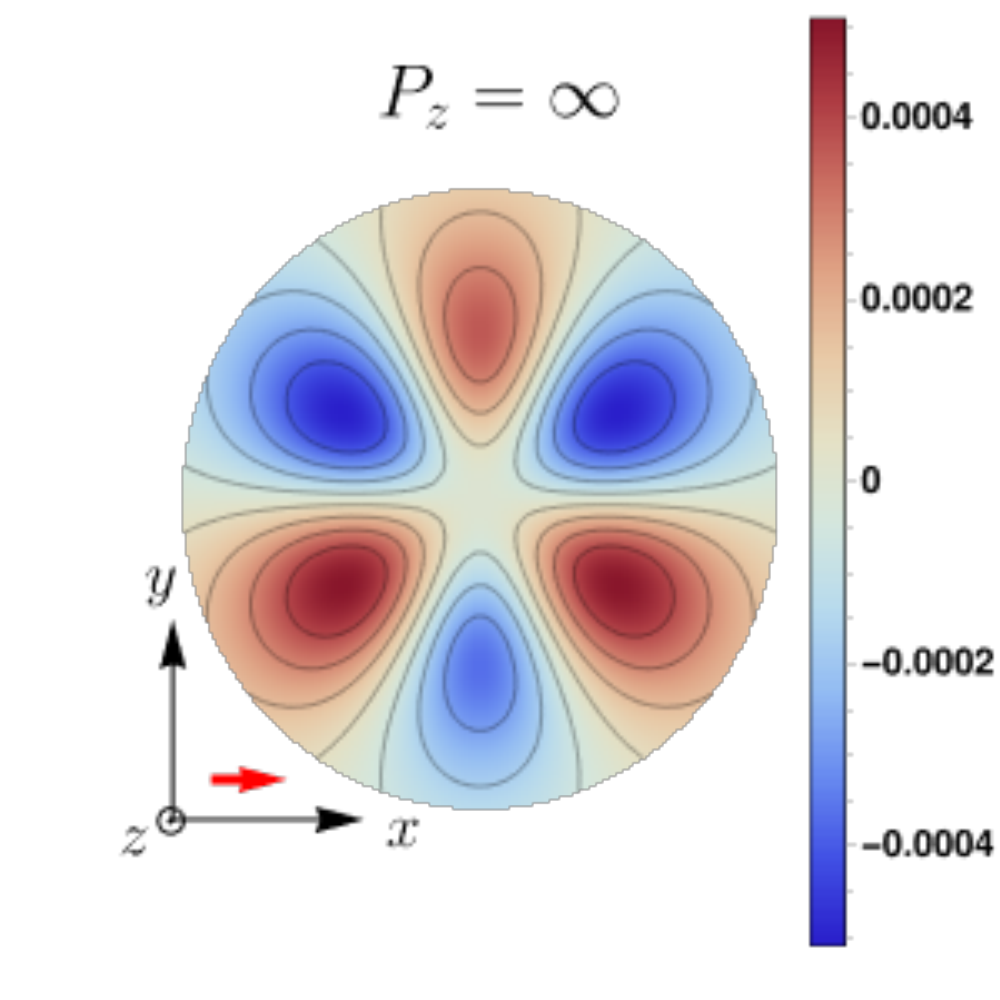}}
    \caption{(a)-(d) monopole, (e)-(h) dipole, (i)-(l) quadrupole, and (m)-(p) octupole contributions of $\Delta^0$ with $s_x=3/2$ to the 2D charge distribution.}
    \label{fig:11}
\end{figure}

\begin{figure}[htpb]
    \centering
    \subfigure[$s_x=1/2$, Monopole]{\includegraphics[width=0.23\linewidth]{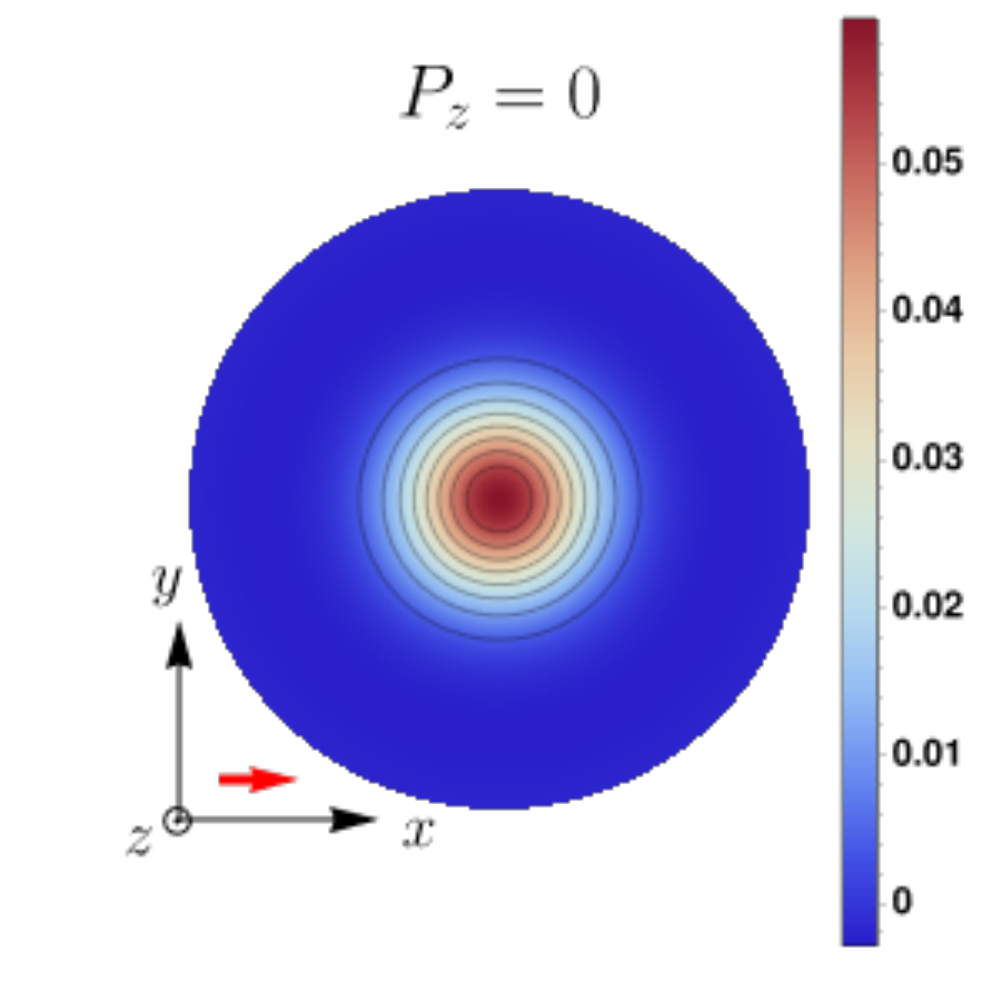}}
    \subfigure[$s_x=1/2$, Monopole]{\includegraphics[width=0.23\linewidth]{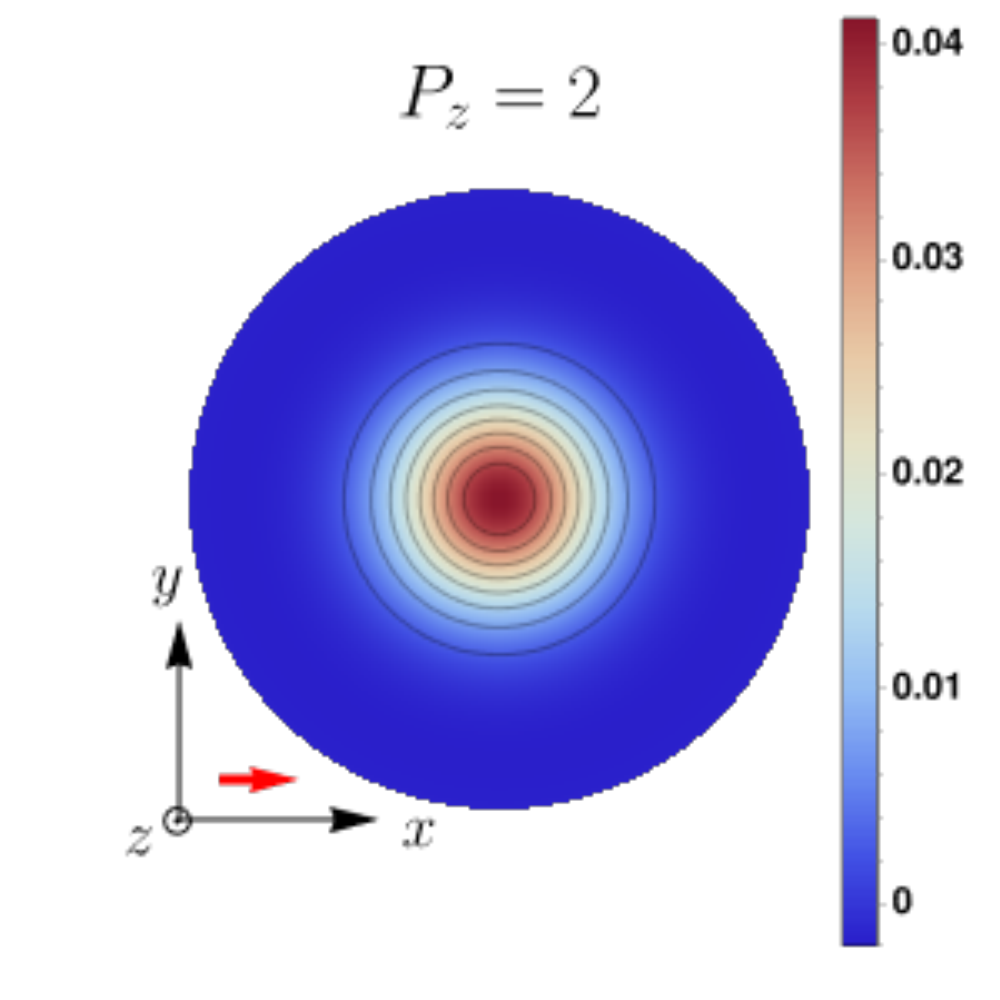}}
    \subfigure[$s_x=1/2$, Monopole]{\includegraphics[width=0.23\linewidth]{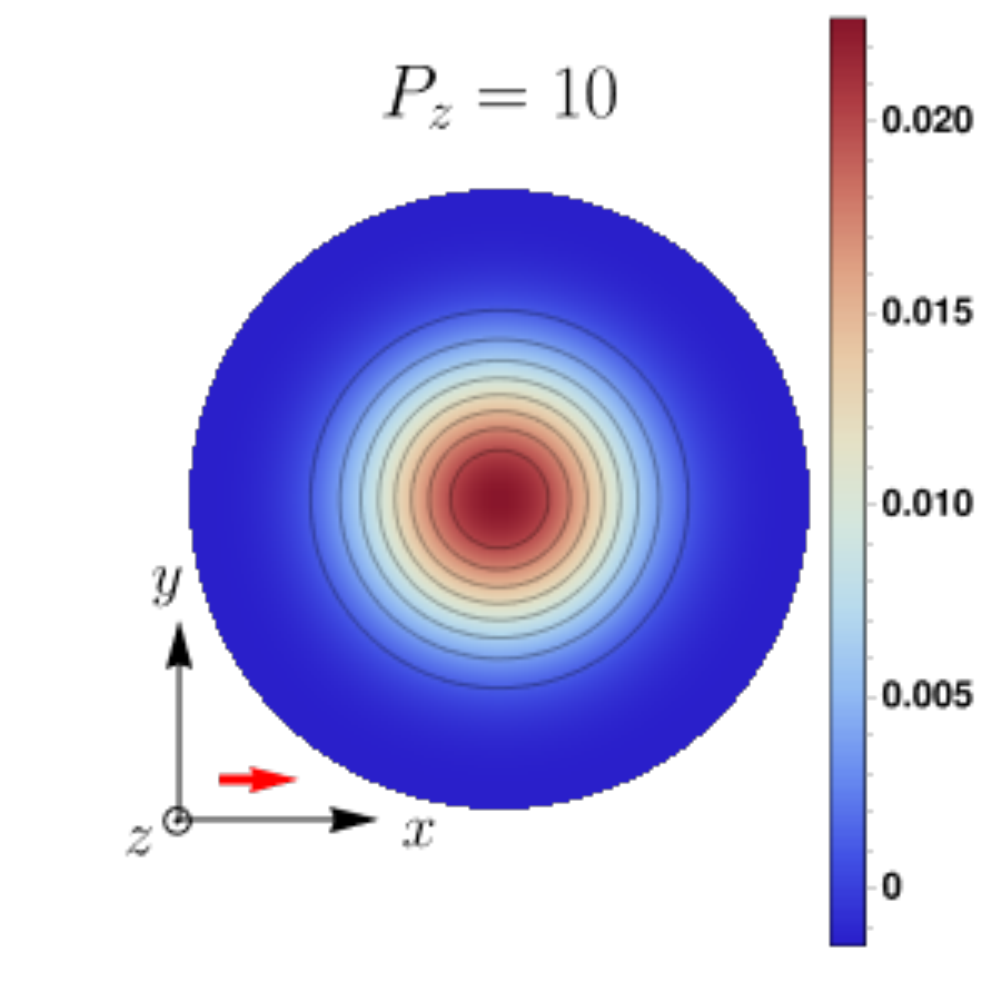}}
    \subfigure[$s_x=1/2$, Monopole]{\includegraphics[width=0.23\linewidth]{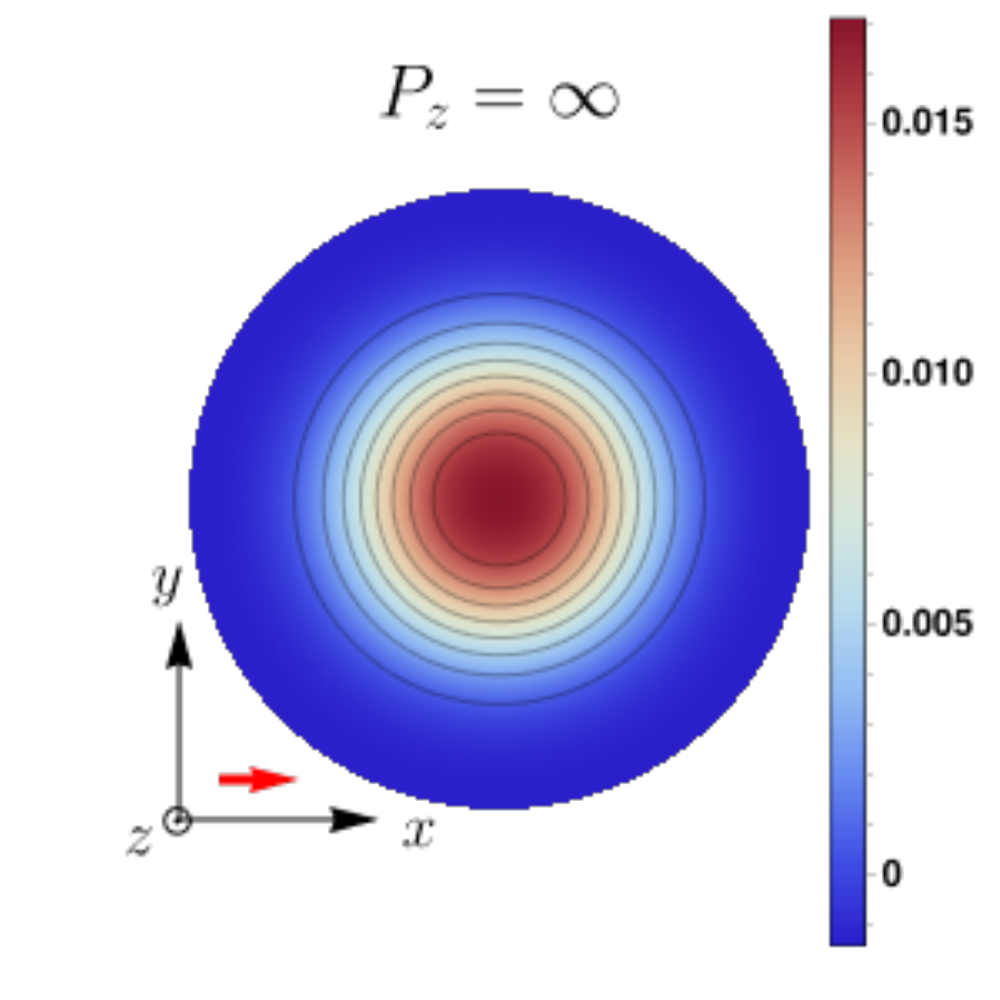}}\\
    \noindent
    \subfigure[$s_x=1/2$, Dipole]{\includegraphics[width=0.23\linewidth]{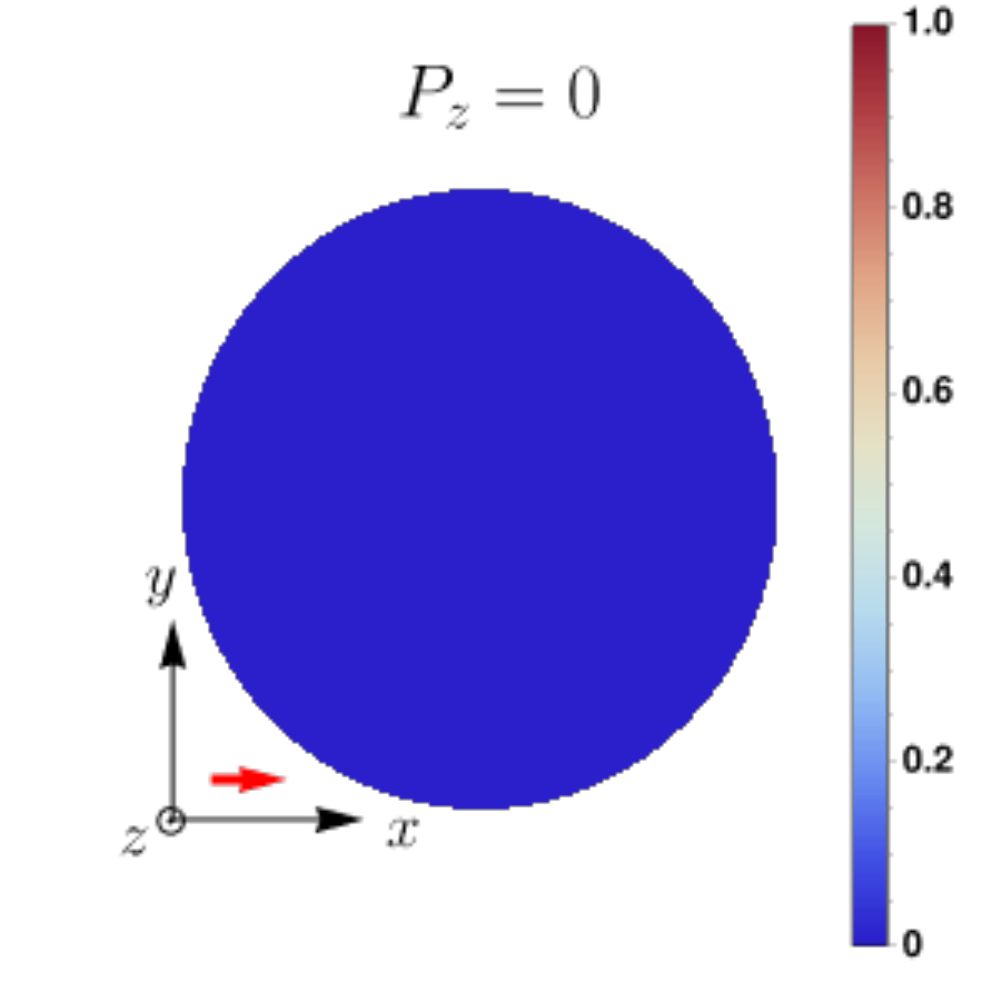}}
    \subfigure[$s_x=1/2$, Dipole]{\includegraphics[width=0.23\linewidth]{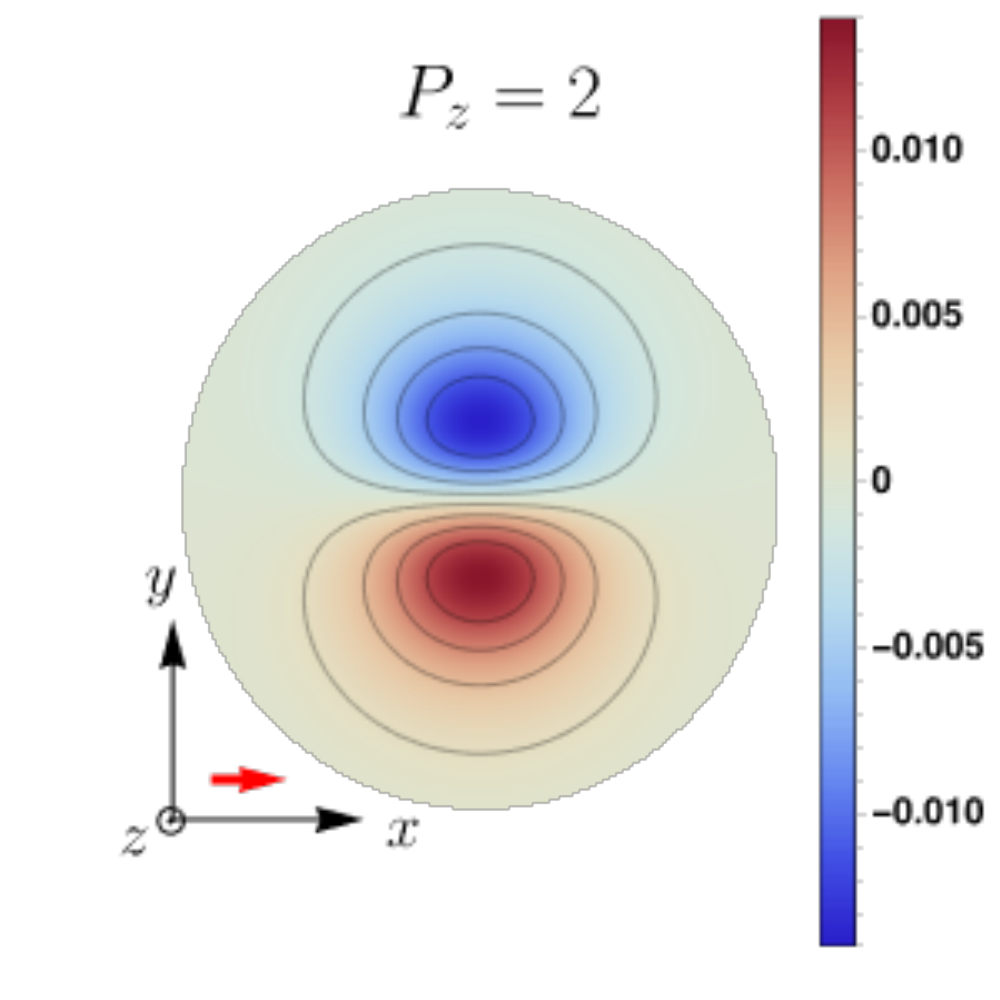}}
    \subfigure[$s_x=1/2$, Dipole]{\includegraphics[width=0.23\linewidth]{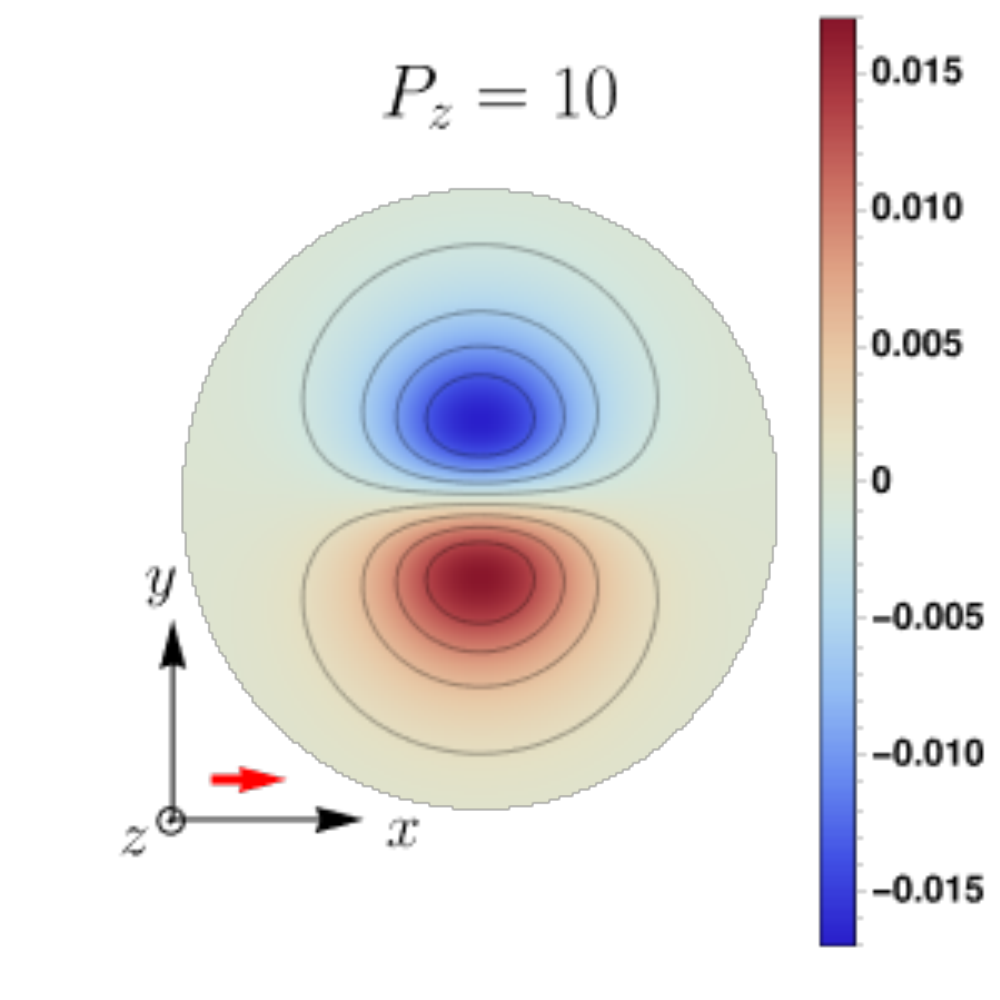}}
    \subfigure[$s_x=1/2$, Dipole]{\includegraphics[width=0.23\linewidth]{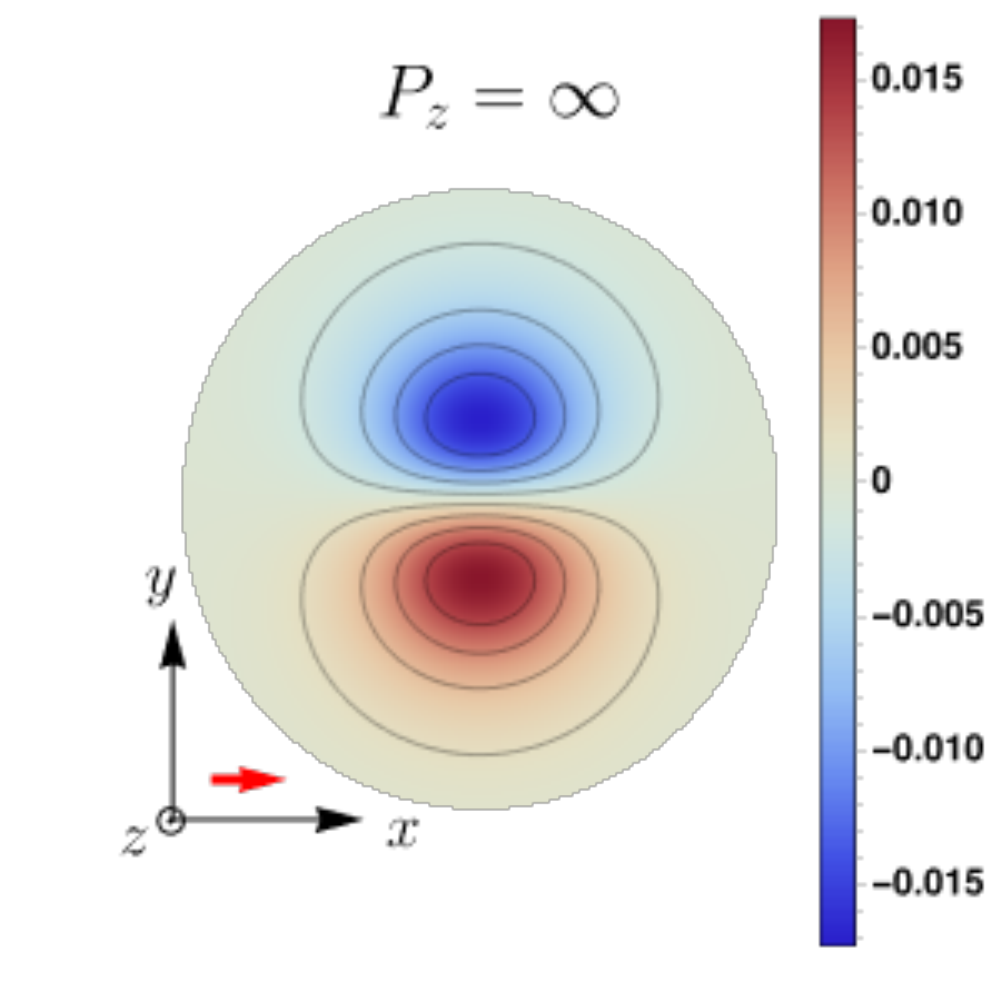}}\\
    \noindent
    \subfigure[$s_x=1/2$, Quadrupole]{\includegraphics[width=0.23\linewidth]{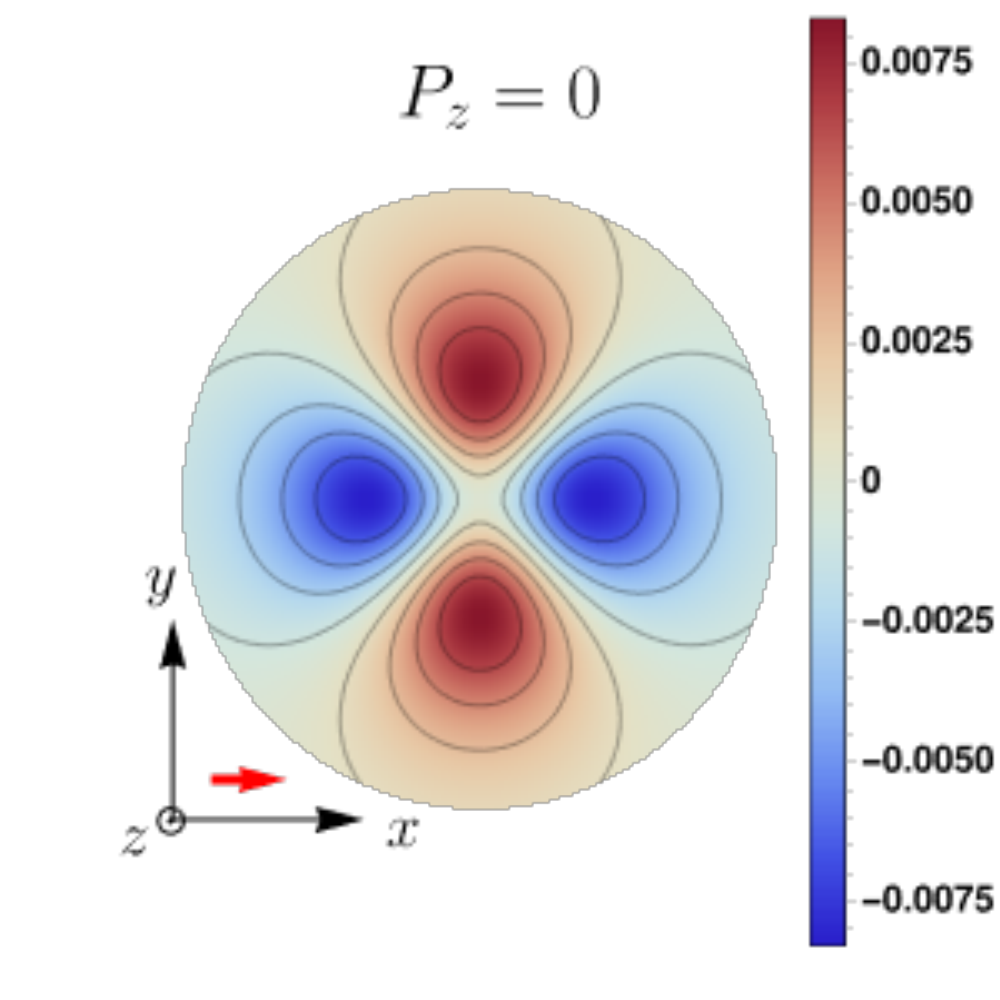}}
    \subfigure[$s_x=1/2$, Quadrupole]{\includegraphics[width=0.23\linewidth]{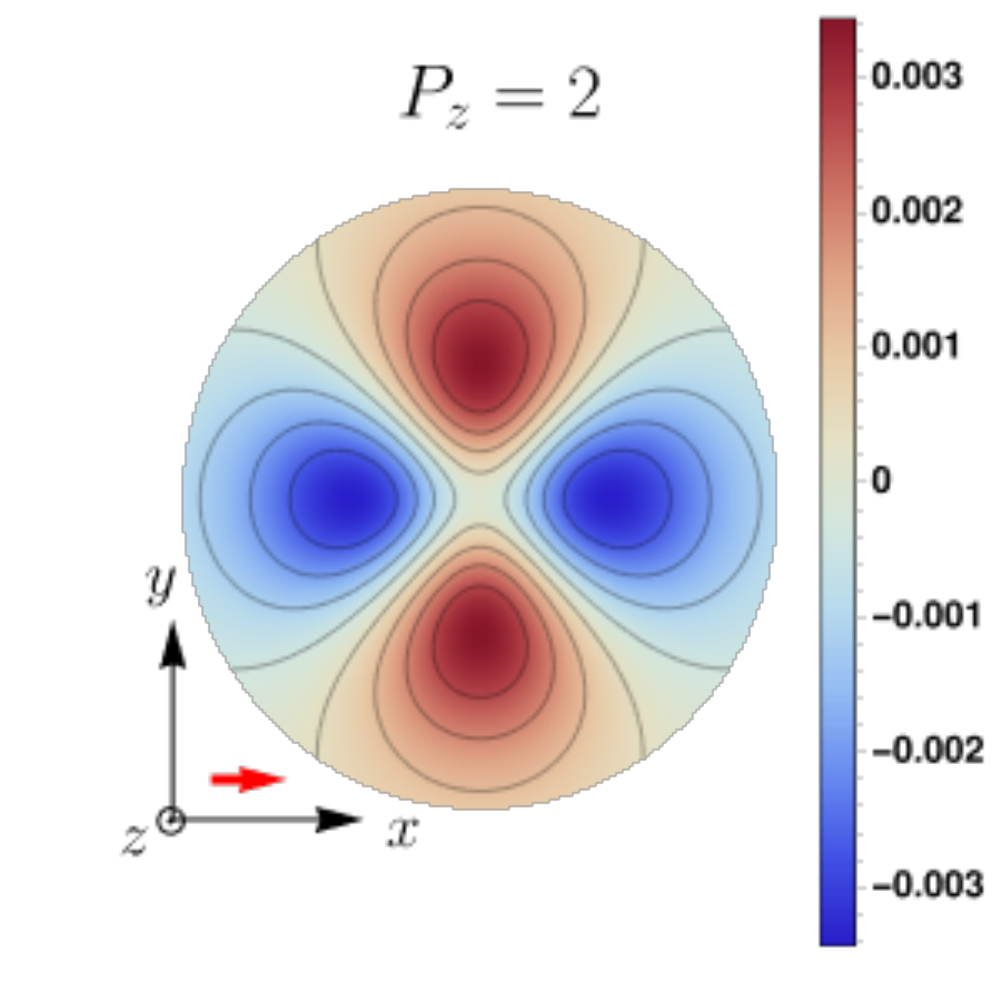}}
    \subfigure[$s_x=1/2$, Quadrupole]{\includegraphics[width=0.23\linewidth]{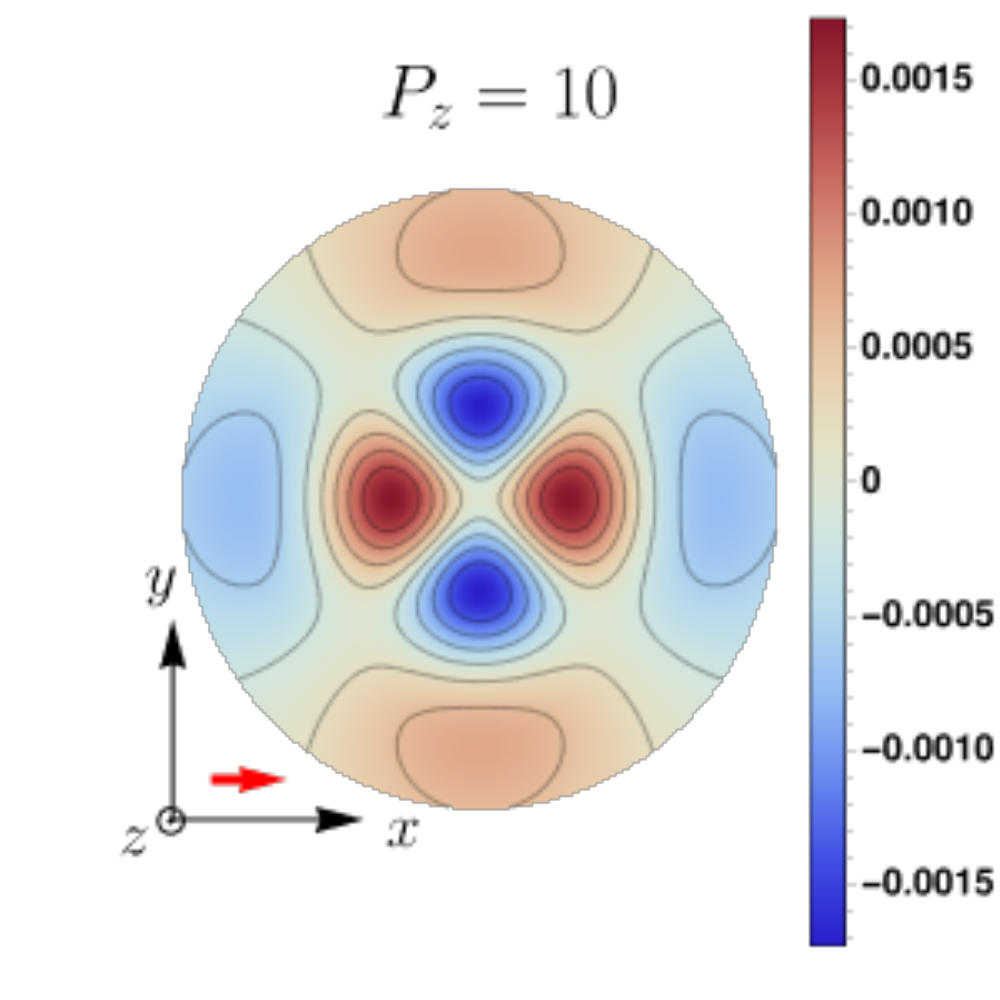}}
    \subfigure[$s_x=1/2$, Quadrupole]{\includegraphics[width=0.23\linewidth]{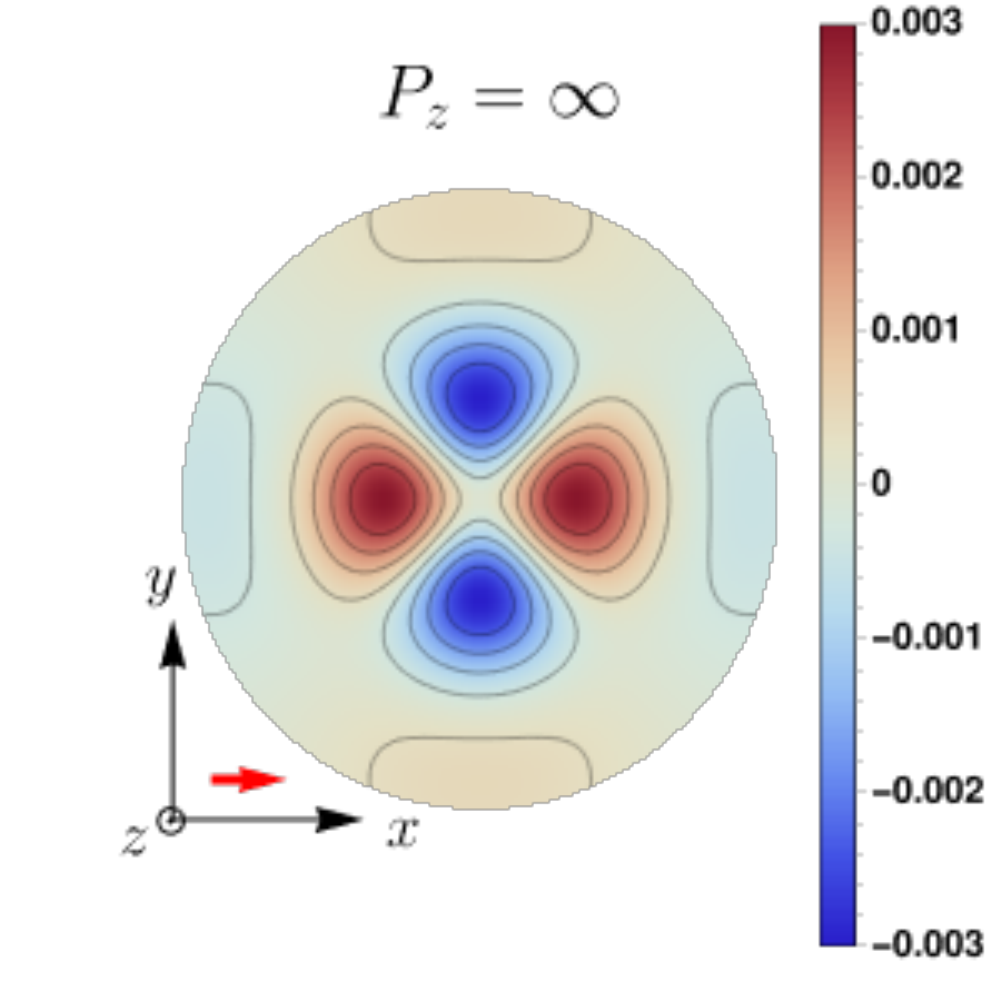}}\\
    \noindent
    \subfigure[$s_x=1/2$, Octupole]{\includegraphics[width=0.23\linewidth]{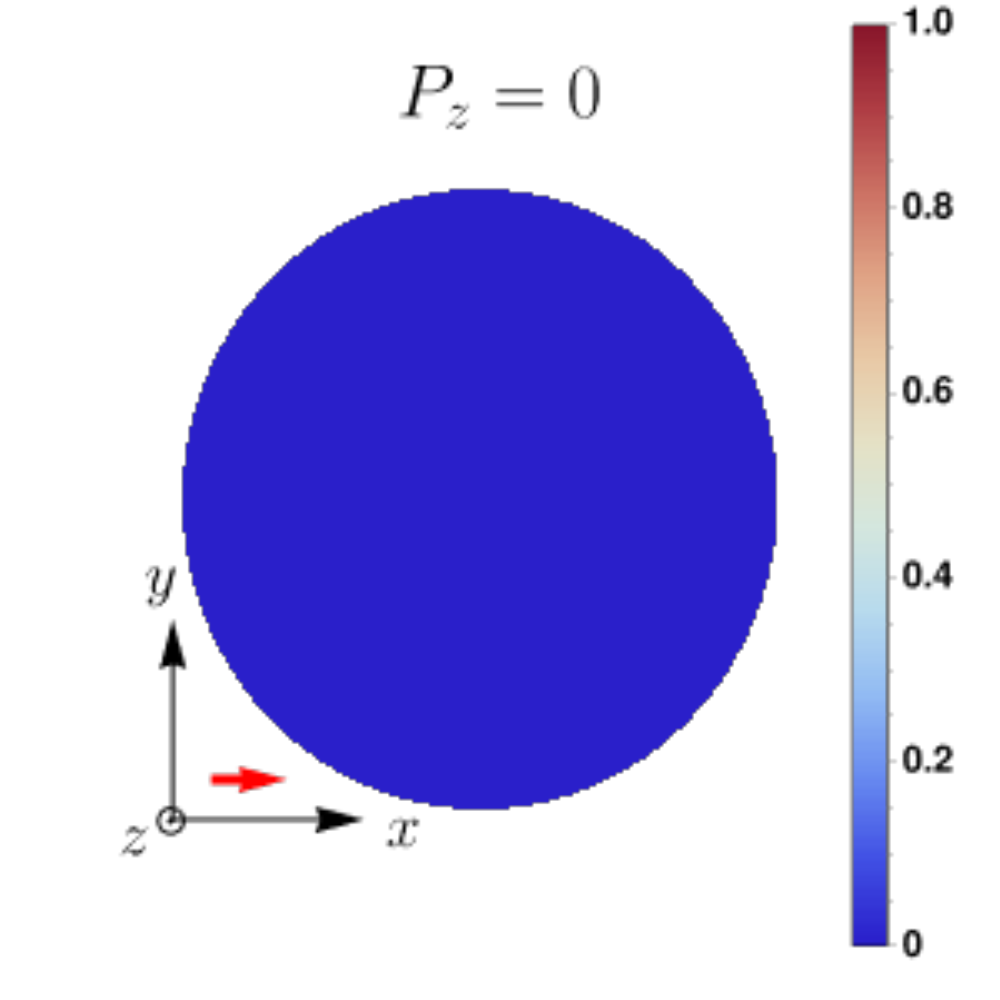}}
    \subfigure[$s_x=1/2$, Octupole]{\includegraphics[width=0.23\linewidth]{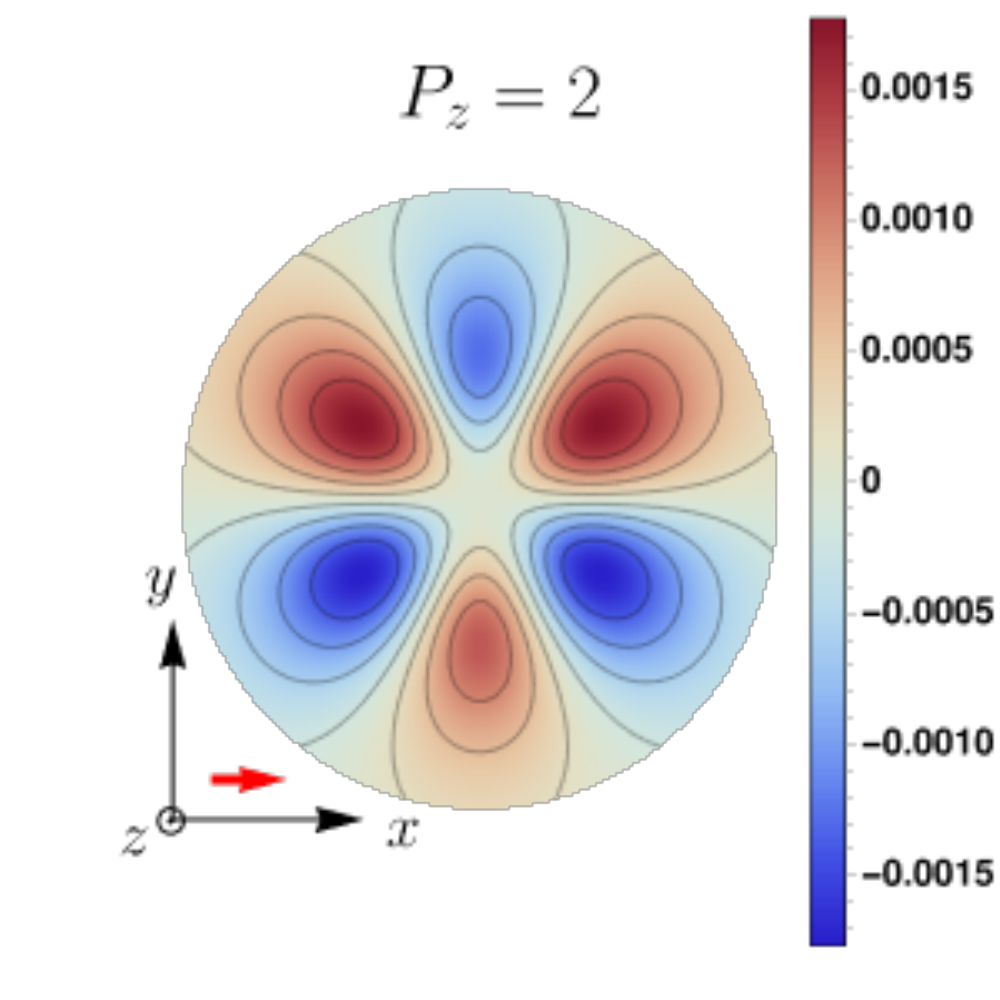}}
    \subfigure[$s_x=1/2$, Octupole]{\includegraphics[width=0.23\linewidth]{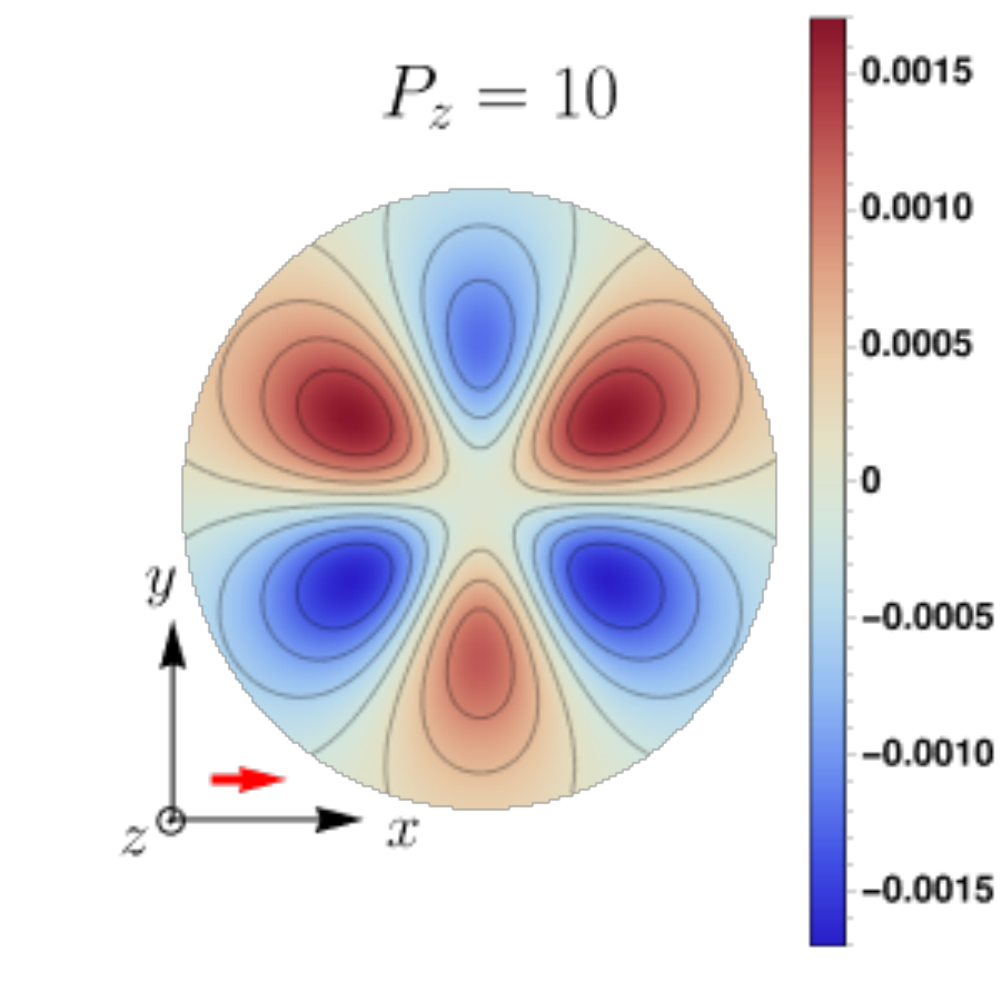}}
    \subfigure[$s_x=1/2$, Octupole]{\includegraphics[width=0.23\linewidth]{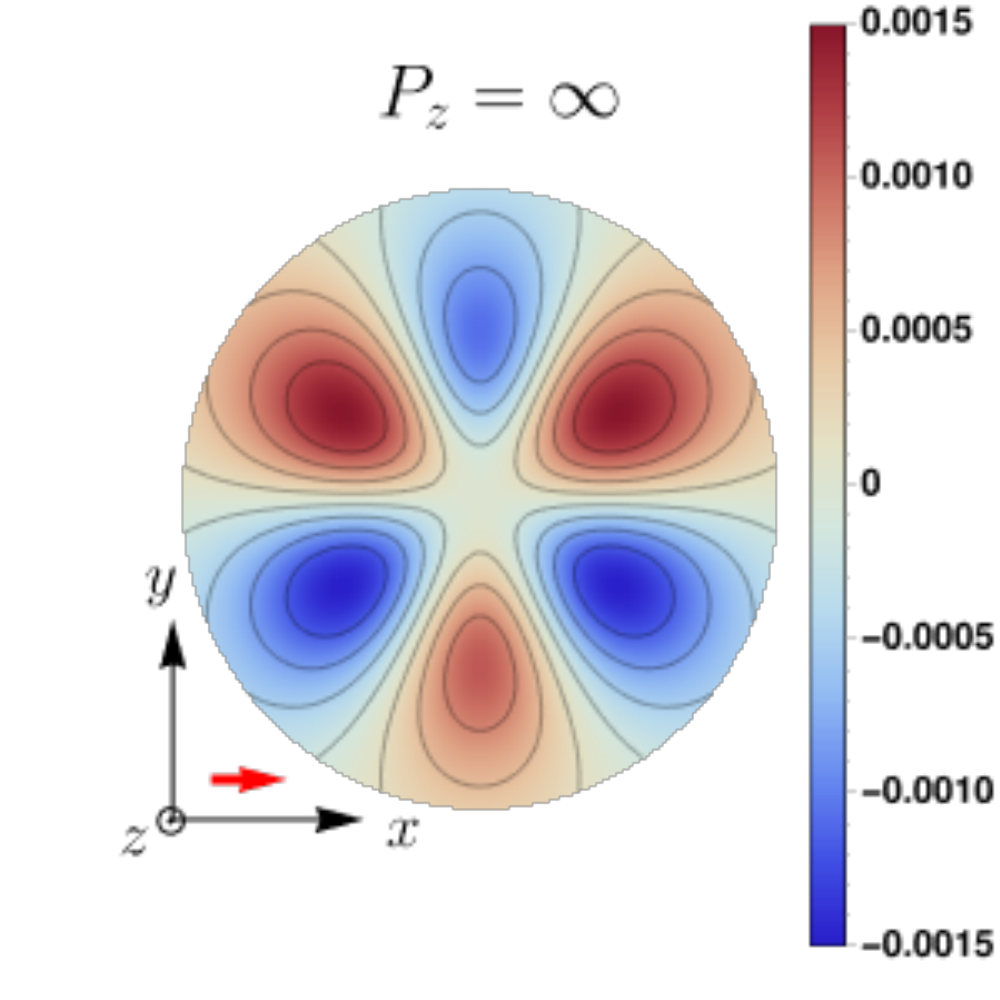}}
    \caption{(a)-(d) monopole, (e)-(h) dipole, (i)-(l) quadrupole, and (m)-(p) octupole contributions of $\Delta^0$ with $s_x=1/2$ to the 2D charge distribution.}
    \label{fig:12}
\end{figure}

In the upper panel of Fig.~\ref{fig:10}, we draw the 2D charge
distributions of the moving $\Delta^0$ baryon transversely polarized
along the $x$-axis with $s_x=3/2$. As shown in Fig.~\ref{fig:8} and
Fig.~\ref{fig:9}, the charge distribution is squeezed along the
$y$-axis due to the presence of the quadrupole contribution at the
rest frame. See the first column in Fig.~\ref{fig:11}
(Figs.~\ref{fig:11}(a), ~\ref{fig:11}(e), ~\ref{fig:11}(i), and
~\ref{fig:11}(m)). If the $\Delta^{0}$ baryon starts to move along the
$z$-axis, the electric dipole is induced and deforms the charge
distribution. So, the charge distribution starts to be tilted to the
negative $y$-direction, and the dipole contribution is saturated to
$G^{\Delta^{0}}_{M}\sim-0.3$ in  the IMF. Together with the quadrupole
and octupole contributions, we obtained a rather complicated structure
of the charge distribution of the $\Delta^{0}$ with $s_x=3/2$ in the
IMF in Fig.~\ref{fig:10}. When it comes $s_x=1/2$, the tendency is
almost kept to be the same as $s_x=3/2$, but the opposite sign of the
quadrupole contribution squeezes the charge distribution along
$x$-axis instead of $y$-axis at the rest frame. See the first column
in Fig.~\ref{fig:12} (Figs.~\ref{fig:12}(a), ~\ref{fig:12}(e),
~\ref{fig:12}(i), and ~\ref{fig:12}(m)). 

\section{Summary and conclusions}
In the present work, we aimed at investigating how the transverse
charge distributions of both the unpolarized and transversely polarized
$\Delta$ baryon change under the Lorentz boost from $P_{z}=0$ to
$P_{z}=\infty$ in the Wigner phase-space perspective. We
first observed that the elastic frame naturally interpolates the
transverse charge distributions between the Breit frame and infinite
momentum frame, even for the spin-3/2 particle. In this elastic frame,
the transverse charge distributions acquire four different
contributions: the monopole, quadrupole, induced dipole, and induced
octupole contributions. To visualize  them in the 2D space, we
employed the electromagnetic form factors of the $\Delta$ baryon
extracted from the SU(3) chiral quark-soliton model. The Loretz boost
and the geometrical projection from the 3D to 2D spaces yield a
split in the spin-polarization of the monopole and induced dipole
contributions. When both the $\Delta^{+}$ and $\Delta^{0}$ baryons are
polarized along the $z$-axis, we found that their charge distributions
are always kept to be spherically symmetric under the Lorentz
boost. For the $\Delta^{0}$, the shape of the transverse charge
distribution was dramatically changed under the Lorentz boost, which
is similar to the neutron case. When the $\Delta$ baryon is
transversely polarized along the $x$-axis, all the multipole
structures start to appear. In the rest frame, the quadrupole
contribution does not vanish and makes the charge distribution
deformed. When $P_{z}$ increases, the dipole and octupole
contributions are induced and cause the asymmetry of the transverse
charge distribution. For the $\Delta^{+}$ baryon with $s_{x}=1/2$ and
$s_{x}=3/2$, the charge distributions start to be shifted to the
positive $y$-direction and reach the maximal values of the electric
dipole moments at around $P_{z}\sim 1.4$\,GeV and gradually
diminish. They turn negative at around $P_{z}=10$\,GeV. As
a result, the transverse charge distributions of the transversely
polarized $\Delta^{+}$ baryon along the $x$-axis is moved to the
negative $y$-direction in the infinite momentum frame. We found that
these results are consistent with the results from the lattice QCD for the
$\Delta^{+}$. For the $\Delta^{0}$ baryon, the positive charges, which
represent the up quark inside the $\Delta^{0}$ baryon, were
displaced to the negative $y$-direction whereas the negative charges
or the down quarks were moved toward the positive
$y$-direction. This is due to the negative values of the electric
dipole moment ($G^{\Delta^{0}}_{M1} \sim -0.3$) of the $\Delta^{0}$
baryon.  

\begin{acknowledgments}
This material is based upon work supported by the U.S. Department of
Energy, Office of Science, Office of Nuclear Physics under contract
DE-AC05-06OR23177 (J.-Y. Kim), and by the Basic Science Research
Program through the National Research Foundation of Korea funded by
the Korean government (Ministry of Education, Science and Technology,
MEST), Grant-No. 2021R1A2C2093368 and 2018R1A5A1025563. 
\end{acknowledgments}

\appendix

\clearpage
\section{frame-dependent functions \label{app:a}}
In this Appendix, We listed the explicit expressions of the frame-dependent functions for both temporal and spatial components of the EM current. To express them in a compact way,  we introduce the following functions:
\begin{align}
    A=\frac{1}{M(M+P_0)},\ \beta_{n}^\perp=1+\frac{\vec{q}_\perp^2}{nM(M+P_0)},\ \beta_{n}^z=1+\frac{P_z^2}{nM(M+P_0)}.
\end{align}
All the frame-dependent functions appearing in the matrix element of the temporal component of the EM current are listed as follows:
\begin{align}
    G_{E0}(t;P_z)&=\frac{1}{2P_0}\Bigg[2M\left(\beta_1^z-\frac{1}{3}At\beta_8^\perp\beta_1^z-\frac{1}{6}P_z^2\beta_4^\perp t A^2\right)F_1^*(t)+\frac{tF_2^*(t)}{2M}\left(1-\frac{1}{3}At\beta_8^\perp+\frac{2}{3}P_z^2A\beta_4^\perp\right)\cr
    &\quad+\frac{\beta_4^\perp t F_3^*(t)}{6M}\left(\beta_1^z\beta_4^{\perp}-\frac{P_z^2A^2 t}{4}\right)+\frac{\beta_4^\perp t^2F_4^*(t)}{8M^3}\left(\frac{P_z^2 A}{3}+\frac{1}{3}\beta_4^{\perp}\right)\Bigg]\cr
    G_{E0}^{a'a}(t;P_z)&=\frac{1}{2P_0}\Bigg[\frac{2}{3}MP_z^2A^2t\left(\beta_1^z-2\beta_4^\perp\right)F_1^*(t)+\frac{P_z^2t^2A^2F_3^*(t)}{12M}\left(\beta_1^z-2\beta_4^\perp\right)\cr
    &\quad-\frac{P_z^2AtF_2^*(t)}{6M}\left(8\beta_4^\perp-At\right)-\frac{P_z^2At^2F_4^*(t)}{48M^3}\left(8\beta_4^\perp-At\right)\Bigg]\cr
    G_{E1}(t;P_z)&=\frac{1}{2P_0}\Bigg[\frac{2P_zM^2AF_1^*(t)}{3}\left(1-2\beta_4^\perp\beta_1^z-\frac{2tA\beta_8^\perp}{5}\right)-\frac{P_ztA\beta_4^\perp F_3^*(t)}{12}\left(2\beta_1^z-\frac{4\beta_4^{\perp}}{5}\right)\cr
    &\quad+\frac{P_zF_2^*(t)}{6}\left(4-2At\beta_4^\perp-\frac{8At\beta_8^\perp}{5}\right)-\frac{P_z t\beta_4^\perp F_4^*(t)}{8M^2}\left(\frac{At}{3}-\frac{8}{15}\beta_4^{\perp}\right)\Bigg]\cr
    G_{E1}^{a'a}(t;P_z)&=\frac{1}{2P_0}\Bigg[\frac{P_z^3t M^2A^3}{3}F_1^*(t)+\frac{P_z^3A^3t^2}{24}F_3^*(t)+\frac{P_z^3A^2t}{3}F_2^*(t)+\frac{P_z^3t^2A^2}{24M^2}F_4^*(t)\Bigg]\cr
    G_{E2}(t;P_z)&=-\frac{3}{2P_0}\Bigg[\frac{2}{3}M^3A\left(\beta_8^\perp\beta_1^z+P_z^2A\beta_4^\perp\right)F_1^*(t)+\frac{M\beta_4^{\perp}F_3^*(t)}{12}\left(P_z^2A^2t-2\beta_1^z\beta_4^{\perp}\right)\cr
    &\quad+\frac{MAF_2^*(t)}{6}\left(t\beta_8^\perp+4P_z^2\beta_4^\perp\right)+\frac{tF_4^*(t)}{24M}\left(2P_z^2A\beta_4^\perp-\beta_4^{\perp2}\right)\Bigg]\cr
    G_{E3}(t;P_z)&=\frac{1}{2P_0}\Bigg[-\frac{2}{3}P_z\beta_8^\perp M^4A^2F_1^*(t)+\frac{1}{6}M^2 AF_3^*(t)P_z\beta_4^{\perp2}-\frac{2}{3}P_z\beta_8^\perp M^2AF_2^*(t)+\frac{1}{6}F_4^*(t)\beta_4^{\perp2}P_z\Bigg],\label{appendix:1}
\end{align}
All the frame-dependent functions appearing in the matrix element of the spatial components of the EM current are listed as follows:
\begin{align}
    G^{3}_{M0}(t;P_z)&=\frac{1}{2P_0}\Bigg[P_z(2-tA\beta_6^\perp)F_1^*+\frac{P_zt\beta_4^\perp}{24M^2}(4\beta_4^\perp-tA)F_3^*+\frac{P_ztA}{6}\left(1-\frac{1}{2}At\beta_4^\perp+2AP_z^2\beta_4^\perp\right)F_2^*\cr
    &\qquad+\frac{P_zt^2A^2\beta_4^\perp}{24M^2}\left(P_z^2-\frac{1}{4}t\right)F_4^*\Bigg]\cr
    G_{M0}^{3,a'a}(t;P_z)&=\frac{1}{2P_0}\Bigg[\frac{2}{3}P_z^3tA^2F_1^*+\frac{P_z^3t^2A^2}{12M^2}F_3^*-\frac{4}{3}P_z^3A^2t\beta_{8/3}^\perp F_2^*-\frac{P_z^3A^2t^2\beta_{8/3}^\perp}{6M^2}F_4^*\Bigg]\cr
    G_{M1}^\perp(t;P_z)&=\frac{1}{2P_0}\Bigg[\frac{2}{3}M\left(1-\frac{2}{5}\beta_8^\perp tA\right)F_1^*+\frac{\beta_4^{\perp2}t}{15M}F_3^*+\frac{1}{3}M\left(1-\frac{2}{5}tA\beta_8^\perp tA\right)(2\beta_2^z-P_z^2A)F_2^*\cr
    &\qquad+\frac{t\beta_4^{\perp2}t}{30M}(2\beta_2^z-P_z^2A)F_4^*\Bigg]\cr
    G_{M1}^3(t;P_z)&=\frac{1}{2P_0}\Bigg[\frac{2}{3}M\left(1-2P_z^2A\beta_4^\perp-\frac{2}{5}\beta_8^\perp tA\right)F_1^*-\frac{\beta_4^\perp t}{6M}\left(P_z^2A-\frac{2}{5}\beta_4^\perp\right)F_3^*\cr
    &\qquad+\frac{M}{3}\left\{P_z^2A\left(1-\frac{7}{5}A t\beta_{14/3}^\perp\right)+2\beta_2^z\left(1-\frac{2}{5}t  A\beta_8^\perp\right)\right\}F_2^*+\frac{\beta_4^\perp t}{30M}\left(P_z^2 A\beta_{2/3}^\perp+\beta_2^z\beta_4^\perp\right)F_4^*\cr
    G_{M1}^{\perp,a'a}(t;P_z)&=\frac{1}{2P_0}\Bigg[\frac{2}{3}MP_z^2A^2t F_1^*+\frac{1}{12M}P_z^2t^2A^2F_3^*+\frac{1}{3}MP_z^2tA^2(2\beta_2^z-P_z^2 A)F_2^*\cr
    &\qquad+\frac{1}{48M}P_z^2A^2t^2(2+2\beta_2^z-P_z^2A)F_4^*\Bigg]\cr
    G_{M1}^{3,a'a}(t;P_z)&=\frac{1}{2P_0}\Bigg[\frac{1}{3}MP_z^2A^2tF_1^*+\frac{1}{24M}P_z^2t^2A^2F_3^*+\frac{1}{6}MP_z^2A^2t(2\beta_2^z+P_z^2A)F_2^*\cr
    &\qquad+\frac{1}{48M}P_z^2t^2A^2(2\beta_2^z+P_z^2A)F_4^*\Bigg]\cr
    G_{M3}^\perp(t;P_z)&=\frac{1}{2P_0}\Bigg[-\frac{2}{3}M^3\beta_8^\perp A F_1^*+\frac{1}{6}M\beta_4^{\perp2}F_3^*-\frac{1}{3}M^3A\beta_8^\perp(2\beta_2^z-P_z^2 A)F_2^*+\frac{1}{12}M\beta^{\perp2}_4(2\beta_2^z-P_z^2A)F_4^*\Bigg]\cr
    G_{M3}^3(t;P_z)&=\frac{1}{2P_0}\Bigg[-\frac{2}{3}M^3\beta_8^\perp A F_1^*+\frac{1}{6}M\beta_4^{\perp2}F_3^*-\frac{1}{3}M^3A\beta_8^\perp(2\beta_2^z+P_z^2A)F_2^*+\frac{1}{12}M\beta_4^{\perp2}(2\beta_2^z+P_z^2A)F_4^*\Bigg]\cr
    G_{M2}^\perp(t;P_z)&=\frac{1}{2P_0}\Bigg[\frac{2}{3}M^2P_zA\beta_4^\perp F_1^*+\frac{1}{12}P_zAt\beta_4^\perp F_3^*+\frac{1}{3}P_zM^2A\beta_4^\perp(2\beta_2^z-P_z^2A)F_2^*\cr
    &\qquad+\frac{1}{24}P_ztA\beta_4^\perp(2\beta_2^z-P_z^2A)F_4^*\Bigg]\cr
    G_{M2}^3(t;P_z)&=\frac{3}{2P_0}\Bigg[-\frac{2}{3}M^2P_zA(\beta_4^\perp+\beta_8^\perp)F_1^*+\frac{1}{12}P_z\beta_4^\perp(2\beta_4^\perp-tA)F_3^*\cr
    &\qquad-\frac{1}{3}M^2P_zA\left(\frac{1}{2}At\beta_8^\perp+2\beta_2^z\beta_4^\perp+P_z^2A\beta_4^\perp\right)F_2^*+\frac{1}{24}P_ztA\beta_4^\perp(\beta_4^\perp-2\beta_2^z-P_z^2A)F_4^*\Bigg].
\end{align}

\end{document}